\documentclass[iop,twocolappendix]{emulateapj}
\usepackage{soul,color}
\usepackage{verbatim}
\usepackage{epstopdf}
\pdfoutput=1
\newcommand{\myemail}{bsalmon@physics.tamu.edu}
\newcommand{\lya}{Ly$\alpha$\ }
\newcommand{\Data}{$D$}
\newcommand{\Ha}{H$\alpha$\ }
\newcommand{\Hb}{H$\beta$\ }
\newcommand{\Muv}{\hbox{$\mathrm{M}_{1500}$}}
\newcommand{\Mstar}{\hbox{$\mathrm{M}_\star$}}
\newcommand{\SFRuv}{SFR$_{\text{UV}}$}

\newcommand{\nd}{number density}

\newcommand{\ks}{ISAAC $K_s$}

\newcommand{\msol}{\hbox{M$_\odot$}}
\newcommand{\Msol}{\hbox{M$_\odot$}}
\newcommand{\Msun}{\hbox{M$_\odot$}}
\def\ergcm2s{\ifmmode {\rm\,erg\,cm^{-2}\,s^{-1}}\else
                ${\rm\,ergs\,cm^{-2}\,s^{-1}}$\fi}

\newcommand{\referee}{}
\newcommand{\refereetwo}{}
\shorttitle{Relation Between SFR and Stellar Mass for Galaxies at $3.5\le \lowercase{z}\le6.5$ in CANDELS}
\shortauthors{Salmon et al.}

\submitted{Accepted for publication by ApJ on 11/25/14}

\begin{document}
\title{The Relation Between Star Formation Rate and Stellar Mass for Galaxies at $3.5\le \lowercase{z}\le6.5$ in CANDELS}
\author{Brett Salmon\altaffilmark{1,$\dagger$}, 
Casey Papovich\altaffilmark{1}, 
Steven L. Finkelstein\altaffilmark{2}, 
Vithal Tilvi\altaffilmark{1}, 
Kristian Finlator\altaffilmark{3}, 
Peter Behroozi\altaffilmark{4,5}
Tomas Dahlen\altaffilmark{5}, 
Romeel Dav{\'e}\altaffilmark{6,7,8,9}, 
Avishai Dekel\altaffilmark{10}
Mark Dickinson\altaffilmark{11}, 
Henry C. Ferguson\altaffilmark{5}, 
Mauro Giavalisco\altaffilmark{12}, 
James Long\altaffilmark{13},
Yu Lu\altaffilmark{4},
Bahram Mobasher\altaffilmark{14},
Naveen Reddy\altaffilmark{14,15},
Rachel S. Somerville\altaffilmark{16}, 
Risa H. Wechsler\altaffilmark{4}
}

\altaffiltext{1}{George P. and Cynthia W. Mitchell Institute for Fundamental Physics and Astronomy, Department of Physics and Astronomy 
Texas A\&M University, College Station, TX 77843, USA}
\altaffiltext{2}{Department of Astronomy, The University of Texas at Austin, Austin, TX 78712, USA}
\altaffiltext{3}{DARK fellow, Dark Cosmology Centre, Niels Bohr Institute, Copenhagen University, Juliane Maries Vej 30, DK-2100 Copenhagen O, Denmark}
\altaffiltext{4}{Physics Department, Stanford University; Particle Astrophysics, SLAC National Accelerator Laboratory; Kavli Institute for Particle Astrophysics and Cosmology Stanford, CA 94305 USA}
\altaffiltext{5}{Space Telescope Science Institute, Baltimore, MD, USA}
\altaffiltext{6}{Steward Observatory, University of Arizona, 933 North Cherry Avenue, Tucson, AZ 85721, USA}
\altaffiltext{7}{Physics Department, University of the Western Cape, 7535 Bellville, Cape Town, South Africa}
\altaffiltext{8}{South African Astronomical Observatories, Observatory, Cape Town 7925, South Africa}
\altaffiltext{9}{African Institute for Mathematical Sciences, Muizenberg, Cape Town 7945, South Africa}
\altaffiltext{10}{Racah Institute of Physics, The Hebrew University, Jerusalem 91904, Israel}
\altaffiltext{11}{National Optical Astronomy Observatories, Tucson, AZ, USA}
\altaffiltext{12}{Department of Astronomy, University of Massachusetts, Amherst, MA 01003, USA}
\altaffiltext{13}{Department of Statistics, Texas A\&M University, College Station, TX 77843-3143, USA}
\altaffiltext{14}{Department of Physics and Astronomy, University of California, Riverside, 900 University Avenue, Riverside, CA 92521, USA}
\altaffiltext{15}{Alfred P. Sloan Fellow}
\altaffiltext{16}{Department of Physics \& Astronomy, Rutgers University, 136 Frelinghuysen Road, Piscataway, NJ 08854, USA}
\altaffiltext{}{$\dagger\ $\myemail}
\begin{abstract}

Distant star-forming galaxies show a correlation between their
star formation rates (SFR) and stellar masses, and this has deep
implications for galaxy formation.   Here, we present a study on the
evolution of the slope and scatter of the SFR--stellar mass relation
for galaxies at $3.5\leq z\leq 6.5$ using multi-wavelength photometry
in GOODS-S from the Cosmic Assembly Near-infrared Deep Extragalactic
Legacy Survey (CANDELS) and \emph{Spitzer} Extended Deep Survey.    We
describe an updated, Bayesian spectral-energy distribution
fitting method that incorporates effects of nebular line emission,
star formation histories that are constant or rising with time, and
different dust attenuation prescriptions (starburst and Small Magellanic
Cloud).      
From $z$=6.5 to $z$=3.5
star-forming galaxies in CANDELS follow a nearly unevolving
correlation between stellar mass and SFR that follows SFR $\sim$
$M_\star^a$ with \mbox{\refereetwo{$a$}$  = 0.54 \pm 0.16$ at $z\sim 6$ and $0.70
\pm 0.21$} at $z\sim 4$.    This evolution requires a  star formation
history that increases with decreasing redshift (on average, the SFRs
of individual galaxies rise with time).   
The observed scatter in the SFR--stellar mass relation is tight, $\sigma(\log
\mathrm{SFR}/\mathrm{M}_\odot$ yr$^{-1})<  0.3\ - $ 0.4 dex, for
galaxies with $\log M_\star/\mathrm{M}_\odot > 9$ dex. 
Assuming that the SFR is tied to the net gas inflow
rate (SFR $\sim$ $\dot{M}_\mathrm{gas}$), then the scatter in the gas
inflow rate is also smaller than 0.3$-$0.4 dex for star-forming galaxies in
these stellar mass and redshift ranges, at least when averaged over
the timescale of star formation. We further show that the
implied star formation history of objects selected on the basis 
of their co-moving number densities is
consistent with the evolution in the SFR--stellar mass relation.

 \end{abstract}

\section{Introduction} \label{sec:intro}

Modern broadband photometric surveys (e.g. the Cosmic Assembly
Near-infrared  Deep Extragalactic Legacy Survey, \refereetwo{hereafter} CANDELS) now
routinely identify  thousands of galaxies at redshifts greater than
$z\sim4$  \citep[e.g.,][]{Dickinson98, Steidel99, Giavalisco02,
Stark13, Reddy12}.  Such projects are able to probe the high-redshift
\refereetwo{galaxy spectral energy distribution (SED) from the rest-frame UV to 
the optical for galaxies with redshifts out to ${z   >  7}$.} This
information allows us to characterize galaxies by their physical
properties such  as stellar mass (\Mstar) and star formation rate
(SFR) \citep{Sawicki98, Papovich01, Shapley01, Giavalisco02, Stark09,
Forster04, Drory04, Labbe06, Maraston10, Walcher11,
Curtis-Lake13, Lee12}. 

\referee{A correlation between the SFRs
and stellar masses of galaxies exposes interesting mechanisms of the
star formation history: a high scatter in this correlation implies} a 
stochastic star formation history with many discrete
``bursts'', while a tighter correlation implies a star formation history that traces
stellar mass growth more smoothly \citep{Noeske07, Daddi07, Renzini09,
Finlator11, Lee12}. The level of scatter between the SFR--stellar mass relation 
can be attributed to \referee{differences in the star formation histories of galaxies, which 
can be caused by} the variation in their gas accretion rates (SFR $\sim$ $\dot{M}_\mathrm{gas}$)
\refereetwo{and feedback effects,} 
assuming the timescale for gas to form stars is small \citep{Dutton10,Forbes13}. 

While the SFR--stellar mass relation has been well studied out to
$z\lesssim 2$ \citep{Daddi07, Noeske07, Dunne09, Oliver10, Rodighiero10},
\refereetwo{divergent} results have  been observed in the literature for higher
redshift ($z>2$) galaxies \refereetwo{\citep[see][for a detailed comparison of many recent
studies]{Speagle14}.}    Many studies have argued  that the
correlation is tight \citep{Daddi07, Pannella09, Magdis10, Lee11,
Sawicki12, Steinhardt14}, implying smooth gas accretion.  This agrees with results
from hydrodynamic simulations, which predict a tight relation between
SFR and stellar  mass \citep{Finlator06, Finlator07,
Finlator11, Neistein08, Dave08}, due in large part to  their consensus
that mergers are subdominant to galaxy growth at high redshift $z>2$
\citep{Murali02, Keres05} and the SFR tracks the gas
accretion rate \citep{Birnboim03, Katz03, Keres05, Dekel09, Bouche10,
Ceverino10, Faucher11}.   In contrast, other studies find no correlation or high
scatter in the SFR--stellar mass relation \citep{Shapley05, Reddy06, Mannucci09, Lee12, Wyithe13},
implying bursty star formation. As \refereetwo{suggested} by \cite{Lee12},
these differences may be physical or a result of systematics in the
data analysis. If the latter, then the differences likely arise from
biases in the methods of deriving stellar masses and SFRs 
or from inconsistent sample selections (i.e., UV color, \referee{stellar mass, 
flux, photometric redshift, or spectroscopic redshift selections).}  If physical, these
differences may be due \refereetwo{to stochasticity in the star formation history}
or a more complicated galaxy evolution that
changes with halo mass and rest-frame UV luminosity
\citep[see][]{Renzini09, Lee09, Wyithe13}.

\referee{Inferring stellar masses  and SFRs from
broadband photometry can be a convoluted process, and careful
attention to the methods of SED fitting 
could be the key to resolve
discrepant results in the SFR--stellar mass relation. }
Many studies have already recognized the sensitivity
of  the SED fitting process to assumptions on metallicity,
dust attenuation prescription, nebular emission,  and choice of initial mass
function \citep[IMF;][]{Papovich01, Zackrisson01, Zackrisson08, Wuyts07,
Conroy09b, Marchesini09, Ilbert10, Maraston10, Michalowski12,
Banerji13, Moustakas13, Schaerer13,  Stark13, Mitchell13, Buat14}.  In
particular, much attention has been given to varying the dust
attenuation prescription beyond the typically assumed ``starburst''-like
attenuation \citep{Calzetti00}.  For example, 
the starburst attenuation has been known to produce unphysically
young stellar  population ages for UV selected samples, 
with best--fit ages often at the edge of the parameter space 
\citep{Fontana04, Reddy12, Oesch13, Kriek13,
Chevallard13, Buat14}.

This work aims to address the discord in the results on the
scatter in the SFR and stellar mass relation, the
redshift evolution of  the SFR per unit stellar mass (the specific
SFR, sSFR),  and, in general, the nature of
the star formation history at high redshift. The new Bayesian fitting
method used in this work is able to recover stellar masses and SFRs  
of simulated galaxies with complex star formation
histories, while at the same time producing realistic distributions of
stellar population ages (as predicted by semi-analytic models). \referee{Thus, this
work shows there is an observed relation between SFR and stellar mass with low scatter
and an evolution of the sSFR that increases with redshift.}
Furthermore, the star formation history inferred from the
progenitor-to-descendant evolution of galaxies selected by their
co-moving number densities reproduces the observed SFR--mass relations
over the redshift range of this work.  This provides a 
self-consistent check on the derived star formation history.
%

This paper is outlined as follows. In \S~\ref{sec:data} we
describe the CANDELS survey data, sample selection, and the simulated
and mock catalogs from models used in this work.  In 
\S~\ref{sec:sed_methods}, we define our SED fitting assumptions,
including our choices of dust--attenuation prescription,  and we introduce our
method to include nebular line emission to stellar population
synthesis (SPS) models.   In \S~\ref{sec:PDF}  we discuss our
Bayesian method to  derive our stellar mass and SFR estimates
from the full posterior of each galaxy, marginalizing over other
nuisance parameters.  We show that the quantities derived by fitting
to synthetic photometry from models agree well with the true model values.
In \S~\ref{sec:SFRMassM1500} we show 
the inferred SFR--stellar mass relation at $z\sim$ 4, 5, and 6. We compute 
the slope and scatter in the SFR--mass relation, and we compare it to recent
theoretical simulations.  In \S~\ref{sec:discussion} we discuss the implications of the SFR--stellar mass 
relation, use an evolving number density to 
track the progenitor-to-descendant evolution within our sample,
and measure the redshift evolution of the sSFR.   Finally,
in \S~\ref{sec:conclusions} we summarize our
conclusions.  We also provide Appendices that support assuming a constant star formation history in the SED fitting process over histories
that exponentially rise (Appendix A), argue how results using best-fit parameters from the SED fits provide less reliable conclusions
due to best-fit results being (more strongly) affected by model assumptions (including nebular emission and dust attenuation, 
Appendix B), and outline how the adopted prior does not significantly influence the results of this work (Appendix C).
Throughout, we assume a \cite{Salpeter55} IMF. Switching Salpeter to a
\cite{Chabrier03} IMF would require reducing in \refereetwo{log scale} both the SFR and
stellar mass by 0.25 dex. Throughout, we assume a cosmology with
parameters, $H_0$ = 70 km s$^{-1}$ Mpc$^{-1}$, $\Omega_{\text{M,0}}$ =
0.3 and $\Lambda_0$ = 0.7. All magnitudes quoted here are measured
with respect to the AB system, $m_{\text{AB}}$ = 31.4 -- 2.5
$\log$($f_{\nu}/1$ nJy) \citep{Oke83}. 

\begin{figure}[t!]
\epsscale{1.1}
\centerline{\includegraphics[scale=0.4]{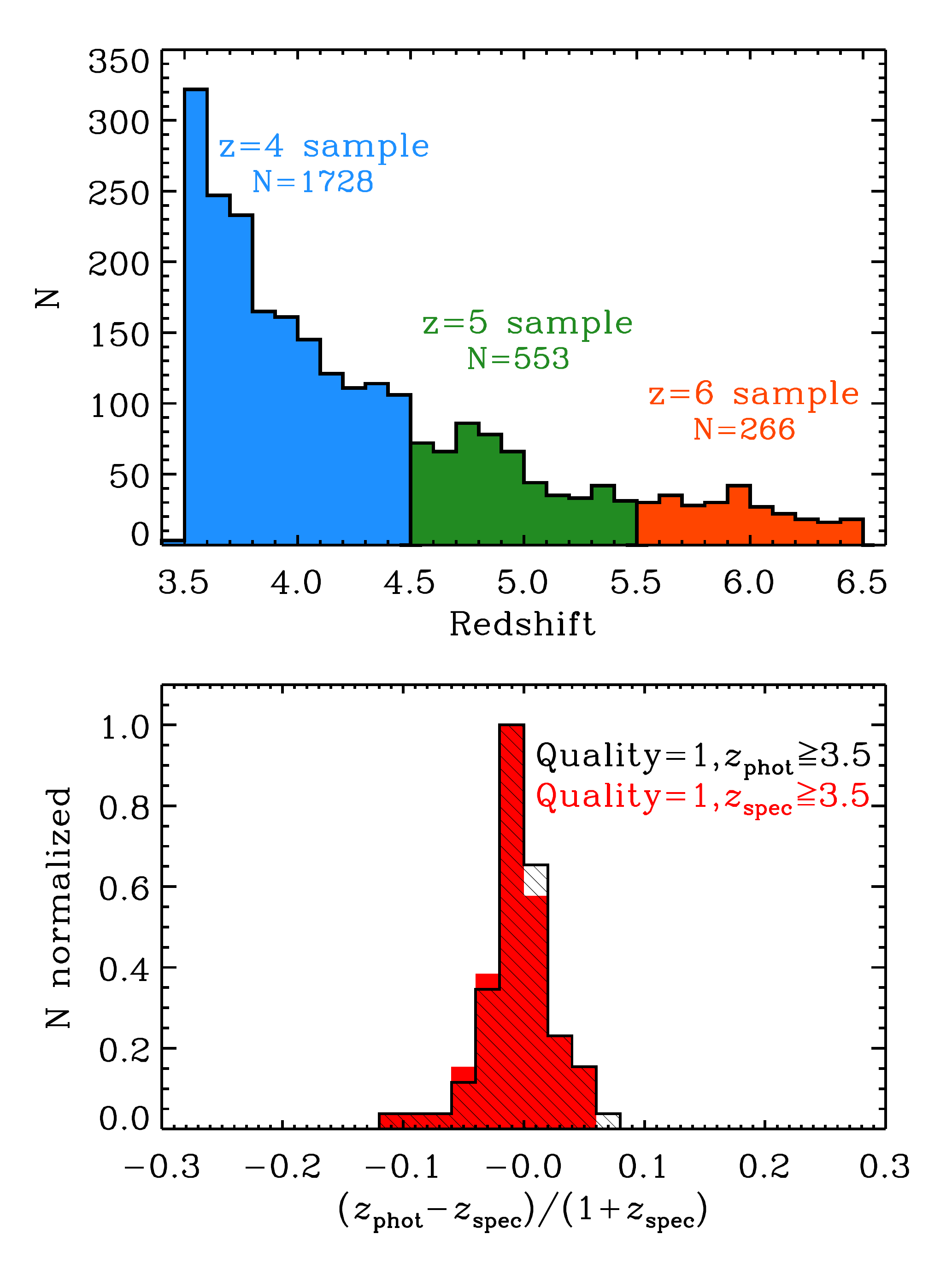}}
\caption{\emph{Top:} photometric redshift distributions for the objects 
used in this work. Throughout, the blue, green, and orange colors represent
 objects in our $z\sim4, 5,$ and 6 sample, respectively. 
 \emph{Bottom:} histograms of the photometric-redshift accuracy as 
 compared to known highest quality (quality=1) spectroscopic redshifts.  
 This figure shows that our $z_{\text{phot}}$ catalog well represents 
 the ``true'' best quality $z_{\text{spec}}$ objects.  Formally, the scatter 
 in the photometric redshift accuracy is approximately 
$\sigma_\mathrm{MAD}/(1+z) =$  0.016 at $z \sim 4$ to 0.028 at $z \sim 6$. 
\vspace{3mm}} 
\label{fig:redshifts}
\end{figure}

\section{\referee{Observational Data and Simulations}} \label{sec:data}

\subsection{CANDELS GOODS-S Multi-wavelength Data} \label{sec:candelsData}

This work uses multi-wavelength photometry from the CANDELS  GOODS-S
field \citep{Grogin11, Koekemoer11}. In addition to CANDELS, this work
includes the Early Release Science (ERS), Hubble Ultra Deep Field
(HUDF),  and deep IRAC imaging in all four IRAC channels
(3.6--8.0~$\mu$m) from  the \emph{Spitzer} Extended Deep Survey
\citep{Ashby13} programs.     Throughout, we denote magnitudes
measured by \emph{HST} passbands  with the ACS F435W, F606W, F775W,
F814W and F850LP as $B_{435}$, $V_{606}$,  $i_{775}$, $I_{814}$, and
$z_{850}$,  and with the WFC3 F098M, F105W, F125W,  F140W, and F160W,
as $Y_{098}$, $Y_{105}$, $J_{125}$, $JH_{140}$, and  $H_{160}$,
respectively.   Similarly, bandpasses acquired from ground-based
observations include the  CTIO/MOSAIC U-band; VLT/VIMOS U-band; the
VLT/ISAAC $K_s$;  and VLT/HAWK-I $K_s$.


We use \referee{fluxes from the catalog} constructed by \cite{Guo13}.  Guo
et al. selected objects via SExtractor in  dual-image mode with
$H$-band as the detection image.  As described in Guo et al., two
versions  of the catalog were constructed using SExtrator parameters
that were (1)  optimized  in detection threshold and object deblending
to identify  faint, small galaxies (the ``hot'' catalog) and (2)
optimized  to keep large, resolved galaxies from being subdivided into
multiple objects  (the ``cold'' catalog).   Both catalogs are then
merged whereby any object  in the ``hot'' catalog that falls within the
isophote of a galaxy in the ``cold''  catalog is removed in favor of
the ``cold''-catalog object. 

The \emph{HST} bands were point spread function (PSF)-matched and the
photometry is measured  on the \emph{HST} bands using the SExtractor
double-image mode described  above.  For the ground-based and IRAC
bands, the catalog uses  TFIT \citep{Laidler07} to measure photometry
of these  lower-resolution  images using the \emph{HST} WFC3 imaging as a
high-resolution template for the galaxies. We use the \referee{final} version 
of the GOODS-S TFIT catalog which includes the new $I_{814}$ (CANDELS) and
$JH_{140}$ (HUDF12)  photometry. 

In addition to the flux densities and
uncertainties  provided in this catalog, we include an additional
uncertainty,  \referee{defined to be 10\% of the flux density of each object in
band.    This additional uncertainty accounts for any systematic
uncertainty  that may be related to the source fluxes themselves. 
This includes,  for example, flat-field variations, PSF and aperture
mismatching, and local background subtraction, many of which will be
(to first order) proportional to the flux itself. The value of
10\% was chosen such that the distribution of reduced $\chi^2$ is $\geq 1$,
and is justified based on arguments in \cite{Papovich01}.}
 We add this uncertainty  in quadrature to the measured uncertainties to estimate a
total  uncertainty on the flux density in each band for each object.

\subsection{CANDELS GOODS-S Redshifts} \label{sec:selection}

We \refereetwo{use results from} the recent CANDELS GOODS-S photometric-redshift 
project \citep{Dahlen13} which  we briefly summarize here. A team of eleven 
investigators tested their individual photometric redshift fitting
codes on blind control samples provided by the CANDELS team. A hierarchical
Bayesian approach was then performed to combine the seven investigators' 
individual $P(z$) distributions to a final $P(z$) distribution for each object. 
The photometric-redshift ($z_{\text{phot}}$) is thereafter derived as the weighted mean of this 
distribution. Another sample was constructed as the median $z_{\text{phot}}$ 
of all eleven \refereetwo{individual results.} The $z_{\text{phot}}$ distributions 
from the medians and the combined $P(z$) methods both retained a lower 
scatter and outlier fraction than the results of any single investigator. Tests 
by \cite{Dahlen13} showed that the hierarchical Bayesian $z_{\text{phot}}$ 
method produces the best (smallest) scatter between the $z_{\text{phot}}$ 
and spectroscopic redshifts. Finally, these methods were applied to the 
same CANDELS TFIT catalog \citep{Guo13} 
from which our data were obtained. 

Figure~\ref{fig:redshifts} compares redshifts from the combined $P(z$) 
method with their highest-quality spectroscopic counterparts. The top 
panel exhibits a histogram of the number of objects used in our samples 
as a function of their photometric redshift. The bottom panel shows the 
ability of the photometric redshifts to recover known spectroscopic redshifts
in the redshift range of this work. 

\referee{Unless otherwise specified, we use the median absolute deviation (MAD) to compute
the equivalent standard deviation,  $\sigma_\mathrm{MAD}$, as the measure of
scatter in given quantities \citep{Beers90}, including the quoted scatter for \refereetwo{redshift,}
stellar mass, and SFR.}  The
$\sigma_\mathrm{MAD}$ is an analog for the 68\% confidence, $\sigma$, if the
error distribution were Gaussian and is therefore less sensitive to
outliers  \citep[see][]{Brammer08}. The MAD standard deviation in the photometric
redshift  accuracy ranges from $\sigma_\mathrm{MAD}/(1+z)$ = 0.016 at $z \sim
4$ to  0.028 at $z \sim 6$, indicating that these photometric
redshifts reliably  recover known spectroscopic redshifts at high
redshift. 

\refereetwo{Even} in the highest quality spectroscopic redshift 
sample, there is a non-zero chance that some objects will  have a 
misidentified $z_{\text{spec}}$ due to a misinterpreted emission line or 
Lyman break \citep[see discussion in][]{Dahlen13}. 
So it is likely that some outliers are actually due to a misidentified 
$z_{\text{spec}}$ rather than a poorly fit $z_{\text{phot}}$ fit. \refereetwo{The} 
number of outliers where $| z_{\text{spec}}- z_{\text{phot}}|/(1 + z_{\text{spec}})>$ 
0.1 are 2, 1, and 1 for $z\sim$ 4, 5, and 6, respectively (only 5, 5, and 11\% of each sample). For the 
remainder of this work we use the $z_{\text{phot}}$ catalog derived from 
the combined $P(z$) method, and substitute for high-quality 
($z_{\text{qual}} = 1$) spectroscopic redshifts when available.

\subsection{Sample Selection} \label{sec:redshift}
We selected objects according to their photometric-redshift 
($3.5\le z_{\text{phot}} \le 6.5$). This redshift range was chosen to be close 
to the redshift range of the traditional $B$, $V$, and $i$ -drop samples. \referee{The lower redshift bound was chosen
to avoid higher photometric redshift uncertainties, which may be due to a weaker Lyman break signal at $z<3.7$ \citep[see][their Fig. 11]{Dahlen13}.} Our samples have been cleaned from a total 
of 46 objects from X-Ray \citep{Xue11}, IR \citep{Donley12}, 
and radio \citep{Padovani11} detected AGN, as flagged by the \cite{Dahlen13} photo-$z$ catalog. 

Objects with a best-fit SED with $\chi^2 > 50$ are omitted from all samples. 
This cut removes objects with particularly poor fits, which comprise 
less than 4\% of all objects. We interpret these objects as poor
detections that do not well represent the data, and note that the removal
 of these objects do not impact the results of this work. 
The final sample includes 1728 objects with $3.5 < z < 4.5$,
 553 objects with $4.5 < z < 5.5$, and 266 objects with 
$5.5 < z < 6.5$, as illustrated in Figure~\ref{fig:redshifts}. 
We refer to these as the $z\sim  4$, 5, and 6 samples, 
respectively.

\subsection{Galaxy Photometry from Models and Simulations} \label{sec:describeModels}

This work takes advantage of recent mock catalogs with
synthetic photometry for galaxies from semi-analytic models (SAMs),
as well as a semi-empirical dark matter \refereetwo{and hydrodynamic} simulation. 
We collectively refer to these as ``the models''.  The benefit of comparing our 
derived results against these model galaxies is that the models incorporate
realistic star formation histories and galaxy physics.  Here we use
these models for two comparisons.  First, in \S~\ref{sec:SAMcompareSEC} we
derive stellar population parameters (SFRs and stellar masses) from
the synthetic photometry for the model galaxies and compare to their
``true'' values as a test of our SED fitting procedures. Second, in
\S~\ref{sec:SFRmassSAMs} we use our derived SFRs and stellar masses
from the CANDELS samples to compare to the models and
interpret the SFR--mass relation and its scatter. 

\subsubsection{SAMs of Somerville et al. and Lu et al.}\label{sec:SomervilleYu}

This work uses the results of two SAMs that were specifically 
designed for the CANDELS GOODS-S field 
\citep[][\refereetwo{hereafter referred to as Somerville et al. and 
Lu et al. respectively}]{Somerville08, Somerville12, Lu13a, Lu13b}, 
which we summarize here. Areas where the two SAMs differ are highlighted
to emphasize the assumptions that lead to different SFR and stellar mass results.
\referee{A more detailed comparison of the Somerville et al. and Lu et al.  models can be found
in \cite{Lu13b}.}

The mock catalogs produced by
the SAMs are based on the Bolshoi $N$-body simulation 
\citep{Klypin11} for the same field-of-view size and geometry as the 
CANDELS \refereetwo{GOODS-S} field. The two SAMs are applied on the halo merger trees for 
halos in the mock catalogs. The models adopted a cosmology favored by 
\emph{WMAP7} data \citep{Jarosik11} and \emph{WMAP5} data 
\citep{Dunkley09, Komatsu09} with $\Lambda$CDM 
cosmology.
The mass resolution of the simulation is 
$1.35\times10^8 h^{-1}\msol$, which allows the SAMs to track halos 
and subhalos with mass $\sim 2.70\times 10^9  h^{-1}$ \msol. 

The SAMs make explicit predictions for gas cooling rates, 
star formation, outflows induced by star formation feedback, and 
galaxy-galaxy mergers for every galaxy in the mock catalog. Both 
models assume that gas follows dark matter to collapse into a 
dark matter halo. When the gas collapses into the virial radius 
of the halo, it is heated by accretion shocks and forms a hot 
gaseous halo that cools radiatively\referee{. If} the 
cooling timescale is longer than the halo dynamical time, both 
models follow the treatment that the halo gas cools gradually 
and settles on a central disk. The central disk of cold gas in both 
models is assumed to have an exponential radial profile, where 
stars form in regions where the surface density of the cold gas 
is higher than a threshold. 

In the Somerville model, the SFR is predicted 
based on the cold gas surface density using the Schmidt--Kennicutt 
law \citep{Kennicutt98} explicitly. In the Lu et al. model, the star formation 
efficiency is assumed to be proportional to the total cold gas mass for 
star formation and inversely proportional to the dynamical time-scale 
of the disc, with an overall efficiency that matches observations \citep{Lu13a}.

Star formation feedback is assumed in both models, 
but the implementations are slightly different. Both models assume that the 
feedback reheats a fraction of the cold gas in the galaxy and a fraction 
of the reheated gas leaves the host halo in a strong outflow. However, the Lu et al.
model allows a fraction of the kinetic energy of supernovae (SN) to drive an additional 
outflow to expel a fraction of hot halo gas. Nevertheless, the mass loading 
of the outflow in both models is assumed to be proportional to the 
SFR, and inversely proportional to a certain power of the halo maximum 
velocity. In the Somerville model,  the mass-loading factor is assumed 
to be inversely proportional to the second power of the halo maximum velocity, 
mimicking the so-called ``energy driven wind.'' In the Lu et al. model, a much 
stronger power law of the halo circular velocity dependence is adopted. 
Both models assume a fraction of the ejected baryonic mass comes back 
to the halo as hot halo gas on a dynamical time-scale with different efficiencies. 

The model parameters governing Star formation 
and feedback are tuned to match the local galaxy stellar mass function 
\citep{Moustakas13}. The Lu et al. model is tuned using a 
\refereetwo{Markov Chain Monte Carlo (MCMC)} algorithm 
to find plausible models in the parameter space. The model precisely 
reproduces the local galaxy stellar mass function between $10^9$ and 
$10^{12}$ \msol, within the observational uncertainty. The Somerville
model is further tuned based on a previously published model 
\citep{Somerville08} against the new data. 
In spite of different parameterizations adopted by the
two models, they yield qualitatively similar predictions for the
assembly histories of galaxy stellar mass and SFR over
cosmic time. 

\subsubsection{Semi-empirical Matching of Observed Galaxies to Dark Matter Halos of Behroozi et al.}\label{sec:Behroozi}
The semi-empirical model employed by \cite[][BWC13 hereafter]{Behroozi13d} uses 
a flexible fitting formula for the evolution of the stellar mass--halo mass relation 
with redshift, ($SM(M_h,z)$; \refereetwo{see BWC13 for further definition).}  
This formula includes parameters for the characteristic stellar and halo 
masses, faint-end slope, massive-end shape, and scatter in stellar mass at fixed 
halo mass, as well as the redshift evolution of these quantities.  Given halos 
from a dark matter simulation, each point in the $SM(M_h,z)$ function parameter 
space represents an assignment of galaxy stellar masses to every halo at every 
redshift; the simulation and halo catalogs used by BWC13 are detailed by 
\cite{Klypin11} and \cite{Behroozi13a, Behroozi13b}.  The abundance of halos 
as a function of redshift can then be used to calculate the implied stellar 
mass function; the buildup of stellar mass over time in halos' main progenitor 
branches can be used to calculate implied galaxy SFRs.  
BWC13 compares these predicted observables to published results from 
$z$ = 0 to $z$ = 8 and employs an \refereetwo{MCMC} algorithm 
to determine both the posterior distribution for $SM(M_h,z)$ and the implied 
$SFR(M_h,z)$.  The resulting best-fits are consistent with all recent published 
observational results in this redshift range, including galaxy stellar mass 
functions, cosmic SFR, and sSFRs.  
Full details, including comparisons with other techniques for deriving 
the stellar mass--halo mass relation, are presented by BWC13.

\subsubsection{Hydrodynamic Simulation of Dav\'e et al.}\label{sec:Dave}
This simulation was run with an extended version of the cosmological 
smooth particle hydrodynamic
code Gadget-2 \citep{Springel05} described by \cite{Oppenheimer08}.
The simulation includes metal cooling and heating following \cite{Wiersma09}, 
star formation and a multi-phase interstellar medium model following \cite{Springel03}, 
and galactic outflows assuming momentum-driven wind scalings which 
have been shown to be crucial for providing a reasonable match to a variety 
of intergalactic medium (IGM) and galaxy properties from $z\sim 0-4$ and beyond \citep[see][]{Dave11a, Dave11b}.  
The simulation employs a \emph{WMAP-7} concordant cosmology within a 
co-moving cube of length 48 Mpc/h per side with $2\times 384^3$ particles 
and 2.5 kpc/$h$ (co-moving) resolution. Mass growth in 
galaxies is resolved down to stellar masses of approximately 
$10^9  \mathrm{M}_\odot$. See \cite{Dave13} for a full description.

\begin{table*}[t!]
\caption{SED Fitting Parameters} 
\centering  
\begin{tabular}{c  c  c  l} 
\hline                        
Parameter & Quantity & Prior & Relevant Sections \\ [0.5ex] 
\hline                  
Redshift & fixed & photometric redshifts, 3.5 $ \leq z  \leq $ 6.5 & \S~\ref{sec:selection} \\
Age & 74 & see equation \ref{equ:prior} [$\log$, 10 Myr - $t_\mathrm{max}$] 
              \footnote{The lower end of this range represents the
                minimum dynamical time of galaxies in our redshift
                range up to $t_\mathrm{max}$, which is the age of the
                Universe for the redshift of each object.}  & \S~\ref{sec:sed_methods}, Appendix C \\ 
Metallicity & fixed & 20\%\ Z$_\Sun\ (Z  =  4\times10^{-3})$ & \S~\ref{sec:sed_methods} \\
Star formation history
              \footnote{Star formation history is defined as $\Psi(t)
                = \Psi_0 \exp(-t/\tau)$ such that an SFR
                that increases with cosmic time has a negative
                $\tau$.   To ensure that the constant star formation models
are treated the same way as our $\tau$ models in the BC03 software, we
approximate a constant star formation history as having a very long $e$-folding time, $\tau \sim 100$ Gyr.}
              & fixed & 100 Gyr (constant) & \S~\ref{sec:sfh}, Appendix A \\ 
$f_{\text{esc}}$ & fixed & 0 or 1 & \S~\ref{sec:nebular}, Appendix B \\
$E(B-V)$ \footnote{We fit to a range of selective extinctions, $E(B-V)$,
  but throughout this work we primarily refer to the total extinction,
  $A_V = R_V \cdot\ E(B-V)$, where $R_V$ is the total-to-selective
  extinction ratio determined by the attenuation prescription. The
  attenuation enters the modeling as $A(\lambda) = k(\lambda) E(B-V)$
  where $k(\lambda) = E(\lambda - V)/E(B-V)$ is the attenuation prescription for each model. Where applicable, we refer to the attenuation at 1500~\AA\ as A$_\mathrm{UV}$.} & 29 & see equation \ref{equ:prior} [Linear,  0.0 - 0.7]  & \S~\ref{sec:SMC}, Appendix C \\
Attenuation prescription & fixed & starburst (Calzetti et al.\ 2000) or SMC
(Pei et al.\ 1992) & \S~\ref{sec:SMC}, Appendix B \\
Stellar Mass & -- & \Mstar $>0$, see equation \ref{equ:prior} & Appendix C \\ [1ex]  
\hline 
\end{tabular}
\end{table*} \vspace{3mm}

\section{Stellar population synthesis fitting: Methods} \label{sec:sed_methods}

This section describes the methods and assumptions used by our SED fitting procedure
to derive physical quantities. This work uses a 
custom fitting procedure, using an updated version of the methods 
described by \cite{Papovich01, Papovich06}. 

We utilize the Bruzual \& Charlot (2011, private communication)
SPS models, which are created with an updated
version of the \cite{Bruzual03} source code (BC03 hereafter) modified
to accept rising star formation histories, $\Psi \sim \exp(+t/\tau)$, 
where $\tau$ is the $e$-folding timescale.
 We opt to use the
libraries included with the BC03 models, as recent results have suggested the
alternative 2007 libraries  (similar TP-AGB contribution as
\cite{Maraston05})  overestimate the contribution from TP-AGB
stars in the near infrared (NIR), and the original BC03 version 
\refereetwo{is likely to be more realistic} \citep{Kriek10, Conroy10, Melbourne12,
Zibetti13}. Therefore, the remainder of this work uses the BC03
models.  As mentioned above, we use a \cite{Salpeter55} IMF
throughout which ranges in mass from 0.1--100 \Msol.

Although we include the effects of \ion{H}{1} absorption  from IGM
clouds along the line-of-sight to each galaxy (using  the prescription
of \cite{Meiksin06}), the true contribution of \ion{H}{1} clouds to
each galaxy will be  highly stochastic.  Therefore, we only include
bands with  wavelengths red-ward of the observed wavelength of
Lyman-$\alpha$, given the galaxy redshift in our SED modeling.  The
redshift is fixed to the photometric redshift (or spectroscopic if
available; see \S~\ref{sec:selection}), so fitting to bands
blue-ward of Lyman-alpha offers no improvement in determining
redshift.

Table~1 shows a list of the explored parameter space, as well as the
degree to which each parameter is explored. The
metallicity of all objects is fixed as 20\% solar metallicity, partly
due to a lack of confidence to accurately fit to this parameter given
the degeneracies between fits to age and attenuation. The choice of
20\% $Z_\Sun$ is supported by recent work that suggests the
metallicity of high-redshift ($z>2$) galaxies is low 
\citep[][see also \cite{Mitchell13} for a discussion on the
effects of metallicity in SED fitting]{Erb06a, Maiolino08, Erb10, Finkelstein11,
Finkelstein12a, Song14}. 

Objects are fit to all available ages between 10 Myr and the age  of
the Universe at the redshift of the object, which is at maximum  1.8
Gyr at $z  =  3.5$. \refereetwo{The age resolution of the BC03 models is quasi-logarithmic,
with an average log difference in age steps of $\Delta t_{\text{age}}$/yr = 0.02 dex.} 
We adopt a lower limit on the stellar population age
of 10 Myr  in order to avoid galaxies with ages younger than  the
minimum dynamical timescale of a galaxy at our  specified redshifts
\citep{Papovich01, Wuyts09, Wuyts11}.  In practice, we find that this
minimum  age has no impact on the fully marginalized parameter
distributions. 

\begin{figure*}[t]
\epsscale{1.1}
\centerline{\includegraphics[scale=0.43]{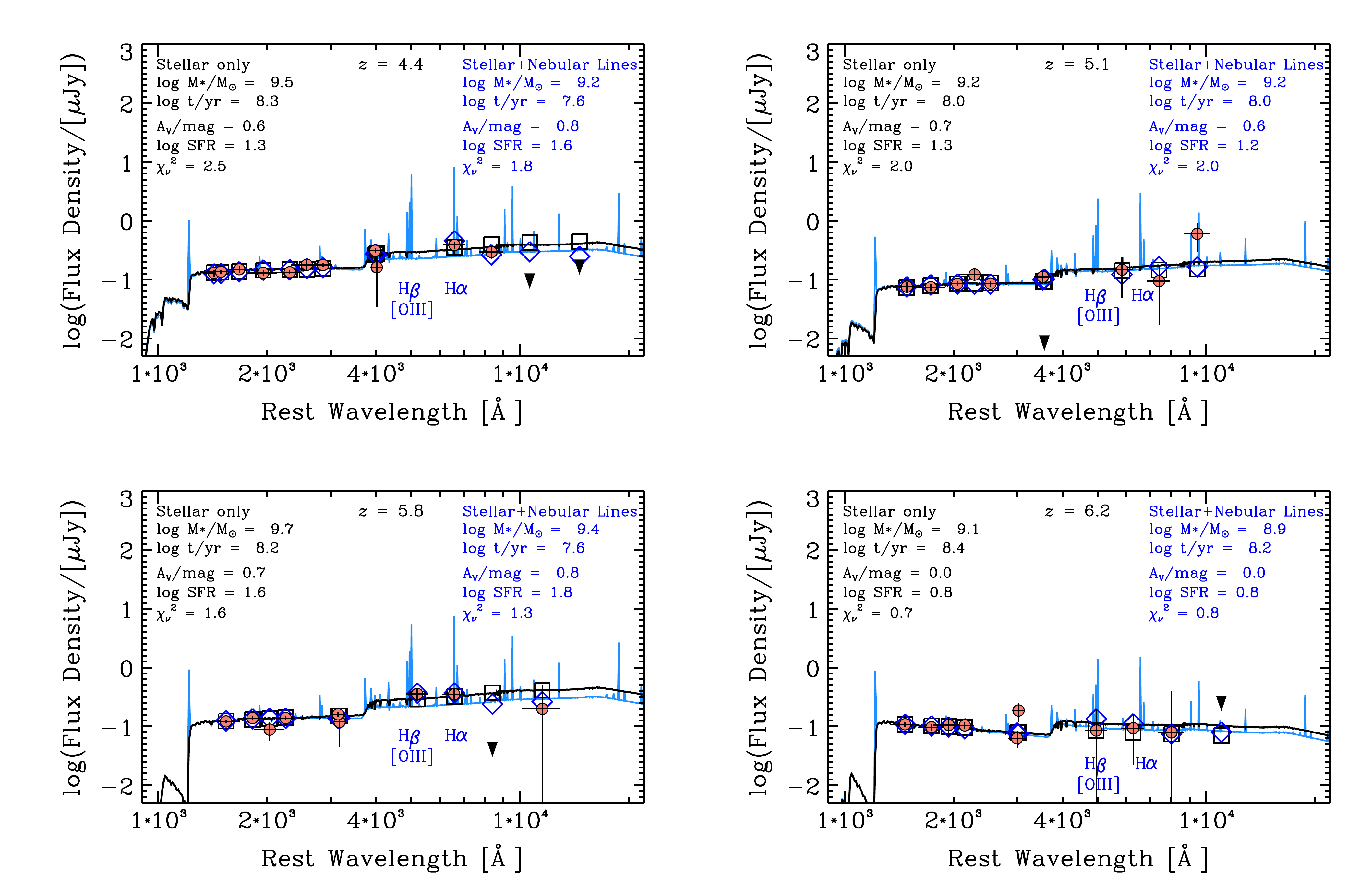}}
\caption{Four example galaxies from our sample with SED fits that do 
include nebular emission lines (blue curves) and do not include emission 
lines (black curves). Circles are the observed photometry and diamonds 
(squares) are the fluxes of the best-fit SED with (without) emission lines. 
The legends indicate the parameters of the best-fit model for both the
case where the nebular emission is excluded and included, as labeled. All objects were fit assuming a 
constant star formation history and starburst-like dust 
attenuation. At certain redshifts (including the objects with $z  =  4.1$,  
5.8, and 6.2), the IRAC 3.6~\micron\ and 4.5~\micron\ bandpasses may 
be enhanced by H$\beta$, [\ion{O}{3}], or H$\alpha$ emission lines, as indicated. 
In contrast, the IRAC bands for the object with $z  =  5.1$  do not include 
of these prominent emission lines.\vspace{3mm}}
\label{fig:examples}
\end{figure*}

\subsection{Star Formation History}\label{sec:sfh}
One of the aims of this paper is to constrain the star formation
history  of the average population of galaxies at high redshift,
$z>3.5$. Previous  works have shown that broadband SED fitting offers
no statistical preference between  constant, rising, or declining
star formation histories, even with broadband coverage  spanning to
the IR \citep{Reddy12}. Furthermore, the star formation histories
assumed in the templates can have non-negligible effects on the
inferred SFRs, stellar masses, and ages
\citep{JLee10}. \refereetwo{We addressed} the shape of the
star formation history as constrained  by individual galaxies by
running three separate fits using templates that  assumed constant,
declining and rising star formation history. The rising  and declining
star formation history templates (``$\tau$'' models) included a suite
of $e$-folding times. We ultimately found no obvious $\chi^2$
preference on the shape of the star formation history for individual
galaxies. 

Qualitatively, there has been some evidence to reject high-redshift star formation
histories that decline with time. Previous studies have shown that
high-redshift declining  star formation histories would under-predict the sSFR 
at lower redshifts \citep{Stark09,Gonzalez10,Maraston10}. In addition, the 
instantaneous SFRs derived when assuming a declining star formation history 
will be under-produced  by a factor of 5-10 as compared to direct estimates
based off of  UV-to-mid-infrared emission \citep{Reddy12}.  Other
evidence against declining star formation histories comes
independently from the SFR
evolution of UV-luminous galaxies selected at fixed number density
\citep{Papovich11}.  Finally,
\cite{Pacifici12}  introduced a state-of-the-art SED fitting procedure
with realistic,  hierarchical mass-assembly histories and showed 
that declining $\tau$-model histories do not
well represent galaxies  \refereetwo{even at ${z  <   2.}$}

Although galaxies at $z > 3$ likely have star formation histories that
increase monotonically with time, we found it was impractical 
to use such models as the derived results are less physical.   Our full
justification for fitting individual galaxies with a constant
star formation history is provided in Appendix A.  \refereetwo{Briefly,} 
the BC03 stellar populations currently only allow for
star formation histories that rise \textit{exponentially} with time
using simple parameterizations.  At late times, such histories increase
their SFR much faster than supported by observations.  
In Appendix A, we show that our modeling of synthetic
photometry for galaxies from semi-analytic models recovers the most
accurate stellar masses and SFRs when we adopt \emph{constant} star formation
histories.  We interpret this as due to the fact that the
SFRs in the models are approximately constant over the
past $\sim$100 Myr \citep[see, e.g.,][]{Finlator06}, and not consistent
with exponentially increasing SFRs.  Therefore, we fix the fitting
templates to have a constant star formation history in our analysis of the CANDELS
data for the remainder of this work.  In a future work, we will
explore possible improvements in parameters using models with
star formation histories that increase as a power law in time ($\Psi
\sim t^\gamma$).


\subsection{Nebular Emission}\label{sec:nebular}
This section presents our method of incorporating nebular emission. 
\referee{Nebular emission is important because} many galaxies at high redshift are 
observed to have intense star formation and high equivalent widths (EW) from 
emission lines \citep{Erb10,Vanderwel11, Atek11,Brammer13}. 
Such strong nebular emission is able to enhance broadband 
flux by up to a factor of $\sim 2 - 3$ in IRAC 3.6 and 4.5~\micron\ bands 
\citep{Shim11}. \refereetwo{Previous} studies have shown that the flux excess from 
high EW emission lines causes a systematic decrement in stellar mass 
and SFR inferred from SED fitting \citep{Schaerer09,Schaerer10,
Ono10, Finkelstein11, deBarros14, Reddy12, Stark13}. 

\begin{figure}
\epsscale{1.1}
\centerline{\includegraphics[scale=0.43]{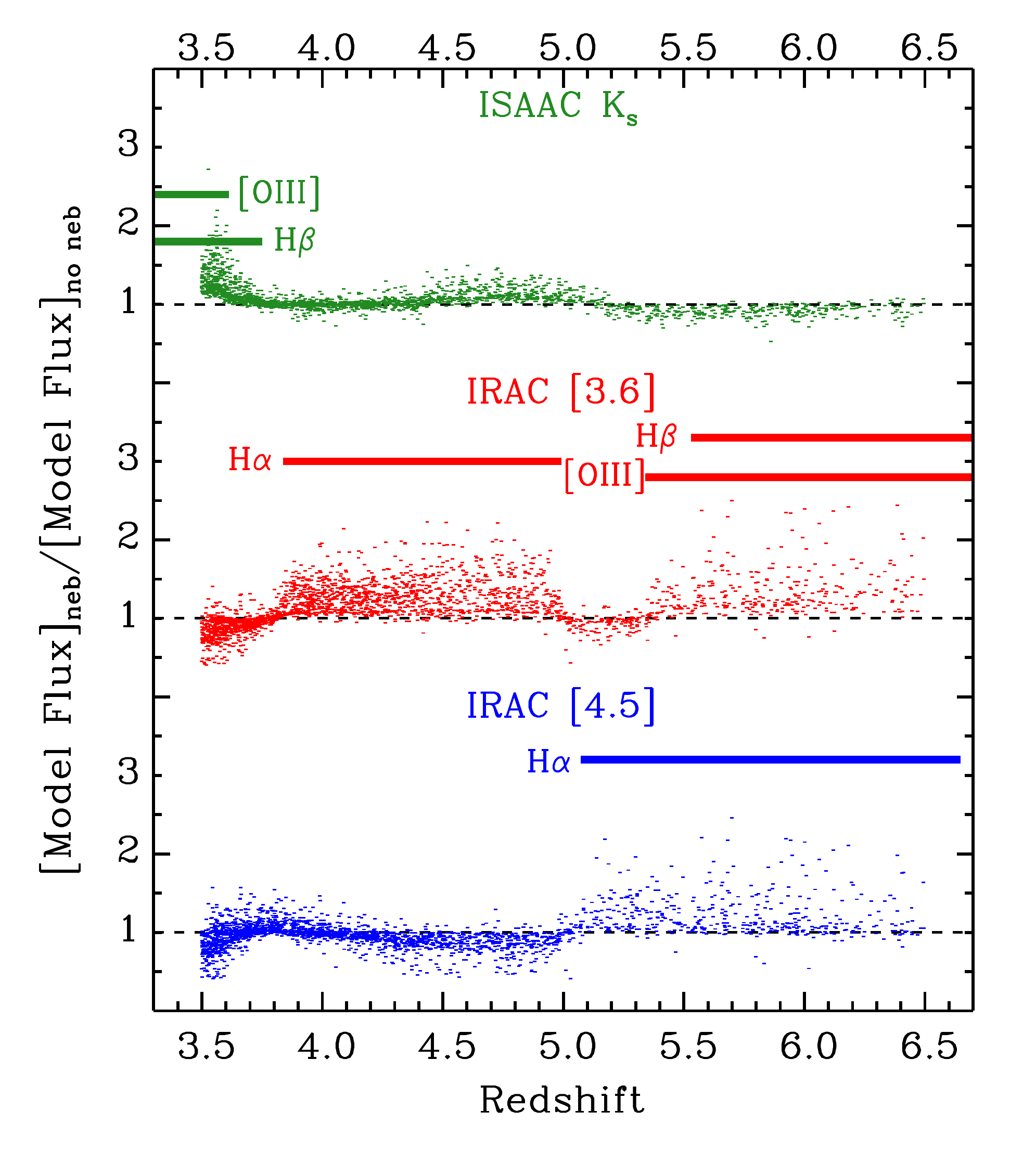}}
\caption{Best-fit SED model fluxes with and without emission lines are 
shown as ratios for ISAAC K$_{\text{s}}$, IRAC 3.6, and IRAC 4.5 
bands as a function of redshift. Horizontal lines describe the redshift 
at which a strong emission is in the bandpass.   The effect of adding 
emission lines is an increase to the model fluxes by as much as a factor 
of 2--3, especially in case of strong emission lines, such as [\ion{O}{3}] 
or H$\alpha$.  \vspace{3mm}}
\label{fig:photbyredshift}
\end{figure}

\subsubsection{Nebular Lines}\label{sec:neblines}
The strength of a given emission line is dependent on the properties 
of both the gas cloud and the incident ionizing source. These properties 
include metallicity, ionization parameter, electron density, and number 
of ionizing photons. \cite{Inoue11} explored these parameters and the 
resulting strength of nebular emission in the regime of high-redshift 
galaxies by utilizing 
\footnotesize CLOUDY \small 08.00\normalsize \citep{Ferland98}, 
which we use in our incorporation method.

After modeling a wide parameter space of seven metallicities, five
ionization parameters and five Hydrogen densities, \cite{Inoue11}
reports 119 sets of metallicity-dependent emission line strength
relative to H$\beta$. These line ratios, ranging from 1216~\AA\ to
1~$\micron$, are in close agreement with empirical metal line ratios
\citep{Anders03, Maiolino08}. We use the \cite{Inoue11} line ratios and include
Paschen and Bracket series lines from \cite{Osterbrock06} and
\cite{Storey95}. Following \cite{Inoue11}, we relate the
H$\beta$ line luminosity to the incident number of Lyman-continuum
photons as
\begin{equation}
\label{equ:neb}
 \text{L}_{\text{H}\beta}\ =\ 4.78 \times 10^{-13} \frac{1-f_{\mathrm{esc}}}{1+0.6 f_{\text{esc}}} \text{N}_{\text{LyC}}\ [\text{erg s}^{-1}],
\end{equation}
where $f_{\text{esc}}$ is the fraction of ionizing Lyman continuum
(LyC) photons escaping the galaxy into the IGM and
$\text{N}_{\text{LyC}}$ is the production rate of Hydrogen-ionizing
photons \citep[see also][]{Osterbrock06, Ono10}.  \referee{The number 
of ionizing Lyman-continuum photons, $N_\mathrm{LyC}$,
depends on the age of the stellar population, and we take $N_\mathrm{LyC}$
 from each BC03 SPS model for each age.}  It follows that
$1-f_\mathrm{esc}$ is the fraction of LyC photons that ionize gas
within the galaxy, which then produce the emission lines. The
additional factor in the denominator of equation \ref{equ:neb} comes
from a ratio of recombination coefficients \citep[see][]{Inoue11}.
Here, we equate the metallicity of the nebular gas to the metallicity
of the SPS template (set as $Z = 20\%\ Z_\Sun$ for all models, see
above).  Following the results
of \cite{Erb06b}, we attenuate both nebular and stellar emission in
the same manner (see \S~\ref{sec:SMC} for details on
attenuation).

The escape fraction has been measured 
to be low, i.e., $f_{\text{esc}} \approx 0$ at low redshift $z \sim $1
\citep[see][]{Malkan03,Siana07,Siana10,Bridge10}.  At $z\gtrsim 4$,
the IGM  imparts a large optical depth to ionizing photons, making it
difficult to constrain $f_{\text{esc}}$. \cite{Nestor11} used $z\sim
3$ Lyman Break Galaxies to study the high-redshift escape fraction, finding it to be
consistent with $f_{\text{esc}} \approx 0.1$. 
\cite{Finkelstein12b} concluded that if galaxies are
the main contributors to reionization, then the escape fraction must
be $f_{\text{esc}} < 0.34$, or $f_{\text{esc}} < 0.13$ (2$\sigma$) at $z\sim 6$ if the
luminosity function extends to fainter galaxies than those observed, 
in order for the inferred ionization from galaxies to be consistent 
with the ionization background inferred from quasar spectra 
\citep{Bolton07}. In addition, \cite{Jones13} reinforced this claim 
by finding the covering fraction of neutral hydrogen in $z \sim$4 galaxies 
to be lower by 25\% compared to $z \sim$2-3.  From
these results, it seems reasonable to assume a low, but non-zero,
escape fraction at high redshift.

\begin{figure*}[t!]
\epsscale{1.1}
\centerline{\includegraphics[scale=0.40]{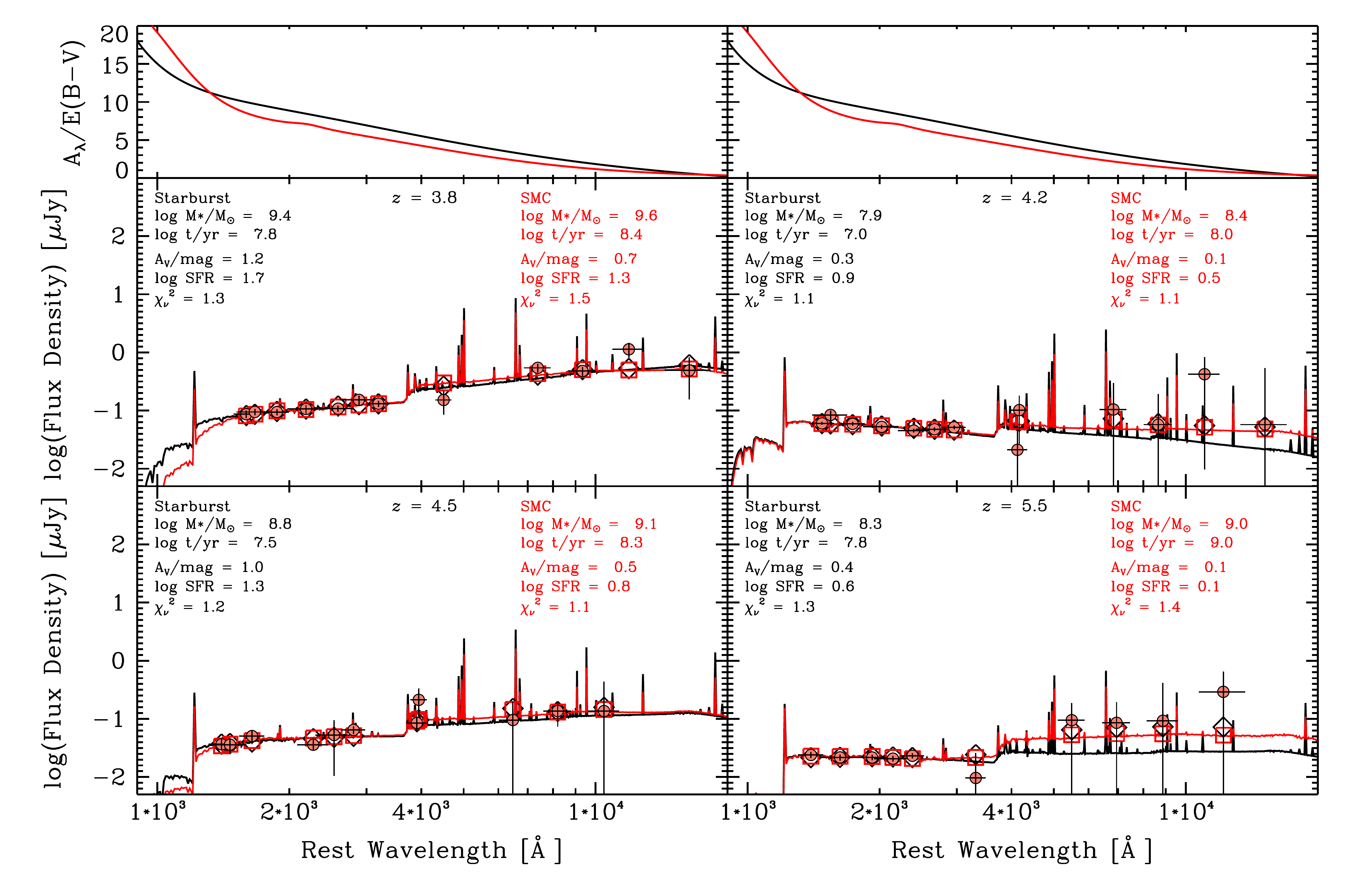}}
\caption{Four example galaxies with the largest differences in the SFR
  derived from the best-fit models using SMC 
and starburst attenuations.  For each galaxy, the best-fit SEDs are shown for SMC (red) 
and starburst (black) attenuations. Circles are the observed 
photometry and squares (diamonds) are the fluxes of the best-fit SED 
with SMC (starburst) attenuation assumed. The legends indicate 
the derived properties when assuming each attenuation. 
All objects were fit assuming a constant star formation history with nebular 
emission lines. Objects may have similarly shaped SEDs, 
but the difference in $A_{V}$ drives the 
change in the inferred parameters. In all cases, the $\chi^2$ values
are equal or exhibit no preference for the SMC or starburst
attenuations. \vspace{0mm}}
\label{fig:SMCexample}
\end{figure*}

Here we consider two limiting cases.  The first has $f_\mathrm{esc} =
1$, for which all LyC photons escape the galaxy, preventing the
creation of nebular emission and reverting the spectrum to the output
of the SPS model.  The second case has
$f_\mathrm{esc} = 0$, for which all LyC photons are absorbed and their
energy is converted into the nebular emission spectrum. These two
cases span the range of possibilities and allow us to study the effects
of nebular emission on the inferred physical parameters.
 Nevertheless, given current constraints of $f_\mathrm{esc} \sim 0.1$ (see above), we
expect the $f_\mathrm{esc}=0$ case will provide a more physical 
model \refereetwo{for galaxies} in our sample. 

To illustrate the effect of nebular emission, Figure~\ref{fig:examples} 
displays four examples that include the best-fit
SED models with and without nebular emission lines for galaxies.
Depending on the redshift, 
emission from H$\alpha$ and/or  [\ion{O}{3}] will enhance the IRAC flux and
can lead to highly different model-parameter values.   The
effect of nebular emission  lines on the inferred stellar mass has a
simple, qualitative explanation:  the flux excess to the optical bands
from nebular emission mimics  a strong Balmer break that is typical
for massive, older stellar  populations. In this sense, when it is
assumed that all of the  observed broadband flux is produced by stars
when much of it  is produced by nebular emission, the inferred stellar mass
will be over-estimated. 

Similarly, Figure~\ref{fig:photbyredshift}  illustrates how the specific emission lines
([\ion{O}{3}], \Hb, and \Ha) affect the bandpass-averaged flux
densities for the observed $K_s$ and IRAC [3.6] and [4.5] bands from
the best-fit SED models. 
The inclusion of [\ion{O}{3}], H$\alpha$, and H$\beta$ lines raise the flux 
of these bands by up to a factor of $\sim$2.    In Appendix B, we
explore how the effects of nebular emission lines change the best-fit
stellar masses and SFRs. However, as we show below, these changes are
largely mitigated using our Bayesian formalism.

\subsubsection{Nebular Continuum}\label{sec:nebcon}
Evolutionary synthesis modeling suggests that nebular continuum
emission can impact broadband photometry
\citep{Leitherer95,Molla09,Raiter10}. In addition, recent
observational  evidence has discovered the presence of strong nebular
continuum in  star-forming galaxies \citep{Reines10}. The inverse
Balmer and Paschen  breaks (Balmer and Paschen ``jumps''), may
contribute additional flux red-ward of rest-frame optical  wavelengths
\citep{Guseva06}. We currently omit  these effects, as the strongest
nebular  continuum is present at wavelengths redder than rest-frame
8~\micron\ (observed-frame $36  -  60~\micron $ for the redshift  range
investigated here), where the objects in this work are not  well
observed \citep[see][]{Zackrisson08}.


\subsection{Dust Attenuation Prescriptions}\label{sec:SMC}
Recent work has suggested that the typically assumed \cite{Calzetti00} 
attenuation prescription for star-forming galaxies is not ubiquitous 
\citep{Reddy12, Oesch13, Chevallard13, Kriek13}. The slope of 
the attenuation curve or presence of the UV dust bump at 2175 
~\AA\ may be dependent on the galaxy type, geometry, metallicity, or 
inclination. However, galaxies at $z>4$ currently lack sufficient 
observations to quantify these effects, so some attenuation 
prescription must be assumed. 
This work aims to test the effects 
of changing the type of assumed attenuation in order to gauge 
its impact on our broadband SED fitting procedure. In this 
subsection, we describe the two different attenuation 
prescriptions used in this work: the \cite{Calzetti00} attenuation 
prescription (``starburst''-like attenuation, hereafter), 
and the \cite{Pei92} attenuation prescription derived for the SMC
(``SMC''-like attenuation, hereafter).

\begin{figure*}[!t]
\epsscale{1.1}
\centerline{\includegraphics[scale=0.40]{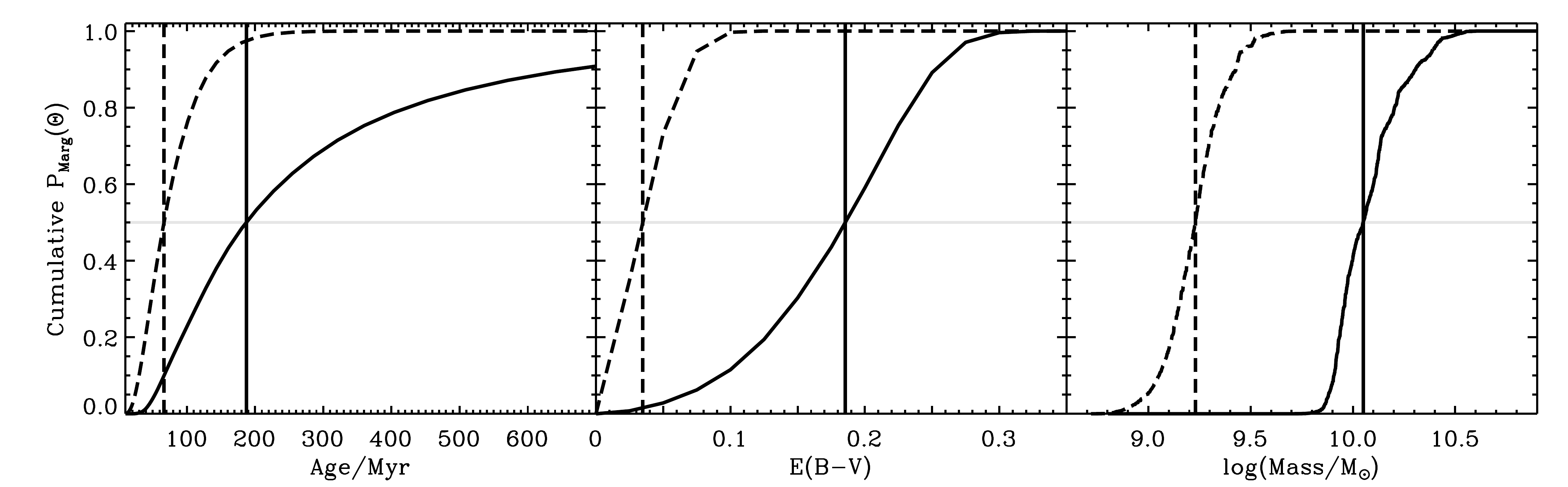}}
\caption{Examples of the posterior cumulative probability densities on a given 
model parameter value, $\Theta$, for a galaxy with higher extinction 
(solid lines) and one with lower extinction (dashed lines),  with 
$z_{\text{spec}}=4.142$ and 3.791, respectively. The posteriors 
in age are often broad, as it is the least  constrained parameter. 
The posteriors in stellar mass are typically narrower.  Throughout, we 
assign the medians of each parameter's posterior (taken as the 50th 
percentile, shown as vertical lines) as the accepted value, with the 
68\% confidence  range as the region that spans the 16th to 84th 
percentiles. \vspace{3mm}}
\label{fig:MargPDF}
\end{figure*}

Figure~\ref{fig:SMCexample} shows four example best-fit  SEDs of
objects that emphasize the difference in SFRs for best-fit models
using the SMC and starburst dust prescriptions.  
\referee{The starburst attenuation has a much ``grayer'' wavelength 
dependence in the UV than the SMC-like attenuation. This means 
the SMC-like attenuation curve has a much  stronger attenuation at
rest-frame, far-UV wavelengths $\lambda \lesssim1200$~\AA, and a
weaker  attenuation across near-UV-to-near-infrared wavelengths
$\lambda \gtrsim1200$~\AA, as shown in the top two panels  of 
Figure~\ref{fig:SMCexample}. As stated 
above, bands  shortward of \lya are omitted in our procedure, so we
do not  fit where the  difference between attenuation prescriptions is
strongest.}

We find no obvious preference in $\chi^2$ between the best-fit models
for an  SMC-like or starburst-like attenuation, and thus cannot  as
yet promote the use of one prescription over the other  from this
data set.  However, we argue that the SMC-like  attenuation could be
invoked as a physical prior to reconcile  the unphysical, extremely
young stellar population ages  that result from assuming a
starburst-like attenuation.  This  method is preferred over, for
example, increasing the minimum allowed age in the models (e.g., from
$\ge$ 10 Myr to $\ge$ 60 Myr), which will not remove the preference of
the fit to choose the youngest available age.  A similar line of
reasoning is used by \cite{Tilvi13} to argue for SMC-like attenuation
over starburst-like attenuation.  Nevertheless, as we will show in
\S~\ref{sec:CompareMargDustNeb}, in our Bayesian formalism these differences arising
from changes in the dust prescription are mitigated, and the dust attenuation \refereetwo{prescription}
has negligible impact on the results here. 


\section{A Bayesian approach to determine physical parameters}\label{sec:PDF} 
This section describes our method to measure the posterior probability 
density for each object and shows how the likelihoods for each stellar population 
parameter were determined during the SED fitting. For the remainder of this work,
we consider the fully marginalized posterior probability density functions 
to derive constraints on physical quantities such as stellar population age, 
galaxy attenuation (i.e., dust extinction), SFR, and stellar mass.

\begin{figure}
\epsscale{1.1}
\centerline{\includegraphics[scale=0.42]{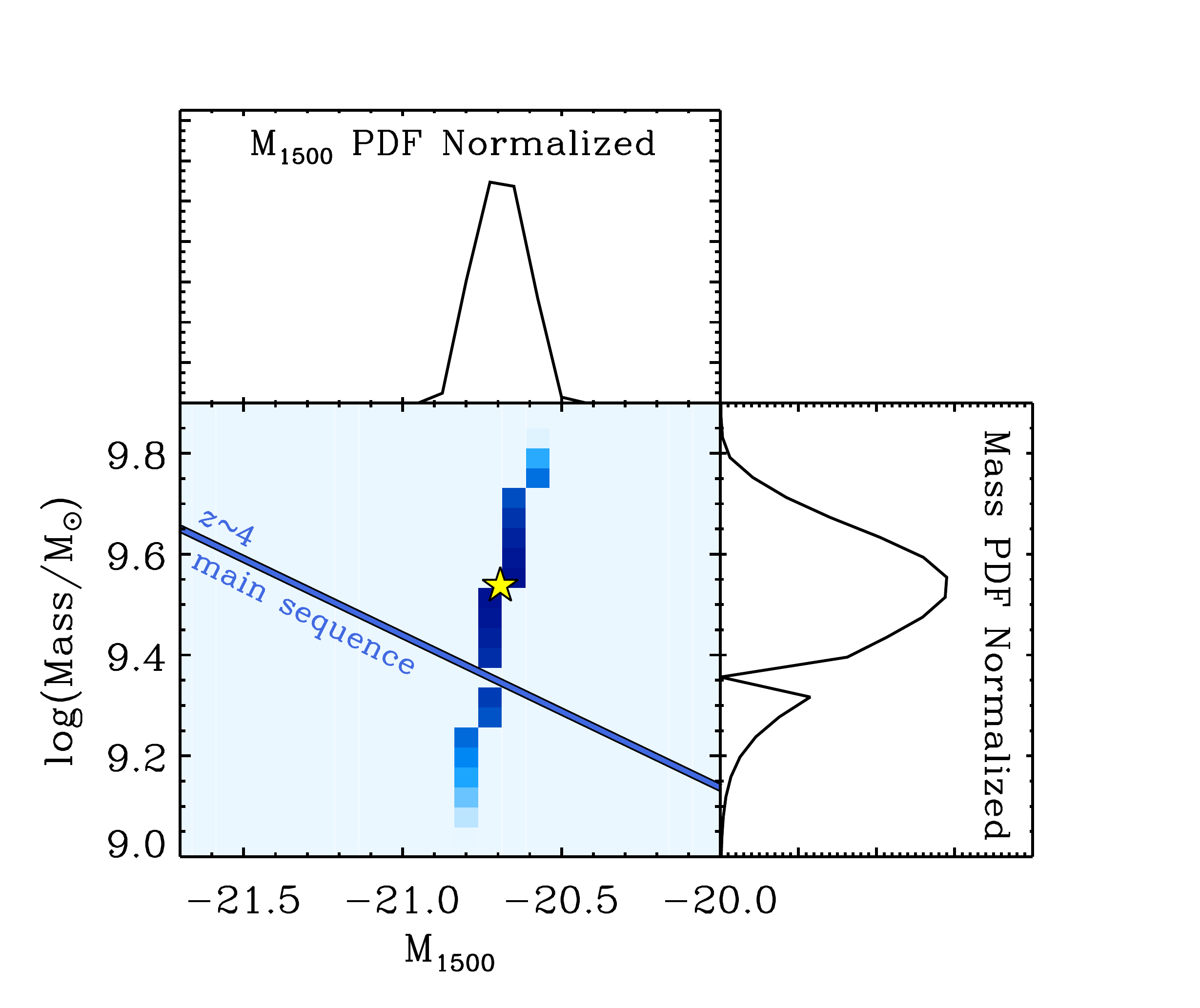}}
\caption{An example of the posterior, joint probability density between
stellar mass and $M_{1500}$ for a single object (where here $M_{1500}$ is
the observed value derived from the model parameters analogous in the
way for the stellar mass, and is uncorrected for dust attenuation).  Darker blue
regions show higher probability density, and the yellow star denotes
the accepted (median) values. The lower left to upper right covariance
is typical for most  objects and results from covariances between the
extinction and age parameters.   It is noteworthy that the scatter in
\Muv-\Mstar\ for a single object is roughly orthogonal  to the
direction of the \Muv-\Mstar\ ``main sequence'' as derived from the full sample
 (see Fig.~\ref{fig:m1500mass}).  
Therefore, this scatter
in \Muv-\Mstar\ likely contributes  to the scatter in the SFR--mass
sequence discussed later. \vspace{3mm}}
\label{fig:2DPDF}
\end{figure}

\subsection{Probability Density Functions: Methods}\label{sec:PDFmethods}

Given a set of data for an individual galaxy which is a function of flux 
densities, $\Data{(f_\nu)}$, we derive the likelihood, 
\begin{equation}
\label{equ:PDF}
P(D| \Theta')\ \propto\ \exp(-\chi^2/2) 
\end{equation}
where $\chi^2$ is measured between the data, $D$, and a model in
the usual way for a given set of  \referee{model stellar population parameters, 
${\Theta' = (\Theta\{t_{\text{age}},  \tau,  A_{\text{V}} \},  \Mstar)}$. Note that 
the likelihood in equation~\ref{equ:PDF} is constructed based on linear fluxes.}
We then find the posterior probability density for any parameter given 
an observed set of data, \Data, and probability density using Bayes'
theorem  \citep[see also,][]{Moustakas13}, 

\begin{equation}
\label{equ:Bayes}
P(\Theta'| D) =\ P(D|\Theta') \times p(\Theta') / \eta
\end{equation}
where $\Theta'$ represents the fitted parameters $\Theta$ and \Mstar, 
and \refereetwo{$\eta$} is a constant such 
that $P(\Theta'|D)$ will normalize to unity 
when integrated over all parameters \citep[see][]{Kauffmann03}.
$p(\Theta')$ represents the priors on the model parameters, and is
\referee{described further in Appendix C.}   As described in 
\S~\ref{sec:sed_methods} and Table~1, we have
adopted a prior (quasi-logarithmic) on the age from 10 Myr to the
age of the Universe for a galaxy's redshift, and we have adopted a prior
(linear) that the attenuation is non-negative up to a maximum value. 
\refereetwo{Further details on these priors and their effect on the fitting 
can be found in Appendix C.}

\begin{figure}
\epsscale{1.1}
\centerline{\includegraphics[scale=0.42]{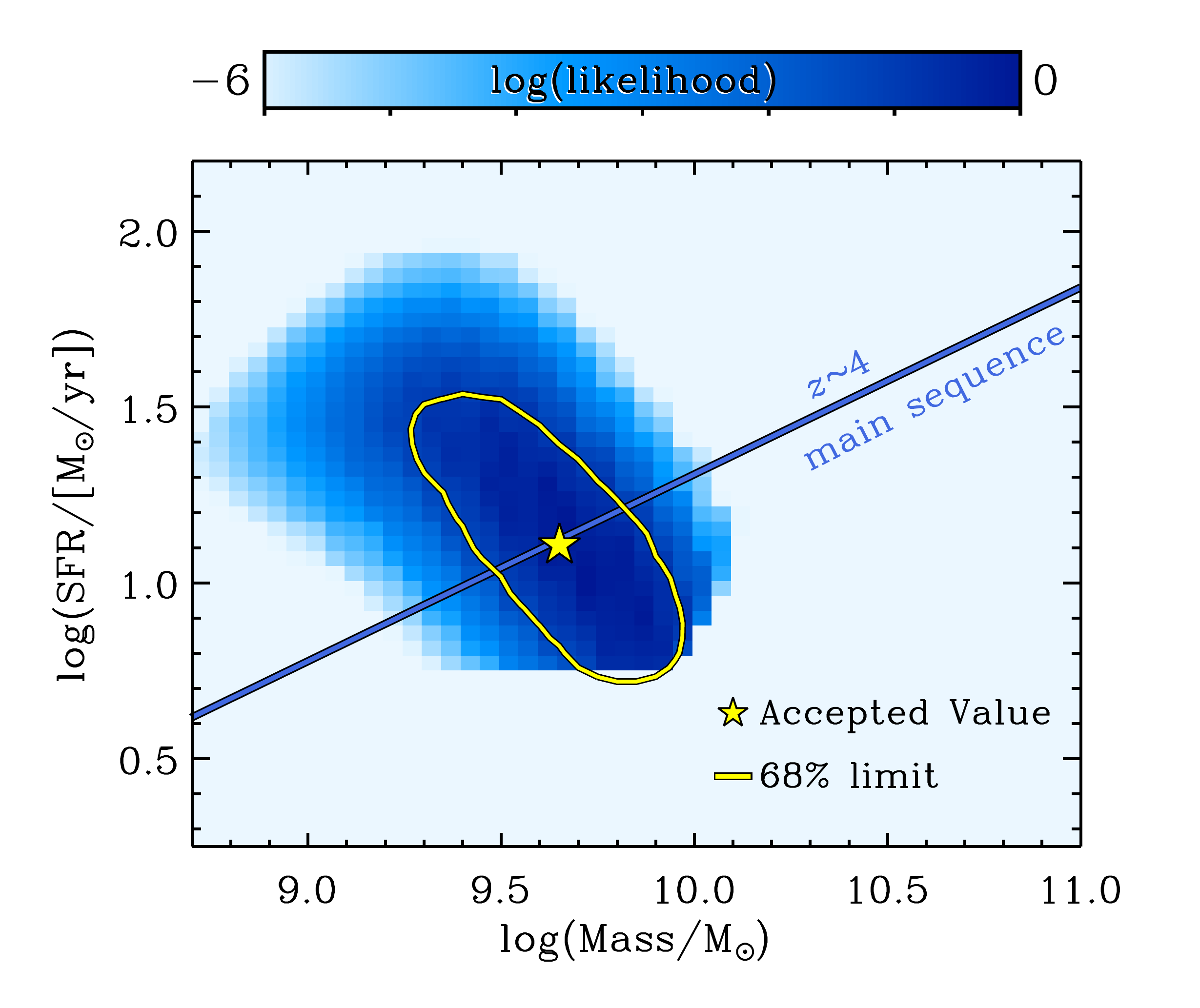}}
\caption{An example of the posterior joint probability between the SFR
 and stellar mass for one object. Darker blue regions show
higher probability density and the yellow star denotes the accepted (median)
values. The range in SFR is driven nearly entirely by the posterior
probability density for attenuation, as described in the text (see
equation \ref{equ:sfr}).  The covariance in SFR and stellar mass is mostly 
orthogonal to the ``main sequence'' as derived from the full sample (see Fig.~\ref{fig:massSFR}), 
which implies that the scatter in the SFR--\Mstar\
relation for individual galaxies translates to scatter in
the SFR--\Mstar\ relation for the galaxy population.  \vspace{3mm}}
\label{fig:2DPDF_sfrmass}  
\end{figure}

We then derive posterior probability densities on individual parameters such as
$t_{\text{age}}$, $A_{\text{UV}}$, etc.  For example, the posterior on the age can be written as,
\begin{equation}
\label{equ:marg}
P(t_{\text{age}}| D)\ =  \int_{A_{\text{UV}},\tau} P(t_{\text{age}},A_{\text{UV}},\tau| D)\ dA_{\text{UV}}\ d\tau ,
\end{equation} 
where the integration is a marginalization over ``nuisance'' parameters,  
dust attenuation, $A_{\text{UV}}$, and possible star formation 
histories/$e$-folding timescales, 
$\tau$\footnote{Here, we ultimately set the star formation history to
be constant (a single value of $\tau$) for the reasons discussed in
\referee{the S~\ref{sec:sfh} and Appendix C.}}.  

The stellar mass must be treated differently because it is effectively
a  scale factor in the fitting process.  In order  to derive its
posterior probability density we must integrate over all parameters,
$ P(\Mstar|\Data) \propto \int P(\Data|\Theta') * p(\Theta')d\Theta .$   The
mean and variance of the stellar mass can be computed as the first and
second moments of the posterior. Similarly, the median stellar mass is 
defined as the value of \Mstar\ such that the integral over the posterior 
from negative infinity to \Mstar\ is equal to 50\%,
while the 68\% confidence range can be calculated by
integrating the posterior from the 16th to 84th percentiles.

\subsection{Probability Density Functions: Results}  \label{PDFresults}

We computed the posterior probability densities for all galaxies in
our sample, including posteriors for the stellar mass, age, and attenuation,
using the methods described above.   Figure~\ref{fig:MargPDF} shows
examples of the cumulative posteriors on age, attenuation (or color excess, $E(B-V)$), 
and stellar mass for two galaxies in our sample: a relatively 
un-extincted and a relatively extincted galaxy.    These objects are typical 
of those in the sample, where the posterior for age is typically broad, while 
that for the stellar mass is relatively tighter.



\begin{figure*}[!t]
\epsscale{1.1}
\centerline{\includegraphics[scale=0.58]{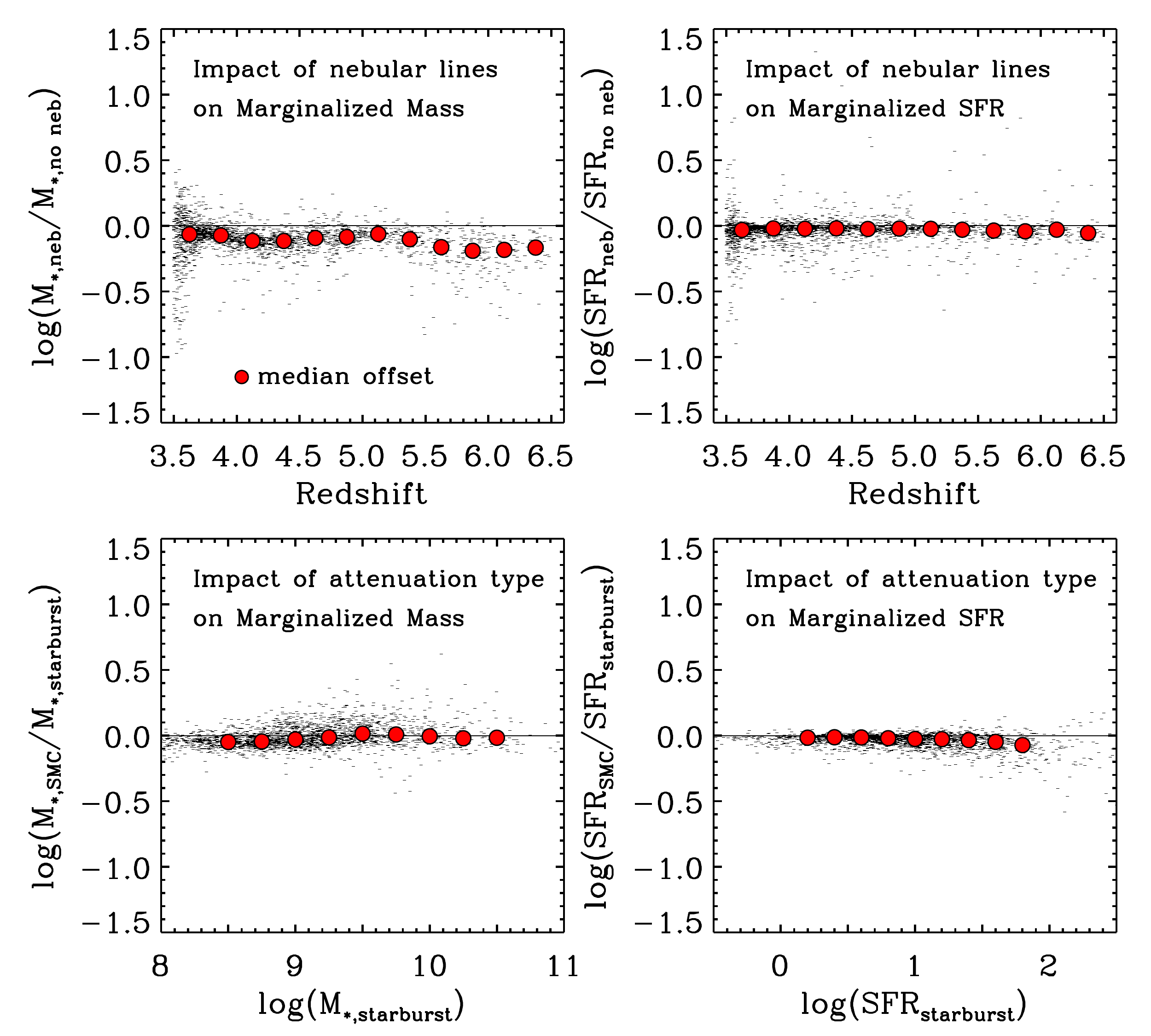}}
\caption{The change in SFRs and stellar masses derived from the galaxy
posteriors using different model assumptions. The top panels
compare the SFRs and stellar masses derived using models that include
and exclude nebular emission lines.   The inclusion of nebular
emission lines has minimal effect on the SFRs ($\lesssim 0.1$ dex),
and the stellar masses have a slight decrease ($0.1-0.2$ dex) when
nebular emission lines are included.  The bottom panels
compare SFRs and stellar masses derived using models with SMC-like and
starburst-like dust attenuation.    Here, varying the dust attenuation \refereetwo{prescription}
has a negligible impact on the derived SFRs and stellar masses.  These
results can be directly compared to the results derived from best
fits, which show much stronger differences in these quantities derived
from these models (see Appendix B). 
 \vspace{3mm}}
\label{fig:CompareBestFitMarg}
\end{figure*}

In our analysis below, we will consider the relation between
stellar mass and UV magnitude, as well as stellar mass and SFR for our
full galaxy sample.  Here, we discuss the relation between stellar mass
and these quantities for an individual object, as it is illustrative.
Figure~\ref{fig:2DPDF} shows a two-dimensional  probability density
function between stellar mass and UV absolute magnitude.   Here we
take the \Muv\ from the conditional posterior on age and attenuation
(similar to way we derive the posterior for stellar mass given the
model parameters and data).  Figure~\ref{fig:2DPDF} also shows the
posterior for \Muv\ and stellar mass individually.     

 There is a weak covariance between \Muv\ and \Mstar, which results from the
degeneracy in dust attenuation and age.   A galaxy with a redder
rest-frame UV continuum has near equal likelihood to a model with an
older, less extincted stellar population as to a model with younger,
higher extinction.  The two models produce a joint posterior that is
anticorrelated between \Muv\ and stellar mass.    The figure also
shows the ``main sequence'' of the \Muv--\Mstar\ relation as derived
from the full sample (see Fig.~\ref{fig:m1500mass}).  The joint posterior is 
approximately orthogonal to the main sequence,
\referee{which implies that the likelihood scatter in each galaxy's \Muv--\Mstar\ plane will
lead directly to scatter in the main sequence of the sample.} We return to this point 
in the discussion of the SFR-\Mstar\ relation below.

\begin{figure*}
\epsscale{1.1}
\centerline{\includegraphics[scale=0.55]{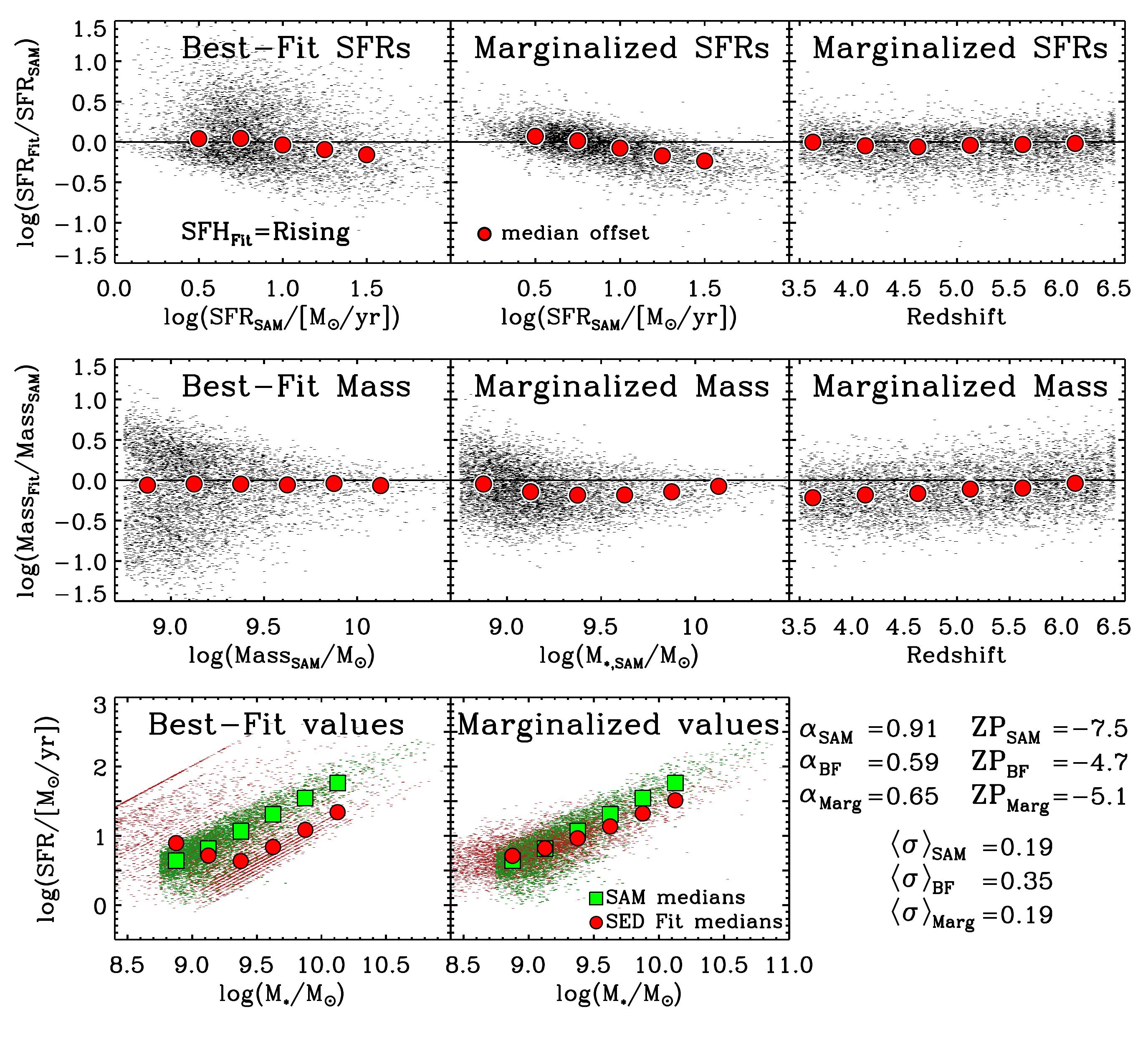}}
\caption{\referee{Tests of the derived SFRs and stellar masses from the
posteriors of our SED fitting to synthetic photometry of mock catalogs
from the SAM of Somerville et al.   The top panels show the log 
difference of measured-to-true SFRs.   The derived SFRs show a weak
trend in that our fits overestimate the SFRs of low-SFR objects and
underestimate the SFRs of high-SFR objects.    The middle panels show the ratio of the
measured-to-true stellar masses.    The scatter in the derived stellar masses 
from their true values likely arises from our simple prescription for the 
star formation histories (similar offsets are observed by \cite{JLee10}).  
 The bottom panels show that our derived SFRs and stellar
masses recover the SFR--mass relation in the models, though with a more shallow
slope. \refereetwo{The legend to the lower right indicates the slope, zero point, and scatter 
of the SFR--mass relation for the SAMs and those recovered using best-fit values or 
our preferred marginalized values.} The main point of the figure is that the bayesian 
method does better at recovering both the SFRs and stellar masses and the SF main 
sequence.}
 \vspace{3mm}}
\label{fig:SAMcompare}
\end{figure*}

We derive the SFR from \refereetwo{the model} parameters
in the following manner.  We first determine the rest-frame UV  luminosity at
1500~\AA. At these redshifts, a large sample of detections  red-ward
of the rest-frame optical are  unavailable, which can cause age and
attenuation inferred from  SED fitting to be degenerate
quantities. These degeneracies  can bias the \Muv\ inferred from the
model parameters, especially if limited to best-fit  values (see
the discussion above and Appendix B).   For this reason, we choose the
closest observed band  to rest-frame 1500~\AA\ as a more
observationally motivated  value of \Muv\ because the broad-band
photometry well samples the rest-frame UV portion of the SED. Galaxies
have  relatively blue rest-frame UV colors, so any corrections between
the band closest to 1500~\AA\ and the interpolated magnitude  at
1500~\AA\ are small (in the ``extreme'' examples of 
Figure~\ref{fig:SMCexample} the differences are $<$ 0.1 mag).   Furthermore,
our tests show that none of our conclusions depend strongly on the
manner we use to obtain \Muv. 

The 1500~\AA\ luminosity is corrected for dust attenuation using
the median from the posterior of the attenuation at 1500~\AA, or
$A_{\text{UV}}$.  
The dust-corrected UV luminosity  is converted to
the  \SFRuv\ using the ratio $\kappa (t,\tau) =$
\SFRuv$/L_{1500}$.   This is similar to the conversion given by
\cite{Kennicutt98}, but we account for variations in \SFRuv$/L_{1500}$
owing to the age ($t$) and star formation history
($\tau$) of the stellar population \citep[see,][]{Reddy12}.  For each
object, we use the median stellar population age from the posterior to
calculate \SFRuv$/L_{1500}$.  However, we note that because most of
these median ages from posteriors are $>$100 Myr for the galaxies in
our sample and we
have assumed constant star formation histories, in most \refereetwo{cases 
$\kappa(t,\tau)$ is very similar} to that of
\cite{Kennicutt98}.  

We summarize the derivation of our \SFRuv\ mathematically as follows.
\begin{equation}
\label{equ:sfr}
\mathrm{SFR}_{\text{UV}} = f_{\text{CB}}\cdot \frac{4\pi D_L^2}{1+z} \cdot  10^{0.4\ A_{\text{UV}}}\cdot \kappa (t,\tau)
\end{equation}
where $f_{\text{CB}}$ is the flux of the closest band to rest-frame
1500~\AA, $D_L$ is the luminosity distance,
$A_{\text{UV}}$ is the median, marginalized  attenuation
at 1500~\AA, and $\kappa$ is the modified  \cite{Kennicutt98}
conversion that depends on age ($t_{\mathrm{age}}$) and star formation
history ($\tau$). 

The differences between the SFRs derived using equation~\ref{equ:sfr}
and other common methods are described in Appendix B. 
In summary, methods that derive the SFR using the best-fit 
or direct UV luminosity slope exhibit higher 
scatter when compared to the marginalized SFR method 
from this work. This scatter stems from degeneracies between the young, dusty
and old, dust-free solutions of a given SED, and photometric
uncertainties (which affect the accuracy of measuring the UV spectral
slope). We find the results of our method more robust as it reproduces SFRs from
SAMs (see \S~\ref{sec:SAMcompareSEC}), and our method is
relatively unaffected by model variations such as extinction prescription
and/or nebular emission lines. 

Figure~\ref{fig:2DPDF_sfrmass} shows the SFR--stellar mass joint
posterior for one object from our sample.   As with the \Muv--\Mstar\
example, the covariance is roughly orthogonal to the star formation
main-sequence, but there is more scatter because of the range in
dust attenuation (and, to a lesser extent, the stellar population age).
The errors on the measured (extrinsic, or attenuated) \Muv\ are
relatively small as they stem from photometric errors only, whereas
the SFR depends on the UV luminosity corrected by the UV extinction,
$A_\mathrm{UV}$. 

%

SED models with higher $A_\mathrm{UV}$ have higher
stellar-mass--to--light ratios (and therefore higher stellar masses at
fixed UV luminosity).   This induces some correlation in the SFR--stellar mass 
plane for each object. However, the covariance is mostly orthogonal to the expected 
direction of the SFR--stellar mass correlation, which implies that it contributes mostly 
to the scatter of the SFR--stellar mass relation and less to the correlation itself.  In our analysis 
we take this covariance into account using Monte Carlo simulations
(see \S~\ref{sec:MC}).

\subsection{Impact of SED Fitting Assumptions on Marginalized Values}\label{sec:CompareMargDustNeb}

Here we discuss our SED model assumptions and their impact on derived
quantities such as SFR and stellar mass using our Bayesian method.  In
Appendix B, we show that these model choices have a significantly
stronger impact on best-fit results, while the results using medians
derived from posteriors  are relatively unaffected.   


The panels of Figure~\ref{fig:CompareBestFitMarg} show that the SFRs
and stellar masses derived from the posteriors for the galaxies in our
sample are rather insensitive to the choice of dust-attenuation prescription or the
presence/exclusion of nebular emission (where we compare the results
with $f_{\text{esc}}$=0 or 1).    In general, varying the assumed dust
attenuation prescription has a negligible impact on the derived SFRs and stellar
masses (differences are $<$0.1 dex over the redshift range of our
sample).    Similarly, including emission lines has minimal effect on
the SFRs ($\lesssim 0.1$ dex).  

There is some evidence that the stellar
masses are reduced slightly when the models include nebular emission
lines.  This is in the same direction but weaker in magnitude as seen
in comparisons of the best-fit models (e.g., see Appendix B).  
However, the effect is only a slight
decrease of $<0.25$ dex, and is typically less than the measurement
errors on mass derived from the posteriors.  Therefore, the inclusion
of nebular emission does not strongly impact the SFRs or stellar
masses.  For this reason, we  have neglected exploring nebular emission
over the full range of $f_\mathrm{esc}$ values in our analysis, and
instead report results assuming $f_\mathrm{esc} = 0$ (all ionizing
radiation is absorbed and produces emission lines).  

 The results from Figure~\ref{fig:CompareBestFitMarg}  can be directly
compared to the results derived from best-fits, which show stronger
differences in these quantities when switching between the above model assumptions
(see Appendix B).  \referee{In contrast to using best-fit values, the Bayesian method uses the likelihood 
of all the models, and so even if there is a ``highest likelihood'' solution of low age, 
there are many good solutions with larger ages, and when marginalized, the latter 
dominate the posterior.}

\begin{figure*}
\epsscale{1.1}
\centerline{\includegraphics[scale=0.52]{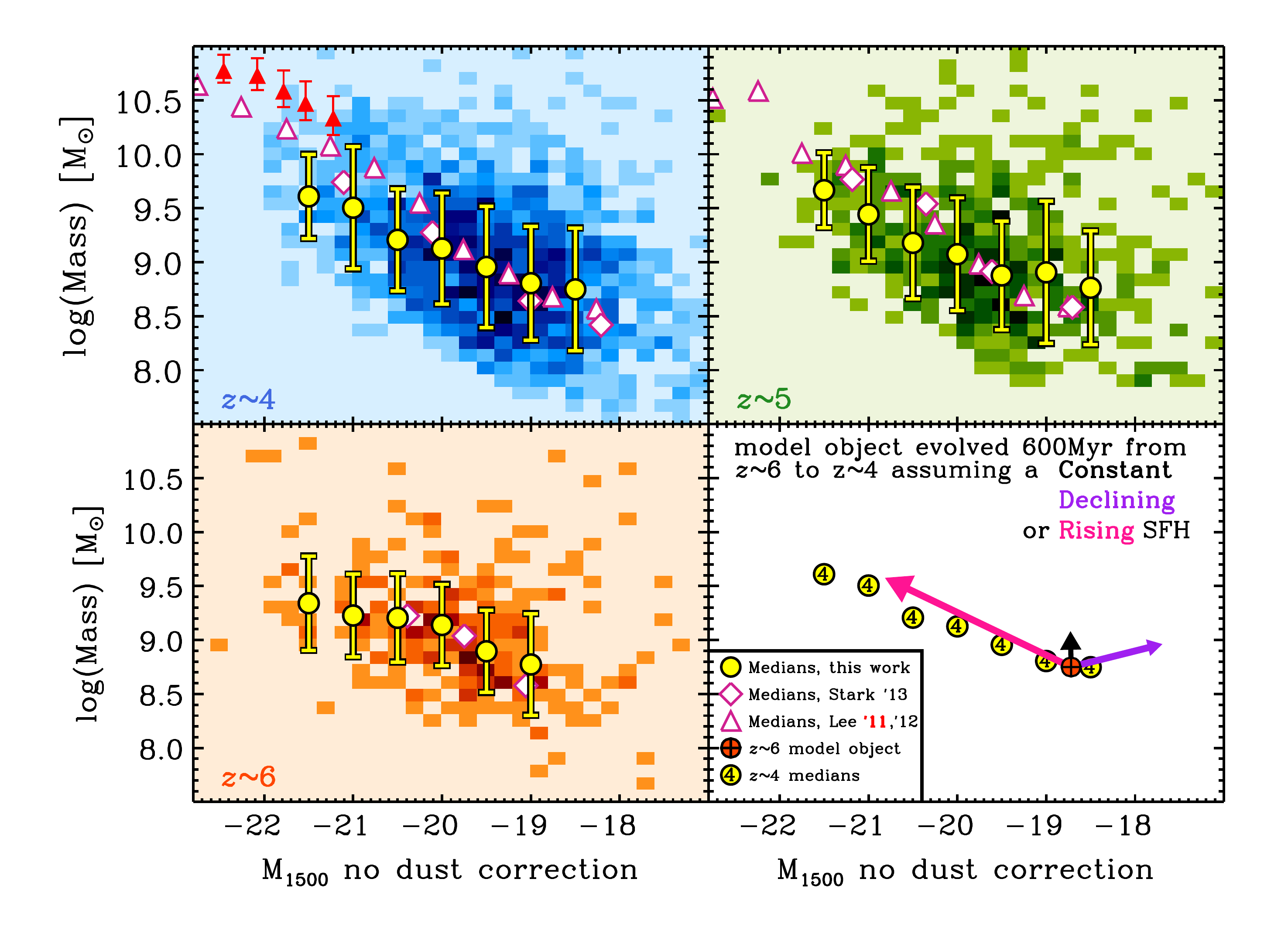}}
\caption{The closest band to rest-frame UV magnitude (\Muv) versus 
stellar mass. Darker-shaded regions indicate a
higher number of individual objects in bins of stellar mass and \Muv.
Yellow circles are the medians of stellar mass in a given \Muv\
bin and error bars are the $\sigma_\mathrm{MAD}$ confidence range (analogous to 
the 68\% confidence, $\sigma$, if the error distribution were 
Gaussian, see \S~\ref{sec:selection}). Red triangles,  white triangles, 
and white diamonds are medians from \cite{Lee11},  \cite{Lee12},  and \cite{Stark13}
respectively. \emph{Bottom right:} For reference, this cartoon shows
the  strength and direction of galaxy evolution over 590 Myr from
$z\sim 6$ to  $z\sim 4$ under an assumed star formation history.  This
plot implies that there is a weak relation between UV magnitude and stellar
mass in place  up to $z\sim 6$ with significant scatter,  and that
rising star formation histories offer a simple explanation of how
galaxies may evolve  along this relation.}
\label{fig:m1500mass}
\end{figure*}

\subsection{Tests of Derived SFR and Stellar Masses using Semi-Analytic Models}\label{sec:SAMcompareSEC}

We tested the ability of our SED fitting procedure to reproduce
accurately the SFRs and stellar masses in mock galaxy catalogs.  We
used synthetic galaxy photometry in the CANDELS bandpasses derived
from the Somerville et al. SAM discussed in \S~\ref{sec:SomervilleYu}.
The advantage of using synthetic
catalogs from a SAM is that the models include realistic (and complex)
star formation histories, as well as more sophisticated treatments of
extinction \citep[e.g.,][]{Charlot00}.   Therefore, the mock catalog
from the SAM acts as a realistic observation where many of the model
parameters (star formation history, extinction prescription) are
known, and are distinct from the simpler models used in SED fitting
to fit to the galaxy photometry.  Comparing to the SAMs therefore provides a
powerful test of our method to recover physical stellar population
parameters, even when the physical details are unknown, such as the
case of our observed galaxies. \referee{A similar but more comprehensive study 
was conducted among many CANDELS team members which compared 
the estimated and expected parameters from galaxies measured using SED fits 
\citep[see][]{Mobasher14}.}

\begin{table*}
\begin{center}
\caption{M$_{1500}$ -- Stellar Mass Relation Median Values}
\vspace{2mm}
\label{tab:M1500Mass_medians}

$z\sim4$\\[1.2mm]
\begin{tabular}{ l | c c c c c c c }
\hline \\[-2mm]
\Muv\ & -21.5 & -21.0 & -20.5 & -20.0 & -19.5 & -19.0 & -18.5 \\[0.5mm] \hline
\\[-1.5mm]
$\log$(Median Mass/\Msol) & 9.61 &  9.50 &  9.21 &  9.13 &  8.96 &  8.81 &  8.75 \\
$\sigma_{\text{MAD}}$\footnote{The $\sigma_{\text{MAD}}$ scatter (see \S~\ref{sec:selection}) in stellar mass for this \Muv\ bin.}
 & 0.39 &  0.57 &  0.47 &  0.51 &  0.56 &  0.53 &  0.57 \\[0.5mm] \hline 
\end{tabular}
\\[1.5mm]

$z\sim5$ \\[1.2mm]
\begin{tabular}{ l | c c c c c c c }
\hline \\[-2mm]
\Muv\ & -21.5 & -21.0 & -20.5 & -20.0 & -19.5 & -19.0 & -18.5 \\[0.5mm] \hline
\\[-1.5mm]
$\log$(Median Mass/\Msol) & 9.67 &  9.44 &  9.18 &  9.07 &  8.88 &  8.91 &  8.76 \\
$\sigma_{\text{MAD}}$ & 0.35 &  0.43 &  0.52 &  0.52 &  0.50 &  0.66 &  0.53 \\[0.5mm] \hline 
\end{tabular}
\\[1.5mm]

$z\sim6$ \\[1.2mm]
\begin{tabular}{ l | c c c c c c c }
\hline \\[-2mm]
\Muv\ & -21.5 & -21.0 & -20.5 & -20.0 & -19.5 & -19.0 & -18.5 \\[0.5mm] \hline
\\[-1.5mm]
$\log$(Median Mass/\Msol) & 9.34 &  9.23 &  9.21 &  9.14 &  8.90 &  8.77 & --- \\
$\sigma_{\text{MAD}}$ & 0.44 &  0.38 &  0.41 &  0.38 &  0.38 &  0.47 & --- \\[0.5mm] \hline 
\end{tabular}
\end{center}
\end{table*}

The mock catalog from the SAM was filtered to a sample of 6000 
simulated galaxies, evenly distributed across the mass and redshift 
range of our CANDELS sample.
\referee{We took the synthetic photometry from 
the models and randomly perturbed them according to a Gaussian error distribution 
with a similar $\sigma$ as the CANDELS data at a given band and magnitude. This
process accounts for any systematic errors in \refereetwo{creating} the mock catalogs, and we 
process these fluxes the same way we process the data, even when fluxes are perturbed 
to negative values.
These fluxes were used as input to the same SED-fitting procedure we 
applied to the real CANDELS samples.
The masses and SFRs in the SAM were scaled to a Salpeter IMF, to match our 
procedure. \refereetwo{We also fit to templates that exclude the effects of nebular emission (using $f_\mathrm{esc}
= 1$) because the SAMs also exclude nebular effects.} }
For the test here, we show only the case where
we fix the star formation history to be constant.  Our tests
(discussed in Appendix A) show that we recover the most accurate
SFRs and stellar masses using constant star formation histories.   
Appendix A discusses how including additional star formation histories
affects the SFRs and stellar masses. 

\referee{Figure~\ref{fig:SAMcompare} compares the ``true'' stellar masses and SFRs
from the SAM with those derived from using either the ``best-fit'' values
or our method of taking the median, marginalized value from the posterior.  
This figure shows that taking advantage of the whole posterior with marginalized values
produces less scatter in the recovering the ``true'' SAM values. This is because the 
maximum likelihood can be more sensitive to template assumptions
than the median posterior values, as shown in Figure~\ref{fig:CompareBestFitMarg}
and discussed in Appendix B. 
We note that there is a weak systematic where the fits slightly
overestimate objects with low SFRs and slightly underestimate objects
with high SFRs.   The effect is mild, ranging by $\pm 0.25$ dex.  This
systematic  could be due to differences in the extinction prescription 
and assumed star formation history between models used in the fit and those 
in the SAM.  }

\referee{The bottom panel of Figure~\ref{fig:SAMcompare} shows that the offsets
in stellar mass and SFRs conspire to reproduce the accurate SFR--mass
relation as expected from the SAMs.  The parameters derived from our
fits better reproduce the zero point, scatter, and slope of the
relation than when using best-fit values.  \refereetwo{We also find no appreciable
difference in the ability to recover the SFR--stellar mass relation across redshift.
Our ability to test the success of our procedure, however, is limited to the maximal
level of stochasticity in the star formation histories of the SAMs and we have not
tested our procedure to observe its sensitivity to very bursty star formation histories.
Nevertheless, these tests give us confidence that even in the presence of realistic 
photometric errors, we are able to derive a SFR--stellar mass relation from 
high-redshift galaxies in the CANDELS data that reproduces the intrinsic relation 
within these $\sim$ 0.25 dex uncertainties in SFR or stellar mass.}

\begin{figure*}[!t]
\epsscale{1.1}
\centerline{\includegraphics[scale=0.49]{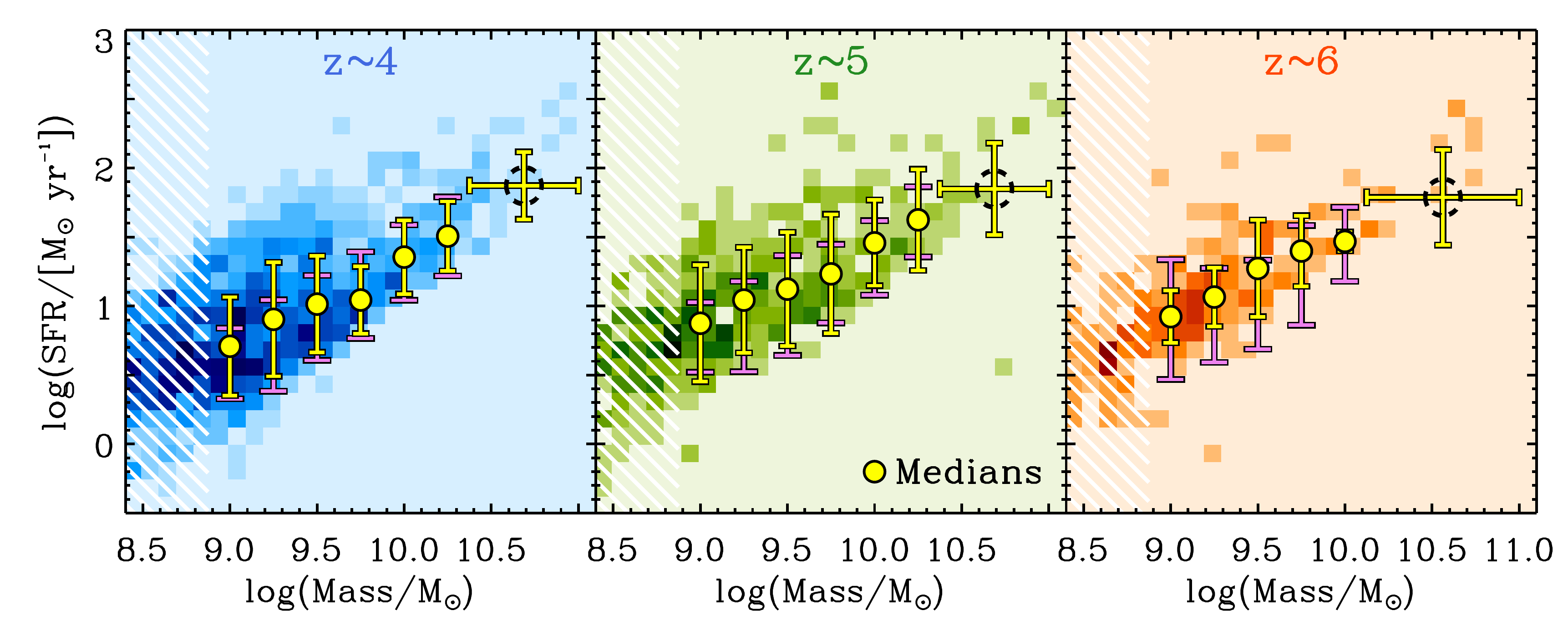}}
\caption{The SFR--stellar mass relation for the CANDELS galaxy
samples.  The darker-shaded regions indicate a higher number of
individual objects in bins of stellar mass and SFR.  Yellow circles
are medians in bins of mass and yellow error bars are their $\sigma_\mathrm{MAD}$ confidence
range (see Table~\ref{tab:SFRMass_medians}). The median SFR of a wider, high-mass bin is also shown by the 
dashed black circle. \refereetwo{The white hatched regions mark the limit 
above which completeness effects become negligible.} 
We measure a slope of $\sim$0.6 (see Table~\ref{tab:SFRMass_fits}),
with no evidence for evolution over the redshift range $z\sim$6 to 4.
The purple error bars show the 68\% range of errors from the Monte
Carlo simulations  described in \S~\ref{sec:MC}.   \vspace{0mm}}
\label{fig:massSFR}
\end{figure*}

\begin{table*}
\begin{center}
\caption{SFR -- Stellar Mass Relation Median Values}
\vspace{2mm}
\label{tab:SFRMass_medians}

$z\sim4$\\[1.2mm]
\begin{tabular}{ l | c c c c c c c }
\hline \\[-2mm]
$\log$(\Mstar/\Msol) & 9.00 &  9.25 &  9.50 &  9.75 & 10.00 & 10.25 & $>$ 10.375\footnote{A larger stellar mass bin from the edge of the previous bin to $\log$(\Mstar/\Msol) = 11}
 \\[0.5mm] \hline
\\[-1.5mm]
$\log$(Median SFR/\Msol\ yr$^{-1}$) & 0.71 & 0.90 & 1.01 & 1.04 & 1.35 & 1.51 & 1.87 \\
$\sigma_{\text{MAD}}\footnote{The $\sigma_{\text{MAD}}$ scatter (see \S~\ref{sec:selection}) in SFR for this stellar mass bin}
$ & 0.36 & 0.41 & 0.35 & 0.24 & 0.27 & 0.25 & 0.24 \\
Monte Carlo $\sigma$\footnote{The average range in the bootstrapped errors calculated by the Monte Carlo on stellar mass and SFR (see \S~\ref{sec:MC}).}
 & 0.26 & 0.33 & 0.31 & 0.31 & 0.27 & 0.29 & --- \\[0.5mm]  \hline 
\end{tabular}
\\[1.5mm]

$z\sim5$\\[1.2mm]
\begin{tabular}{ l | c c c c c c c }
\hline \\[-2mm]
$\log$(\Mstar/\Msol) & 9.00 &  9.25 &  9.50 &  9.75 & 10.00 & 10.25 & $>$ 10.375 \\[0.5mm]  \hline
\\[-1.5mm]
$\log$(Median SFR/\Msol\ yr$^{-1}$) & 0.88 & 1.04 & 1.12 & 1.23 & 1.46 & 1.62 & 1.85 \\
$\sigma_{\text{MAD}}$ & 0.42 & 0.38 & 0.41 & 0.43 & 0.31 & 0.37 & 0.33 \\
Monte Carlo $\sigma$ & 0.25 & 0.33 & 0.36 & 0.28 & 0.27 & 0.25 & --- \\[0.5mm]  \hline
\end{tabular}
\\[1.5mm]

$z\sim6$\\[1.2mm]
\begin{tabular}{ l | c c c c c c c }
\hline \\[-2mm]
$\log$(\Mstar/\Msol) & 9.00 &  9.25 &  9.50 &  9.75 & 10.00 & $>$ 10.125 & \hspace{0.1 cm} ----- \\[0.5mm] \hline
\\[-1.5mm]
$\log$(Median SFR/\Msol\ yr$^{-1}$) & 0.92 & 1.07 & 1.27 & 1.40 & 1.47 & 1.79 & \hspace{0.1 cm} ----- \\
$\sigma_{\text{MAD}}$ & 0.19 & 0.21 & 0.35 & 0.26 & 0.07 &  0.35 & \hspace{0.1 cm}----- \\
Monte Carlo $\sigma$ & 0.43 & 0.34 & 0.32 & 0.36 & 0.27 & ----- & \hspace{0.1 cm} ----- \\[0.5mm] \hline
\end{tabular}
\end{center}
\end{table*}

\section{Evolution of SFR and Stellar Mass at $3.5 <  \MakeLowercase{z} < 6.5$}\label{sec:SFRMassM1500}
This section describes the relations between the observed rest-frame
UV magnitude, \Muv, and stellar mass, and the SFR and stellar mass
of galaxies in our CANDELS sample from $z$ = 3.5 to 6.5. All SFRs and 
stellar masses are derived from the posteriors from the SED fits.

\subsection{$M_{1500}$--Stellar Mass Relation}\label{sec:mass-m1500}
The panels of Figure~\ref{fig:m1500mass} show the relation  between
\refereetwo{the observed magnitude of the band closest to 1500~\AA\ at the redshift 
of the galaxy (the rest-frame UV magnitude)} and stellar mass at each
redshift as two-dimensional histograms for our CANDELS sample. 
\referee{The median and $\sigma$ of stellar mass in each bin of \Muv\ are given
in Table~\ref{tab:M1500Mass_medians}.} Here we show the observed 
UV absolute magnitude with no dust
corrections in order to easily compare against previous studies
\citep{Stark09, Stark13, Lee11, Lee12}.  As discussed above,
quantities used in this figure are median stellar masses derived from
the marginalized PDF of each  object (see \S~\ref{sec:PDF} for
details).
%
%

\referee{We find a correlation between UV absolute magnitude and stellar mass,
though this relation  retains significant scatter. Recent evidence has
suggested a relation with significant scatter between \Muv\ and
stellar mass at high redshift, $z\gtrsim 2$ \citep{Reddy06, Daddi07,
Stark09, Gonzalez11,  Reddy12, Schaerer13, Stark13}. The \Muv--stellar
mass trend in this work is weaker than the literature because
we use an $H_{160}$-band selected catalog, \refereetwo{which is closer to 
stellar mass than optically selected samples, as were used in the previous 
works listed above.} This means that at fixed UV luminosity (or SFR) we are less 
sensitive to blue sources, which have higher mass-to-light ratios. Therefore, 
below our limiting stellar mass ($10^{9}$/\Msol) we \refereetwo{may be} 
missing the bluer sources, as seen in Figure~\ref{fig:m1500mass}.  At 
bright magnitudes (SFRs) our results agree with previous studies (that usually 
used $z_{850}$-band selected catalogs).  It is at fainter magnitudes where our 
sources have fewer low-mass objects compared to the literature. 
Our results are also consistent with an independent analysis of the 
CANDELS catalogs, which used the same $H_{160}$-band
selection \citep{Duncan14}.}

\referee{Regardless of the median relation, we consider the large scatter in 
\Muv\ at fixed stellar mass to mean one or both of the following.} First, it could 
mean gas accretion is low such
that galaxies undergo recurrent and stochastic  star formation that
leads to a range of \Muv\ at a fixed stellar  mass
\citep{Lee06,Lee11}. Second, galaxies at a fixed redshift  and fixed
stellar mass could exhibit a range of $A_{\text{UV}}$
attenuations. In the second scenario, the observed scatter in the
plane of \Muv--\Mstar\ would be largely diminished once we apply
corrections for dust attenuation.  As discussed in the next
subsection, the simulations favor the latter scenario.

The bottom right panel of Figure~\ref{fig:m1500mass} illustrates an
evolutionary cartoon depicting a model z$\sim$ 6 object of a given mass
that is evolved forward $\approx$ 600 Myr to $z  \sim  4$. This is done by assuming
an initial stellar mass (10$^8$ M$_\Sun$), age ($\sim$ 500 Myr, the
average  marginalized age of our entire sample), and zero dust
($A_{\text{UV}}$= 0). The strength  and direction of  three different
star formation histories with the \cite{Bruzual03}
SPS code are shown:  a constant SFR,  a declining SFR
(where $\Psi \sim \exp(-t/\tau)$, with $\tau =$ 1 Gyr), or a rising
SFR (where $\Psi \sim t^\gamma$ using our results derived below from
\S~\ref{sec:cnd}).   As illustrated in Figure~\ref{fig:m1500mass}, only a
rising star formation history naturally evolves galaxies along the
median relation  between stellar mass  and \Muv. Though this simple
explanation does well to explain the UV-faint to UV-bright evolution,
it offers little insight to the fate of the UV-bright galaxies at
later epochs. It remains to be seen if some population of massive,
UV-bright galaxies at $z\sim6$ quench their SFR such that we are
missing a population of massive UV-faint galaxies at
$z\sim4$.


\subsection{SFR--Stellar Mass Relation}\label{sec:sfrmass}
Figure \ref{fig:massSFR} shows the relation between the
(dust-corrected) SFR and the stellar mass, where both parameters are
derived from the fully marginalized probability density functions. 
\referee{Table~\ref{tab:SFRMass_medians} shows the median and 
$\sigma$ scatter of $\log$  SFR  in bins of stellar mass 
from Figure~\ref{fig:massSFR}.} We measure a tight SFR--stellar mass 
relation (a ``main sequence'') for galaxies with $\log M_\star/$\msol $> 9$,  
the mass completeness limit \cite[e.g.,][]{Duncan14}.  \refereetwo{We explore 
how our SED fitting process could contribute to the correlation between 
SFR and stellar mass in Appendix C.} This main-sequence 
in the SFR--mass relation has received much attention in the literature, and
its existence implies that stellar mass and star formation both scale
with the star formation history \citep{Stark09, Gonzalez10,
Papovich11}.   If true, then it follows that the gas accretion onto
dark-matter halos at higher redshift is smooth when averaged over
large timescales and stellar mass  growth at high redshift is not
driven by mergers \citep{Cattaneo11, Finlator11}.  Our results support
this picture. 

We fit a linear relation to the SFR--stellar mass relation as
\begin{equation}
\log( \mathrm{SFR}/\mathrm{M}_\odot\ \mathrm{yr}^{-1}) = a\ \log( M_\star/\mathrm{M}_\odot)  + b
\end{equation} 
where \refereetwo{$a$} is the slope of the relation and \refereetwo{$b$} is a zero point.
The fitted values for \refereetwo{$a$} and \refereetwo{$b$} are given in Table~\ref{tab:SFRMass_fits}.     
\refereetwo{We also show the fitted values for $b$ when the slope is fixed to be $a =  1$, since
the slope and intercept are often degenerate.}
We find that the slope \refereetwo{and normalization} in the SFR--mass relation shows no indication for
evolution, with slopes of \referee{\refereetwo{$a$} $= 0.54 \pm\ 0.16$ at $z\sim 6$ and $0.70
\pm\ 0.21$  at $z\sim 4$.} Furthermore, the scatter in
SFR at fixed stellar mass shows no evidence for evolution, with a
range of $\sigma (\log \mathrm{SFR/\mathrm{M}_\odot  
\mathrm{yr}^{-1}}) = 0.2  -  0.4$ dex from the median.

\begin{figure*}[!t]
\epsscale{1.1}
\centerline{\includegraphics[scale=0.49]{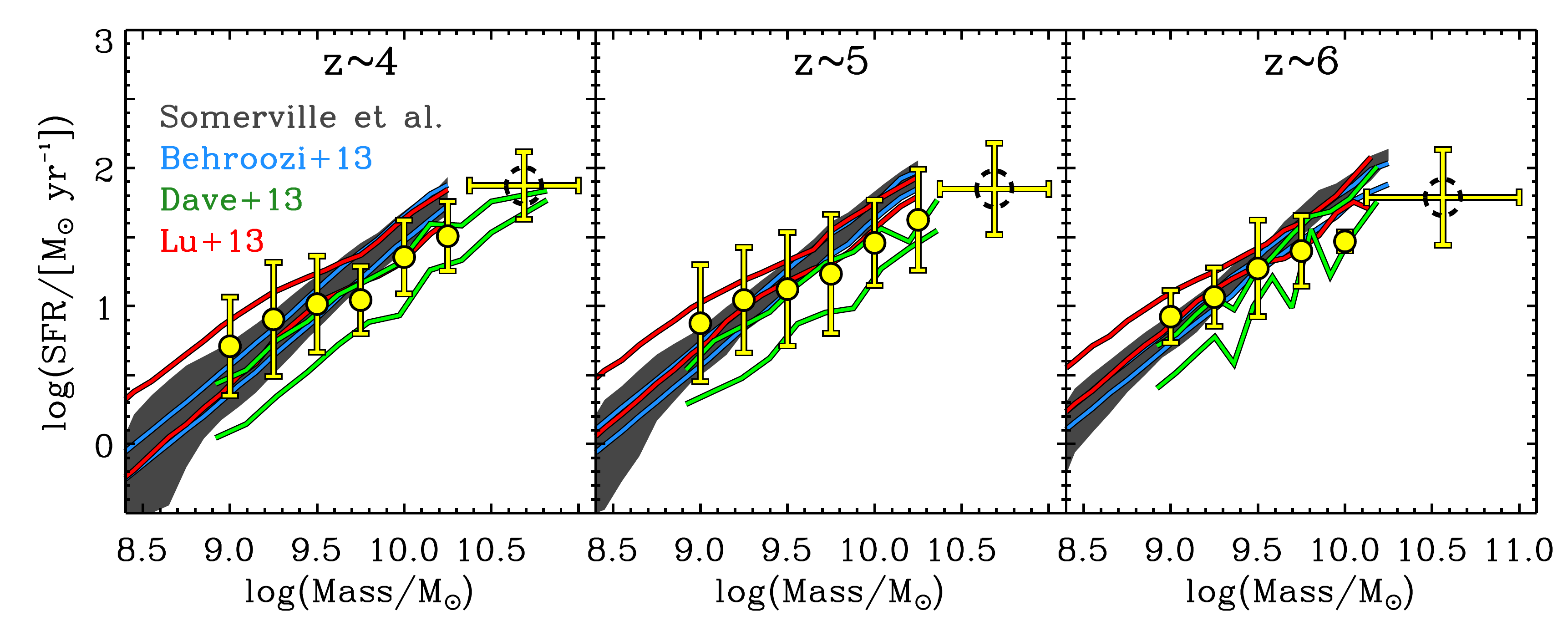}}
\caption{The SFR--stellar mass relation predicted from the \referee{models}.  
Each set of lines or shaded swatch shows the $\pm \sigma$-range of
 galaxies from each \referee{model} as given in the
  inset.   The yellow points and errors bars show the measured
  relation for the CANDELS samples and are identical to the points in
  Figure~\ref{fig:massSFR}.  The zero point, scatter and slope of the
  SFR--stellar mass relation from the models is consistent with the
  measured values over this redshift range. \vspace{1mm}}
\label{fig:massSFRSAMs}
\end{figure*}

\begin{table*}
\begin{center}
\label{tab:SFRMass_fits}
\caption{SFR--Stellar Mass Best-Fit Parameters}
$z\sim4$ \\[0.5mm]
\begin{tabular}{l|ccccc}
\hline \\[-2.9mm]
                       & \refereetwo{Slope $a$}\footnote{Slope of the medians in SFR as a function of stellar mass (Fig.s \ref{fig:massSFR} and \ref{fig:massSFRSAMs}) for Salpeter masses $\log\ M_\star/\mathrm{M}_\odot  >$ 9.}  & \refereetwo{Zero Point $b$} & \refereetwo{$b$ when $a\equiv 1$} & $\langle \sigma_\mathrm{MAD} \rangle$ \footnote{The \refereetwo{observed $\sigma_{\text{MAD}}$ scatter (see \S~\ref{sec:selection} for $\sigma_{\text{MAD}}$ definition)}, averaged over bins of stellar mass, in the SFR--stellar mass relation.} & $\chi^2$ \footnote{Goodness-of-fit of SFR--stellar mass best-fit trend.} \\[1mm] \hline
This work                              & 0.70 $\pm$ 0.21 & -5.7 $\pm$  2.1 & -8.64 $\pm$ 0.11 & 0.35 (0.31)\footnote{\refereetwo{The value in parenthesis is} the average range in the bootstrapped errors calculated by the Monte Carlo on stellar mass and SFR (see \S~\ref{sec:MC}). \refereetwo{Both the observed scatter and the Monte Carlo errors are used to calculate the intrinsic scatter using equation~\ref{equ:quadrature}.}} & 0.38 \\
Somerville et al.                 & 1.1   $\pm$ 0.13 & -9.0 $\pm$  1.3 & -8.47 $\pm$ 0.05 & 0.18 & 0.06 \\
Behroozi et al. 2013           & 1.1   $\pm$ 0.07 & -9.2 $\pm$  0.7 & -8.47 $\pm$ 0.03 & 0.10 & 0.17 \\
Lu et al.                              & 0.80 $\pm$ 0.11 & -6.5 $\pm$  1.1 & -8.45 $\pm$ 0.04 & 0.14 & 0.44 \\
Dave et al. 2013                  & 0.80 $\pm$ 0.05 & -6.8 $\pm$  0.6 & -8.95 $\pm$ 0.03 & 0.16 & 2.1 \\[1mm] \hline
\end{tabular}
\\[1.5mm]
$z \sim$5 \\[0.5mm]
\begin{tabular}{l|ccccc}
\hline \\[-2.9mm]
  & \refereetwo{Slope $a$} & \refereetwo{Zero Point $b$} & \refereetwo{$b$ when $a\equiv 1$} & $\langle \sigma_\mathrm{MAD} \rangle$  & $\chi^2$ \\[1mm] \hline
This work 		& 0.59 $\pm$ 0.26 & -4.4 $\pm$  2.6 & -8.49 $\pm$ 0.14 & 0.41 (0.29)$^{\mathrm{d}}$ & 0.05 \\
Somerville et al.    	 & 1.0   $\pm$ 0.09 & -8.6 $\pm$  0.9 & -8.29 $\pm$ 0.03 & 0.13 & 0.07 \\
Behroozi et al. 2013 	& 1.0   $\pm$ 0.05 & -8.6 $\pm$  0.5 & -8.32 $\pm$ 0.02 & 0.07 & 0.35 \\
Lu et al.    		& 0.79 $\pm$ 0.07 & -6.3 $\pm$  0.7 & -8.29 $\pm$ 0.02 & 0.10 & 0.78 \\
Dave et al. 2013 		& 0.80 $\pm$ 0.07 & -6.7 $\pm$  0.7 & -8.72 $\pm$ 0.02 & 0.15 & 1.5 \\[1mm] \hline
\end{tabular}
\\[1.5mm]
$z \sim$6 \\[0.5mm]
\begin{tabular}{l|ccccc}
\hline \\[-2.9mm]
  & \refereetwo{Slope $a$} & \refereetwo{Zero Point $b$} & \refereetwo{$b$ fwhen $a\equiv 1$} & $\langle \sigma_\mathrm{MAD} \rangle$  & $\chi^2$ \\[1mm] \hline
This work 		& 0.54 $\pm$ 0.16 & -3.9 $\pm$  1.6 & -8.45 $\pm$ 0.06 & 0.21 (0.34)$^{\mathrm{d}}$ & 0.10 \\
Somerville et al.            	& 1.0   $\pm$ 0.06 & -8.5 $\pm$  0.6 & -8.16 $\pm$ 0.02 & 0.10 & 0.47 \\
Behroozi et al. 2013 	& 0.96 $\pm$ 0.05 & -7.8 $\pm$  0.5 & -8.21 $\pm$ 0.02 & 0.07 & 0.81 \\
Lu et al.   		& 0.77 $\pm$ 0.07 & -6.0 $\pm$  0.7 & -8.15 $\pm$ 0.02 & 0.10 & 1.3 \\
Dave et al. 2013 		& 1.1   $\pm$ 0.10 & -9.6 $\pm$  0.9 & -8.29 $\pm$ 0.03 & 0.15 & 6.7 \\[1mm] \hline
\end{tabular}

\end{center}
\end{table*}

We must consider the possibility that the scatter in SFR at fixed mass is
higher, and we are simply missing galaxies with low SFR due to
incompleteness.    We consider this unlikely because even if star formation
ceased in some fraction of the galaxies, the galaxies would require
$0.5  -  1$ Gyr to have their SFR drop below a detectable threshold in the
WFC3 IR data.  These timescales are comparable to the period of time
spanned by our subsamples (i.e.\ the lookback time between $z  =  4.5$ and
3.5 is only 480 Myr), so it seems unlikely galaxies would
``instantly'' move from the observed SFR--mass sequence to undetectable
values.   For example, if such low-SFR objects existed at $z  =  4$ their progenitors 
should be seen at $z  =  5$ and 6 as they are fading, inducing a larger 
scatter in SFR--stellar mass.  This work finds no evidence
for such a population in our sample.  Parenthetically, we note that some studies
report evidence for massive, $\log M_\star/\mathrm{M}_\odot > 10.6$ dex, 
quiescent galaxies at $z\sim 3-4$,  but this population lies at stellar masses 
above those in our sample \citep{Straatman13,Muzzin13,Spitler14}.  

\referee{We note that our SFR--stellar mass relation is tighter than our \Muv--stellar mass relation 
(Fig.~\ref{fig:m1500mass}), and we find \refereetwo{that this can} be explained by 
a correlation between stellar mass and our derived dust attenuation (there is no 
correlation between derived attenuation and \Muv). For example, objects at masses of 
$10^{8.5}$, $10^{9.5}$, and $10^{10.5}$ \Msol\ have median marginalized 
$E(B-V)$ values of 0.05$\pm$0.03, 0.13$\pm$0.07, and 0.32$\pm$0.18 
respectively. This relation accounts for the differences in the scatter seen in  
Figures~\ref{fig:m1500mass} and \ref{fig:massSFR}.}

\subsubsection{Constraints on the Intrinsic Scatter in the SFR--Mass Relation}\label{sec:MC}

Before comparing against models, it is necessary to understand how much of the scatter in the
SFR--mass relation is intrinsic to the galaxy population and how much
is a result of observational errors in SFR and stellar mass.
To a simple approximation, the observed scatter 
\referee{(yellow in Fig.~\ref{fig:massSFR})} is a combination of the
intrinsic (true) scatter and the measurement errors added \refereetwo{in
quadrature, }
\begin{equation}
\label{equ:quadrature}
\sigma_\mathrm{observed} = ( \sigma_\mathrm{intrinsic}^2 +
\sigma_\mathrm{errors}^2)^{1/2} .
\end{equation}

The SFR--mass joint probability density is broad, with covariance
between the SFR and stellar mass (e.g., Fig.~\ref{fig:2DPDF_sfrmass}).   Because we
calculate the posterior probability density functions for both the
stellar mass and SFR for galaxies in our samples, we are able to
estimate how correlations in these parameters contribute to the
scatter and slope of the SFR and stellar mass relation.
\refereetwo{Here we use a Monte Carlo simulation to estimate $\sigma_\mathrm{errors}$, 
and to determine how these errors affect the slope of the SFR--stellar mass relation.}

We set up the Monte Carlo as follows.  As discussed above, the SFR and
stellar mass posteriors are covariant because both involve the dust
attenuation, $A_\mathrm{UV}$, where models with higher
$A_\mathrm{UV}$ have higher SFR from dust corrections, and higher
mass-to-light ratios, which produce  higher stellar masses.    For
each galaxy in each subsample at $z  =  4$, 5 and 6, we randomly sample
the galaxy's posterior density function
of $A_{\text{UV}}$ to find a new UV attenuation, $A_{\text{UV},i}$.
We then compute the conditional posterior for the stellar mass, 
$P($\Mstar $|A_{UV,i})$.    Next, we derive the SFR from equation
\ref{equ:sfr} and the medians of SFR in bins of stellar mass are
re-calculated. This process is repeated $10^4$ times for each galaxy
to generate $10^4$ new realizations of our galaxy sample.  We then
calculate at each stellar mass the median SFR and compute the
$\sigma_\mathrm{MAD}$ from the distribution of medians.   The
scatter in $\log$ SFR from this Monte Carlo is shown in Figure~\ref{fig:massSFR}
and Table~\ref{tab:SFRMass_medians}.

\refereetwo{The $\sigma_\mathrm{MAD}$ scatter in the SFR from the 
Monte Carlo simulations is comparable to the observed SFR scatter, 
$\sigma_\mathrm{observed}$, in most bins of mass and redshift. 
The scatter in the SFRs at fixed stellar mass from the
Monte Carlo are shown in Figure~\ref{fig:massSFR} and given in
Table~\ref{tab:SFRMass_medians}. \refereetwo{We make the approximation 
that $\sigma_\mathrm{errors} = \sigma_\mathrm{MAD}$ from the Monte Carlo. 
We subtract these in quadrature from the observed SFR scatter to estimate the 
intrinsic scatter in SFR at fixed mass using equation~\ref{equ:quadrature}.
We find that} the average intrinsic scatter in SFR across the mass bins to be
$\sigma = 0.26 \pm 0.04, 0.23 \pm 0.10,$ and $
0.34 \pm 0.11$ at $z \sim 4, 5,$ and 6, \refereetwo{respectively.}
In some instances, the measurement errors from the Monte Carlo accounts 
for more scatter than the observed scatter, in which case there is no meaningful constraint 
on the intrinsic scatter.  In this case, we take the Monte Carlo measurement
scatter alone as a conservative limit on the intrinsic scatter (as \textit{some} of the 
errors on the derived quantities must arise from the intrinsic scatter).}
This has implications for the gas accretion rate that we discuss below in \S~\ref{sec:massSFRdiscuss}.

\referee{The above test ignores the effects of our photometric redshift uncertainties
since redshift is a fixed parameter during the fitting process. 
We constructed the following test to determine the effects of redshift
uncertainties on SFR and stellar mass.
We randomly selected 100 objects from each redshift sample and performed a 
Monte Carlo on their redshift uncertainty. In the Monte Carlo,
each object's redshift was re-assigned according to a Gaussian 
error distribution with a sigma equal to the object's 68\% photometric redshift 
uncertainty. Then, we derived the stellar masses and SFRs in the same manner as with the data,
fixing the redshifts to be the new redshift values. We calculated the medians in the SFR--stellar 
mass relation, as in as in Figure~\ref{fig:massSFR}, for each of $10^{4}$ realizations of this process. 
Finally, we found the median and $\sigma$ of SFR in each stellar mass bin from
the distribution of SFR medians that each Monte Carlo realization produced.}

\referee{Redshift errors produce a higher median $\log$ SFR of $<$0.1 dex per stellar mass
bin, and the redshift errors can contribute as much as $\sim$0.1 dex to the scatter in every 
stellar mass bin  (usually it is much smaller, contributing $<$0.03 dex for 50\% of 
the stellar mass bins). Therefore the redshift uncertainties do not significantly 
contribute to the error budget of the SFR--stellar mass relation.}

\refereetwo{We cannot rule out the possibility that a population of dusty,  low-SFR
galaxies are missing from our sample, which would attribute more scatter to the
SFR--stellar mass relation. Indeed, some recent studies find evidence that such a population
may exist at high redshift, at least at high stellar masses \citep{Spitler14, Man14}.  Furthermore, at 
low stellar masses, our sample may be biased toward objects experiencing recent ``bursts''
of star formation \citep{Schreiber14}.  A deep investigation with ALMA is needed for 
further confirmation \citep{Schaerer14}. However, our redshift
range limits the data to rest-frame UV-to-optical wavelengths, and we defer the search for
such a population for future work. }

%
%


\subsubsection{Comparison to the SFR--Stellar Mass Relation in Models}\label{sec:SFRmassSAMs}

This subsection compares the results of the SFR--stellar mass relation in the 
previous section to results of recent SAM (Somerville et al., Lu et al.) 
semi-empirical dark matter abundance matching \citep{Behroozi13d}, and
hydrodynamic \citep{Dave13} simulations. Each of these
simulations were briefly summarized in section
\ref{sec:describeModels} and are collectively referred to as 
``the models''. 

Figure \ref{fig:massSFRSAMs} shows the scatter in the SFR--stellar
mass relation for each of  the models as compared to the observed median
and scatter in the SFR at fixed stellar mass (with data and errors
bars identical to those in Figure  \ref{fig:massSFR}).   For each
model, the median and scatter in the SFR at fixed mass was computed 
in the same way as the data (with all models converted to a
\cite{Salpeter55} IMF, as assumed in this work).   

The SFR--mass relations from each model are in general agreement with
each other and imply a tight relation between SFR and stellar
mass exists for galaxies at high redshift.   The SFR--mass
relations from the models are also very similar to the observed relation 
derived from the data.  Though some models predict a steeper slope of near
unity (\refereetwo{$a$}$  \simeq 1$), higher than measured in this work 
\referee{(\refereetwo{$a$}$ \simeq 0.6$)}, the difference is negligible as it is within the errors.   
The observed offset in the zero point between the models and
observations is likewise insignificant as the zero point is strongly
anticorrelated with the slope in the linear fits of this work.   In addition,
from Figure \ref{fig:SAMcompare} it was shown that the recovered SFRs 
from \referee{tests with the mock data tended to under- (over-) estimate the 
SFR of model galaxies with high (low) SFRs.}  No attempt was made to correct this
for systematic. Therefore the similar offsets seen in Figure~\ref{fig:massSFRSAMs} 
between the high-SFR objects may imply that the data and SAMs are in even 
closer agreement. 

As listed in Table~\ref{tab:SFRMass_fits},
the scatter in the SFR at fixed mass is typically $0.1-0.2$~dex in the
models, whereas the limits on the intrinsic scatter from the data are
\referee{$\sigma(\log \mathrm{SFR}) <0.2-0.3$~dex (see S~\ref{sec:MC}).} Therefore, both the
observations and models support the conclusion that the SFR in
galaxies at $3.5 < z < 6.5$ at fixed mass 
(for $\log M_\star/\mathrm{M}_\odot > 9$ dex) scales nearly
linearly with increasing stellar mass and does not vary
by more than a factor of order 2.   We explore the implication this
has for the net gas accretion rate below in \S~\ref{sec:massSFRdiscuss}.

\section{Discussion}\label{sec:discussion}

\subsection{Implications of the SFR--Stellar Mass Relation}\label{sec:massSFRdiscuss}

The fact that there is a tight SFR--stellar mass relation implies that
the SFR scales almost linearly with stellar mass  for the galaxies in
our sample at $3.5 < z < 6.5$.   Because the SFR is (to a coarse
approximation) the time derivative of the stellar mass, this implies
that the SFR is an increasing function with time.   We explore this
further below in \S~\ref{sec:cnd}.   Furthermore, the tightness of the
scatter indicates that there is little variation in the SFR at fixed
stellar mass.   One caveat is that the SFRs are based on the galaxies'
UV luminosity.  Therefore, the SFRs that we measure are the 
``time-averaged'' over the time it takes for the UV luminosity in galaxies to
respond to changes in their instantaneous SFR.   For the UV luminosity
this timescale is approximately 30-100 Myr \citep[e.g.,][]{Salim09}. 
\refereetwo{Recent simulations have shown that the scatter is highly
sensitive to the timescale of the SFR indicator \citep{Hopkins14,
Dominguez14}. Therefore, the tightness in the SFR--mass relation 
of this work is conditional on the timescale associated with UV SFRs.  
With that in mind, the scatter observed in this work 
implies that galaxies at fixed mass in our sample have similar 
star formation histories when averaged over this timescale.}

The scatter in the SFR--mass relation has important implications for
net interplay between gas accretion into halos and galaxy feedback and
outflows at these redshifts.   The gas-accretion rate is a
crucial piece of physics in galaxy formation models, and measures of the
scatter in the SFR--mass relation are therefore an important test to constrain
simulations and SAMs.   As discussed above, the SFR
is expected to track the net accretion rate/outflow rate of baryonic
gas into the galaxies' dark-matter halos \citep{Keres05, 
Dekel09, Bouche10, Lu13a}.  
Because the SFR--mass relation is tight, it then follows that the gas accretion
rate has little variation ($\sigma \sim 0.2-0.3$ dex, or a factor of
$<$ 2) at fixed stellar mass.    Therefore, this favors a relatively
smooth gas accretion process for galaxies at $3.5 < z < 6.5$, at least
above $\log M_\star/\mathrm{M}_\odot > 9$ dex.

\subsection{Evolution of the SFR}\label{sec:cnd}

\begin{figure}
\epsscale{1.1}
\centerline{\includegraphics[scale=0.35]{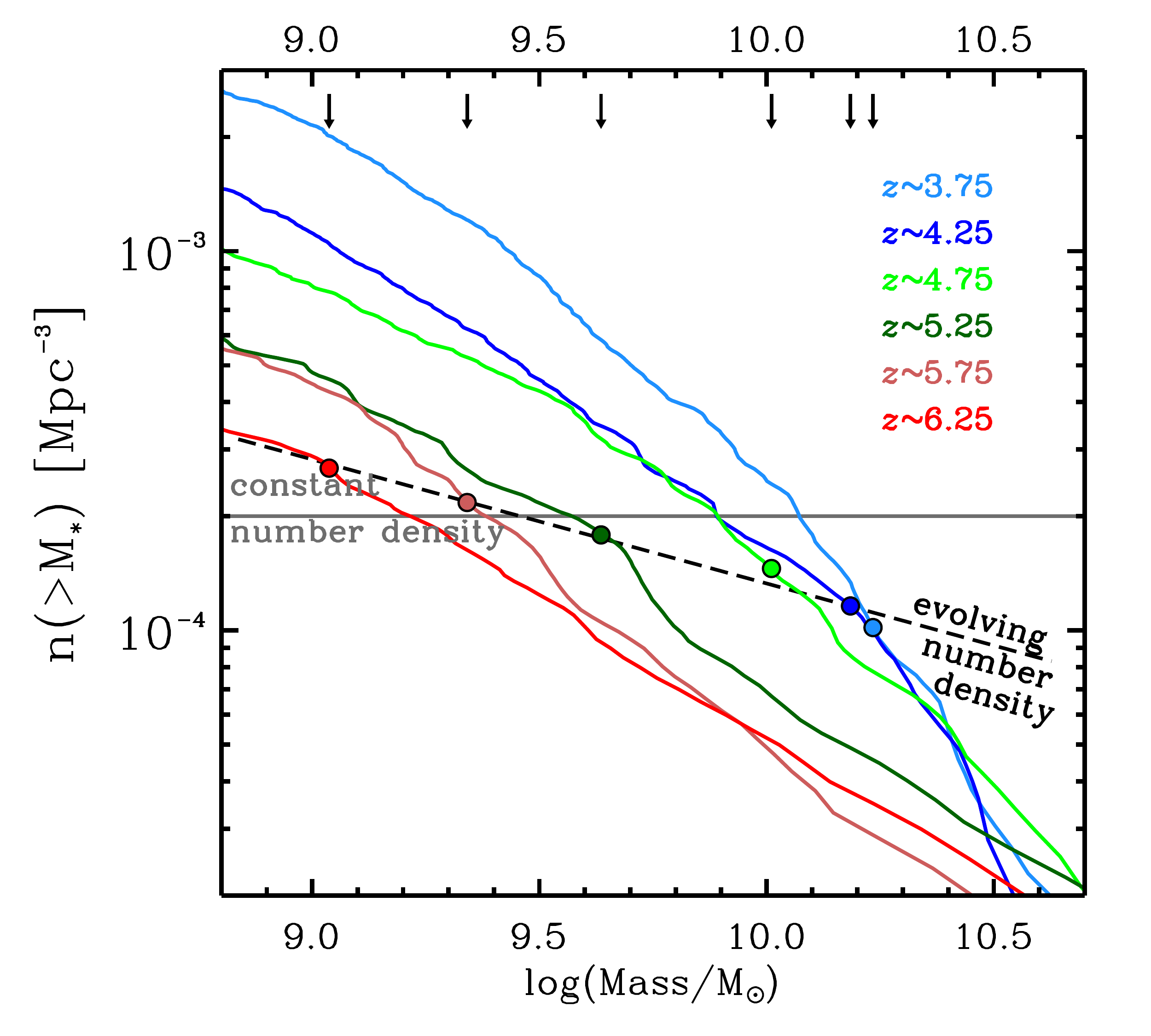}}
\caption{Cumulative stellar mass functions in bins of redshift.  No
  corrections have been applied for completeness, but 
\cite{Duncan14} show these corrections are negligible for $\log M_\star/\mathrm{M}_\odot >
  9$ dex.     The arrows and circles indicate the stellar mass evolution
  of the progenitors of galaxies with an evolving number density with 
$\log  n(>M_\star)$/Mpc$^{-3} = -4$ at $z  = 3.5$ using the evolution
  parameterized by \cite{Behroozi13c}.  We measure the SFR evolution
  of these galaxies.   As we discuss below, we would have
  inferred a very similar evolution in SFR for galaxies selected at
  constant number density, with  $\log  n(>M_\star)/$Mpc$^{-3} = -3.7$, at
  all redshifts. \vspace{3mm}}
\label{fig:MassFunctions}
\end{figure}

\begin{figure}
\epsscale{1.1}
\centerline{\includegraphics[scale=0.35]{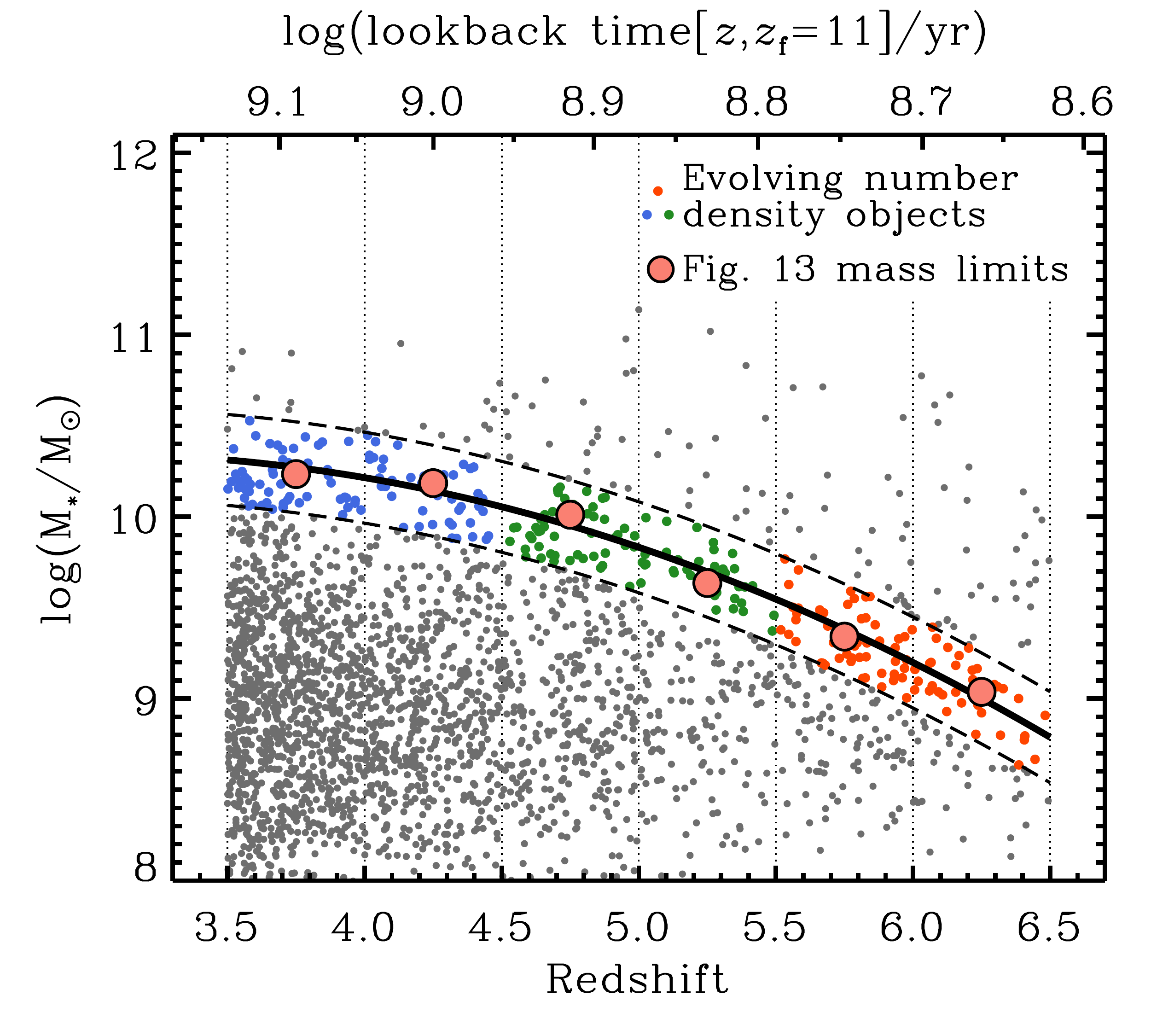}}
 \caption{The selection of galaxies according to the \cite{Behroozi13c} evolving 
number density of $\log n(>M_\star$)/Mpc$^{-3} = -4$ at $z  =  3.5$.   
The large salmon-colored circles show
   the median stellar mass evolution, and the dashed lines illustrate our sample
   selection of $\pm 0.25$~dex in stellar mass about these median
   values.   We select all galaxies within these lines, and use them
   to derive the star formation history. \vspace{3mm}}
\label{fig:CNDselection}
\end{figure}

\begin{figure}
\epsscale{1.1}
\centerline{\includegraphics[scale=0.35]{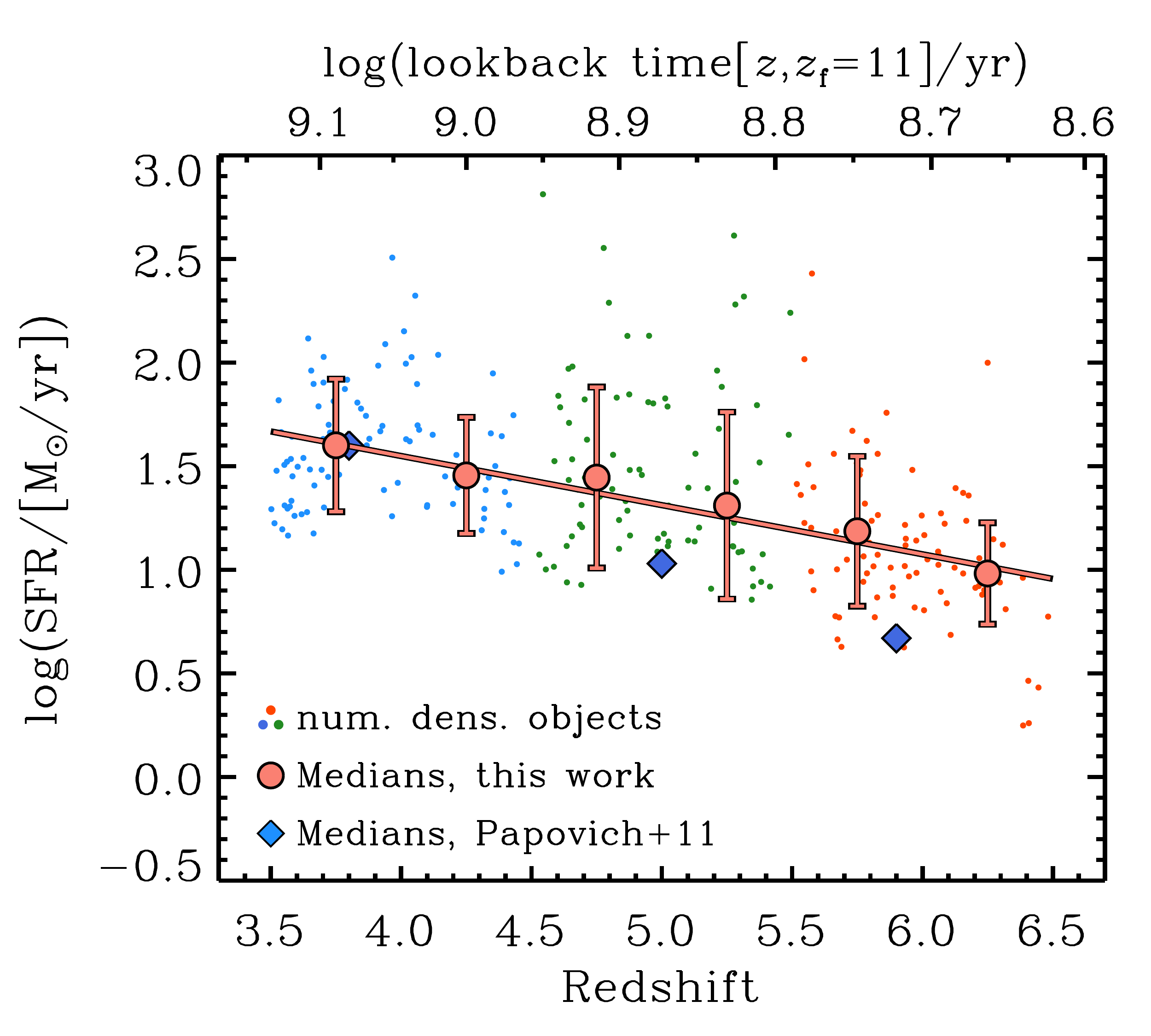}}
 \caption{The SFR history (the SFR as a function of redshift) for
 galaxies selected by their (evolving) number density to track the 
evolution in the progenitors of galaxies at $z  =  3.5$
 with $\log n/\mathrm{Mpc}^{-3} = -4$.  The galaxies from each redshift
 subsample $z  =$ 4, 5, and 6 are indicated as blue, green, and orange
 points respectively.   The larger salmon-colored circles and error
 bars show the median SFR and $\sigma_\mathrm{MAD}$ in bins of $\Delta z =
 0.5$. An average rising star formation history is derived for this
 redshift range  that can be represented by a power law $\Psi=
 t^{\gamma}$ where $\gamma=   1.4  \pm   0.1$.  This evolution is
 somewhat shallower than that found by Papovich et al.\ (2011), but
 consistent within the error budget. \vspace{3mm}}
\label{fig:CNDresult}
\end{figure}

As discussed above, \referee{a tight, linear} relation between SFR and
stellar mass implies an SFR that increases with time.  In this section,
we study the SFR history directly to see \refereetwo{if} it is consistent
with the observed SFR--stellar mass relation. This is achieved by
tracking the  evolution of the progenitors of $z  =  4$ galaxies by 
selecting galaxies at different redshifts based on
their number density. 

\referee{Many studies have shown that a constant (comoving) number-density selection
can trace the progenitor and descendant evolution both to relatively
low and high redshifts} \citep[e.g.,][]{vanDokkum10, Papovich11,
Lundgren13, Leja13a, Patel13, Tal14}.  In addition, recent studies 
have suggested using an evolving number-density selection
to better track the progenitor populations of galaxies
\citep[e.g.,][]{Leja13, Behroozi13c}.    Here, we use the
parameterization of \cite{Behroozi13c}, who provide simple
functions to track the number density evolution of the progenitors of galaxies.  

This number density evolution is used to select the progenitors of
galaxies in our sample.  Figure~\ref{fig:MassFunctions} shows the
cumulative stellar mass functions for the galaxies in our $3.5 < z <
6.5$ CANDLES samples in bins of redshift.   The results of 
\cite{Duncan14} show that the objects in this field are complete for
masses greater than $\log  M_\star/ \mathrm{M}_\odot =$ 8.55, 8.85, 
and 8.85 dex at $z\sim$ 4, 5 and 6 respectively.  We assume a
survey area of  170 arcmin$^2$ as described by \cite{Koekemoer11} and
the co-moving volume is calculated at each redshift assuming an
uniform depth across each field.   These cumulative functions are used
to determine the stellar mass at which the galaxies of that redshift
range achieve a given evolving \nd\ as described by \cite{Behroozi13c}.
As indicated in the figure, we take the galaxies with stellar mass
$\log M_\star/\mathrm{M}_\odot = 10.2$ dex at $z  =  3.75$, which have a number density
\refereetwo{of} $\log n/\mathrm{Mpc}^{-3} = -4$, and identify the galaxies 
at higher redshift that have a stellar mass that corresponds to the
appropriate (de-evolved) number density at that redshift.  

\refereetwo{Figure~\ref{fig:CNDselection} illustrates our criteria to select
objects according to an evolving \nd. We use the Figure~\ref{fig:MassFunctions} 
stellar mass limits} to find a best-fit curve across 3.5 $<  z  <$ 6.5. Then, we 
select objects from our data that are $\pm$ 0.25 dex in stellar
mass about this relation.  For the remainder of this work, we refer to these objects  as
``evolving \nd--selected''.  We will later compare these briefly to
objects selected at ``constant \nd'' as such samples have received
attention in the recent literature.   

\referee{Figure~\ref{fig:CNDresult} shows the average SFR as a function 
of redshift for the evolving \nd--selected galaxies.    
The SFR clearly increases as a function of time (decreasing redshift).
We fit this evolution with a power law, $\Psi (t) \sim
(t/\tau)^\gamma$ where $\gamma$ = 1.4 $\pm$ 0.1 and  $\tau=92 \pm 14$ Myr.
\refereetwo{If our sample is incomplete at low stellar masses
($\log M_\star/\mathrm{M}_\odot < 9.5$ dex), then this would influence 
the measured power, $\gamma$.}  Based on Figure~\ref{fig:MassFunctions}, the lower-mass 
objects will suffer greater incompleteness for objects of low SFR.  This could 
mean that the intrinsic power-law slope is steeper than the one we measure here.
We also note that we observe little difference between this evolution 
and that derived from using a constant-\nd-selection.}

\begin{figure}
\epsscale{1.1}
\centerline{\includegraphics[scale=0.39]{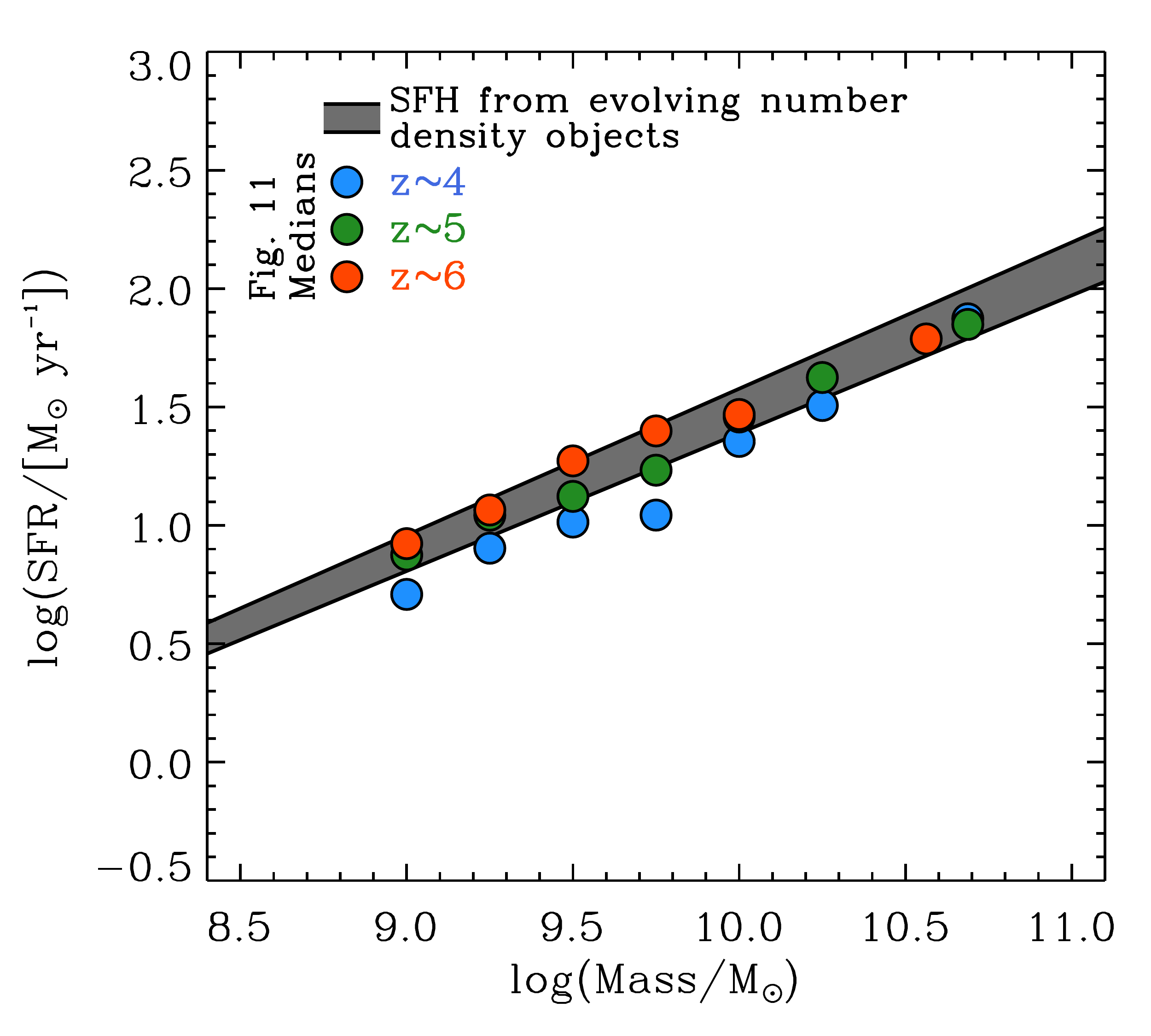}}
\caption{The median values of SFR and stellar mass
relation from Figure~\ref{fig:massSFR} are shown, color coded by
redshift. The gray region line is not a fit to the points, but is instead
the implied relation (with errors) using the measured SFR history (from Fig.~\ref{fig:CNDresult}) 
 derived by integrating that SFR history with the Bruzual \&
Charlot SPS models.  For the $z  =  4$ galaxies with $\log
M_\star/\mathrm{M}_\odot = 10.2$ dex there is good agreement between the observed
SFR--mass relation and the implied value from the SFR history.  This is
reassuring as the derived star formation history corresponds to the
progenitors of galaxies of this mass at this redshift.  At lower
stellar masses, the SFR--mass relation implied from the derived
star formation history underproduces the SFR, but we
attribute this to the fact that the SFR history of lower-mass galaxies
evolves less steeply with time.   \vspace{1cm}}
\label{fig:massSFRCND}
\end{figure}

Lastly, we can explore whether the SFR evolution derived above produces
an average SFR--mass relation as measured from the data.  \refereetwo{We took}
 the SFR history derived above and compute the resulting
stellar mass and SFR for a stellar population starting at $z  =  6$ and evolving
to $z  =  4$ using the BC03 SPS models (see
discussion in \S~\ref{sec:mass-m1500} and bottom right panel of 
Figure~\ref{fig:m1500mass}).    We plot the
resulting SFR--mass relation in Figure~\ref{fig:massSFRCND} along with
the medians derived above in Figure~\ref{fig:massSFR}.   Formally, the
star formation history derived here corresponds to a galaxy with
stellar mass $\log M_\star/\mathrm{M}_\odot = 10.2$ dex at $z  =  3.75$.  
Looking only at that point, the
SFR--mass relation inferred from the SFR history matches the SFR--mass
relation at $z\sim 4$ remarkably well.  (This is not circular because the
stellar mass evolution is measured from the SFR evolution and
\textit{not} from the data itself.)  

\subsection{Evolution of the sSFR}\label{sec:ssfr_sec}

\begin{figure}
\epsscale{1.1}
\centerline{\includegraphics[scale=0.40]{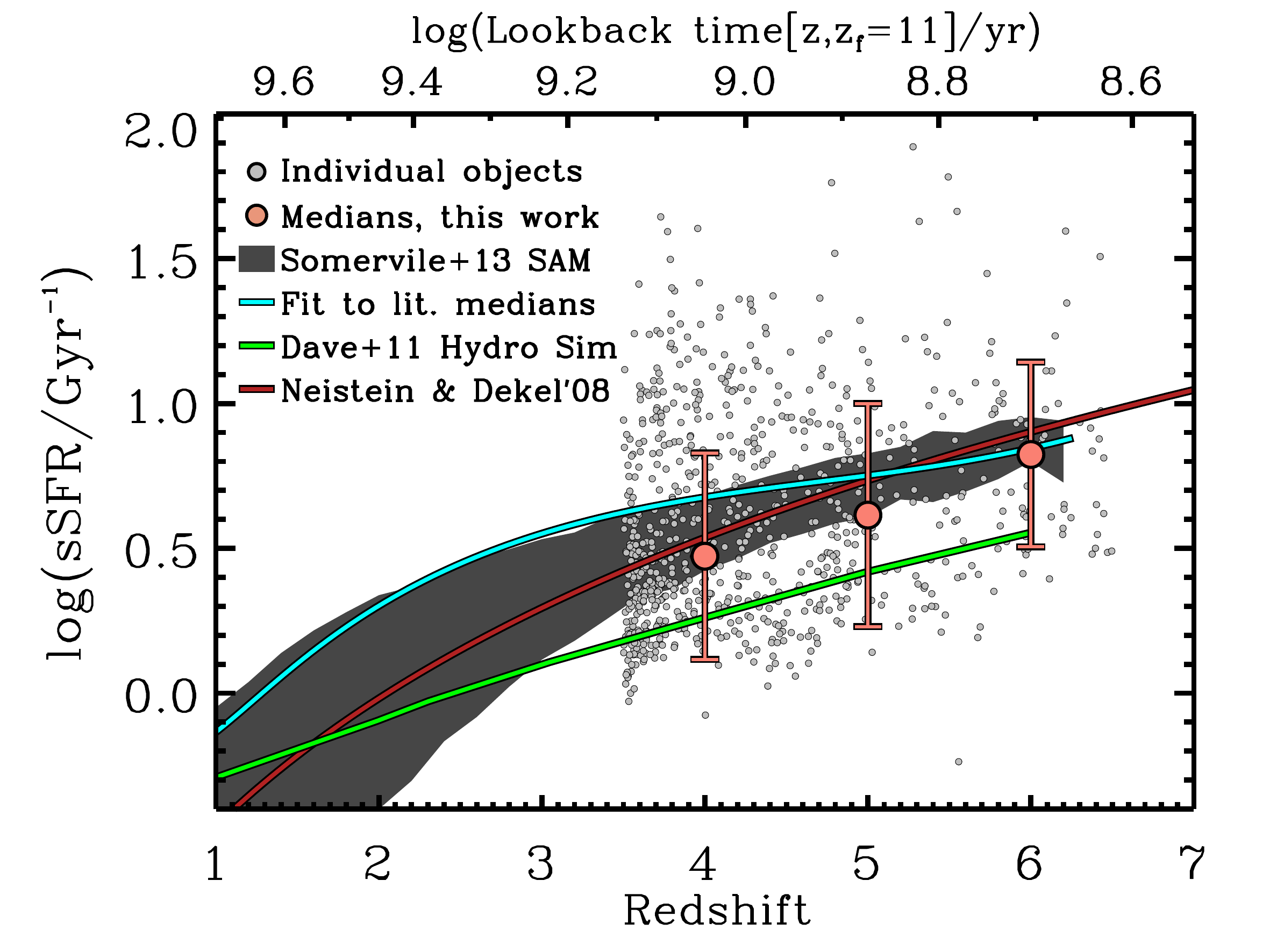}}
 \caption{The evolution of the sSFR as a function of
redshift for galaxies selected with constant stellar mass ($\pm$0.5
dex of  $\log{} M_\star/\mathrm{M}_\odot  = 9$ dex). Grey points represent individual
objects, while salmon-colored circles are medians  in $\Delta z  =  0.5$
bins of redshift. The cyan curve
shows a fit to the literature medians  across all redshift.   We find
that at constant stellar mass, the sSFR gradually decreases over our redshift
range.   The other lines and shaded region correspond to model
predictions as listed in the legend.  The zero point and rate of
evolution between the models and data are similar.   \vspace{3mm}}
 \label{fig:sSFRzSMC}
\end{figure}


Lastly, we use our derived SFRs and stellar masses to study the
evolution of the sSFR. 
We first explore the evolution with redshift at fixed stellar mass as
this has received attention in the literature (even though it is clear
that galaxies with such high SFRs will not remain at the same stellar
mass over this redshift range).  We then consider the evolution
in sSFR for the evolving \nd-selected sample described above, as this will better
track the evolution in galaxy progenitors across this redshift range.

Figure~\ref{fig:sSFRzSMC}  shows the evolution of the sSFR for objects
selected with constant stellar mass:  within $\pm$0.5 dex of
$\log M_\star/\mathrm{M}_\odot$ = 9.    The sSFR for objects at this stellar
mass increases with increasing redshift, with high scatter.  \refereetwo{This} evolution
is consistent with hydrodynamic simulations \citep{Dave11a}, SAMs
(Somerville et al., Lu et al.), \refereetwo{and other} recent
observational results from the \referee{ literature \citep{Stark13,Gonzalez14}. }

\begin{figure}
\epsscale{1.1}
\centerline{\includegraphics[scale=0.40]{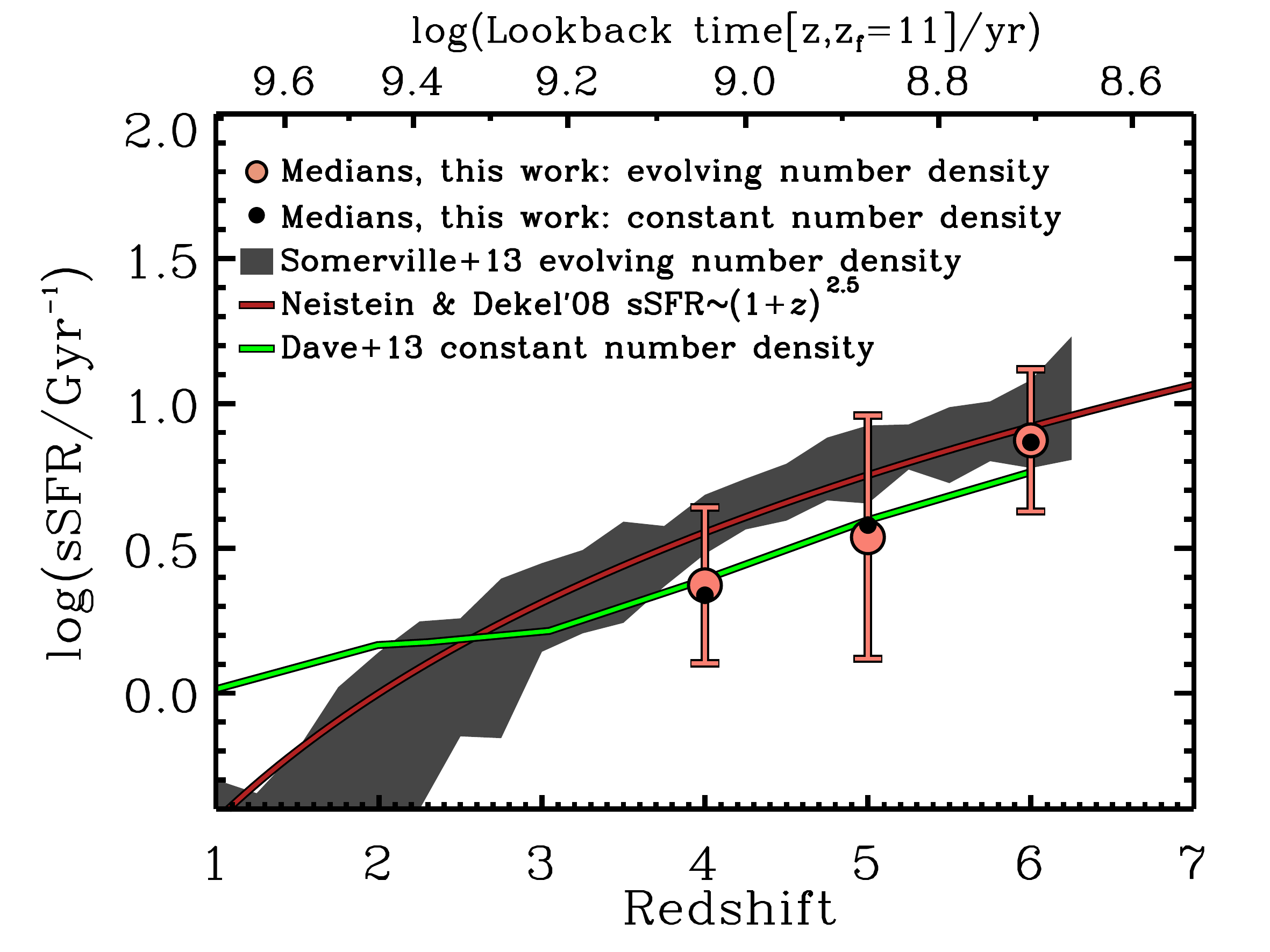}}
 \caption{The evolution of the sSFR as a function of redshift
for  galaxies selected by their evolving number density to track the
progenitors of galaxies at $z  =  3.5$ with $\log n/\mathrm{Mpc}^{-3} = -4$.
 The larger salmon-colored circles and error bars show the medians of
$\log$ sSFR in bins of redshift.  The other lines and shaded region correspond to model
predictions as listed in the legend$^{19}$.
\vspace{3mm}}
 \label{fig:sSFRzCND}
\end{figure}

\referee{Figure~\ref{fig:sSFRzCND}} shows that the sSFR increases with redshift
when galaxies are selected with an evolving \nd.    
The evolution of the simple cosmological accretion models\footnote{Note the 
\cite{Neistein08} model has a halo-mass dependence and tracks cosmological 
accretion, not a \nd--selected evolution.}
where sSFR$\sim\ (1+z)^{\phi=2.5}$ \citep[e.g.,][]{Neistein08} 
is consistent with the sSFR evolution found in this work (Fig.~\ref{fig:sSFRzCND}), 
which has $\phi = 3.4\pm 2.5$ from $z$=3.5 to $z$=6.5. 
There is a weak difference between the evolution of the evolving \nd-selected sample
and the sample at constant mass in Figure~\ref{fig:sSFRzSMC};  the
sSFR evolution for the evolving \nd--selection is shifted to lower
values as a function of redshift.      This is basically a \refereetwo{confirmation}
of the fact that the SFR--mass relation produces an sSFR that is nearly
independent with stellar mass:  sSFR = SFR / $M$ $\sim$
$M^{a - 1}$, where sSFR $\sim M^{-(\approx 0.4)}$ using our results in
Table~\ref{tab:SFRMass_fits}.   This implies that samples selected at constant
stellar mass or an evolving stellar mass (from a number density
selection) return only slightly different sSFR values because the sSFR
is a slowly changing function with stellar mass, \refereetwo{as recently found by 
\cite{Kelson14}.}  \referee{Consequently,
this permits mass-independent modeling of the specific 
accretion rate, and therefore the sSFR \citep[see][for a detailed discussion]{Dekel14}.}

One way to explain the slight offset of the observed sSFR
to lower values than the predictions from the SAMs is that some
feedback  mechanism may be present to hinder SFR from tracing halo or
stellar  mass growth. \cite{Gabor14} recently showed that cold streams
of gas can increase the velocity dispersion in star-forming
disks. They  show that at $z >$ 3 this increased turbulence causes the
gas-mass to  stay higher, but reduces the SFR / gas-mass by a factor
of two.   Assuming gas-mass traces the baryonic-mass in galaxies at $z
>$ 3, then this could explain why the observed median sSFRs of this work
are lower by about a factor  of order two compared to other
models. 

%

%


\section{Conclusions}\label{sec:conclusions}
In this work we use a photometric-redshift sample selected from the
CANDELS GOODS-S field to study the average  population of galaxies
across the redshift range $3.5<z<6.5$.  We present a Bayesian SED
fitting procedure that takes advantage of the full posterior to
determine the physical properties (stellar mass, SFR) of each galaxy.
Our method incorporates effects of nebular emission lines and
different dust attenuations, although we show that the effects of
these  different models are largely mitigated when the parameters
are derived from posterior probability densities.  This method is shown to have several
advantageous over using best-fit parameter values from SED fits,
including the fact that our method recovers stellar masses and SFRs
from a SAM mock catalog.    

We use the stellar masses and SFRs derived from the CANDELS photometry
to study the evolution in the SFR--mass relation, SFR evolution, and
sSFR evolution for galaxies at $3.5 < z < 6.5$.   Our conclusions are
the following:

\begin{itemize}
\item \refereetwo{The ability to recover the slope, normalization, and scatter of 
the SFR--stellar mass relation is tested by taking advantage of mock 
catalogs and synthetic photometry of SAM galaxies. With a control
sample of simulated data, we show that our Bayesian SED fitting
procedure can well recover the SFR--stellar relation from complex 
star formation histories, although these tests are limited to the stochasticity
of the histories in the simulations. Moreover, our procedure is less sensitive to 
stellar population template assumptions (such as the inclusion of nebular emission 
and the type of assumed attenuation prescription) than traditional methods. }

\item We find that from $3.5 < z < 6.5$ the star-forming galaxies in
CANDELS follow a nearly unevolving correlation between SFR and stellar
mass, parameterized as SFR $\sim$ M$^a$ with \referee{\refereetwo{$a$} = $0.54 \pm 0.16$ 
at $z\sim 6$ and $0.70 \pm 0.21$  at $z\sim  4$}. The observed
scatter in the SFR--stellar mass relation is small for galaxies with
$\log M_\star/\msol > 9$ dex at all redshifts, \refereetwo{at least on
timescales associated with UV SFRs}.  This evolution requires a star formation history
that increases with decreasing redshift (on average, the SFRs of
individual galaxies rise with time).  \refereetwo{We note that our redshift range
limits the data to cover rest-frame UV-to-optical wavelengths, and we 
defer the search for an underlying dust obscured population for future work.}

\item \referee{Comparing the observed $\log$ SFR scatter at fixed stellar mass
with the scatter due to measurement uncertainties, the true intrinsic scatter
is as much as  $\sigma(\log$ SFR$)=$ 0.2$-$0.3 dex at all 
masses and redshifts.  Assuming that the
SFR is tied to the net gas inflow rate (including gas accretion and
feedback), then the scatter in the gas inflow rate for star-forming
galaxies over our stellar mass and redshift range is equally smooth,
$\sigma$($\log$ \.{M}$_{\text{gas}}) < 0.2-0.3$ dex,  at least when
averaged over the timescale of \refereetwo{UV SFRs} ($\sim$ 100 Myr)}. 

\item We measure the evolution in the SFR for galaxies from $z  =  6.5$ to
3.5 using an evolving \nd\ selection to measure the evolution in
galaxy progenitors that accounts for mergers between halos and
variations in halo growth factors.    For galaxies with $\log
M_\star/\mathrm  {M}_\odot = 10.2$ dex at $z  =  4$, the star formation history follows 
\referee{$\Psi$/\Msol\   yr$^{-1}$  =  $ (t/\tau)^{\gamma}$ with 
$\gamma = 1.4 \pm 0.1$ and $\tau = 92 \pm 14$ Myr
from $z  =  6.5$ to $z  =  3.5$.} We further show that this star formation history reproduces the
measured SFR--mass relation for galaxies at this mass and redshift.  

\item We show that the sSFR gradually increases with increasing redshift from
$z  =  4$ to $z  =  6$, with only small qualitative differences if 
galaxies are selected at fixed stellar mass or by using an evolving \nd.  Broadly,
the evolution in the sSFR is consistent with the theory of cosmological
gas accretion where the SFR follows the net gas accretion rate.

\end{itemize}

\section*{Acknowledgements}


We acknowledge our colleagues in the CANDELS collaboration for very
useful comments and suggestions.  We also thank the great effort of
all the CANDELS team members for their work to provide a robust and
valuable data set.   This work is based on
observations taken by the CANDELS Multi-Cycle Treasury Program with
the NASA/ESA \emph{HST}, which is operated by the Association of Universities
for Research in Astronomy, Inc., under NASA contract NAS5-26555. This
work is supported by \emph{HST} program No. GO-12060. Support for Program
No. GO-12060 was provided by NASA through a grant from the Space
Telescope Science Institute, which is operated by the Association of
Universities for Research in Astronomy, Incorporated, under NASA
contract NAS5-26555.  This work is based on observations made with the
\emph{Spitzer Space Telescope}, which is operated by the Jet
Propulsion Laboratory, California Institute of Technology under
contract with the National Aeronautics and Space Administration
(NASA).  The authors acknowledge the Texas A\&M University Brazos 
HPC cluster that contributed to the research reported here. 
URL: http://brazos.tamu.edu. PB and RW
received funding from HST-AR-12838.01-A, provided by NASA through a
grant from the Space Telescope Science Institute, which is operated by
the Association of Universities for Research in Astronomy, Inc., under
NASA contract NAS5-26555. PB was also supported by a Giacconi
Fellowship through the Space Telescope Science Institute. 

{\it Facilities:} \facility{HST (ACS, WFC3)}, \facility{Spitzer (IRAC)}, \facility{VLT (ISAAC, HAWK-I)}

\appendix \label{sec:AppendixA}

\section{A. Changing the Assumed Star Formation
  History}\label{sec:AppendixSFH}

\begin{figure*}[!h]
\epsscale{1.1}
\centerline{\includegraphics[scale=0.55]{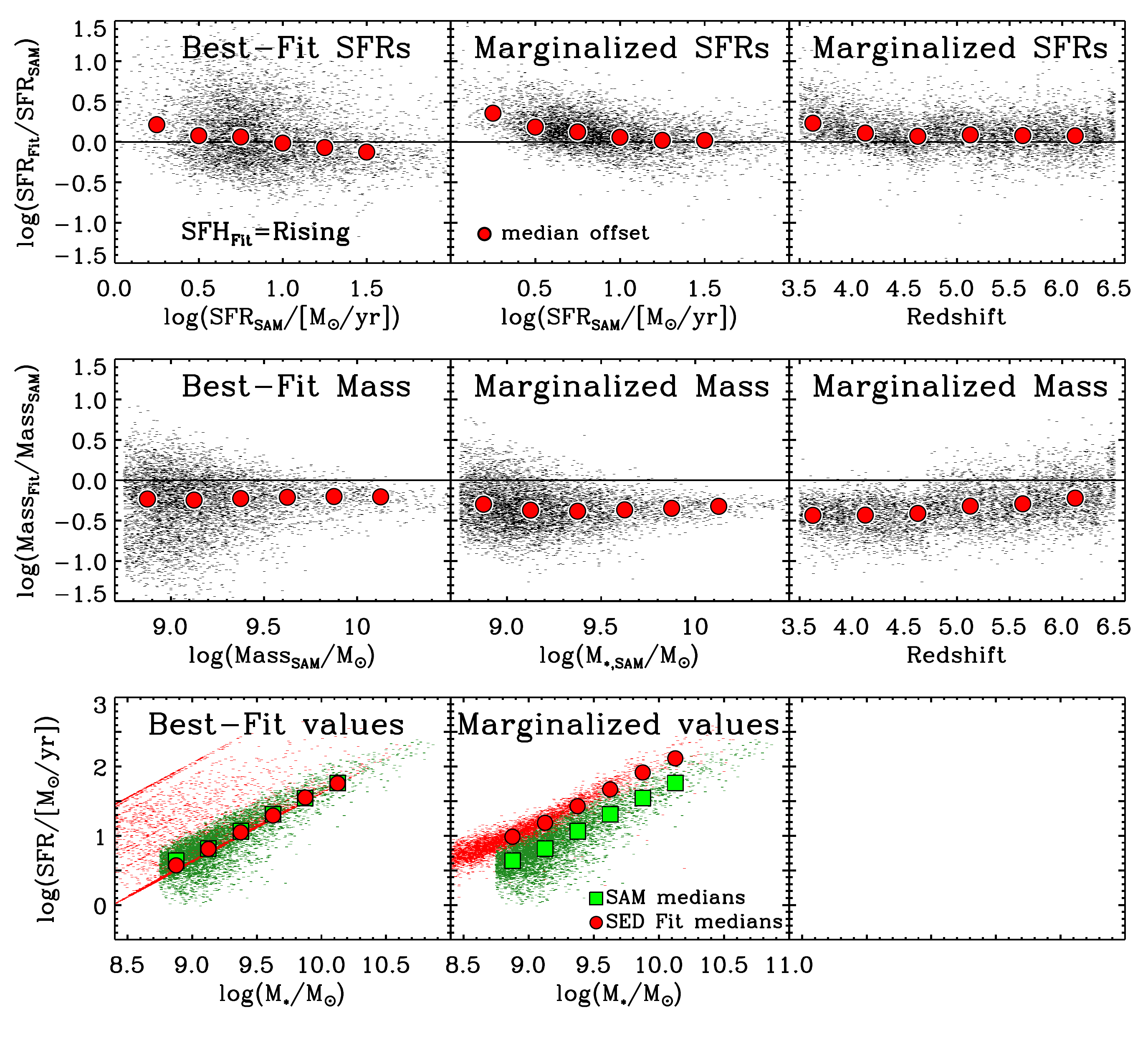}}
\caption{Results of SED fitting to synthetic photometry of  recent SAM
mock catalogs (Somerville et al).,  now fitting to templates with
exponentially rising star formation histories (similar to Fig.~\ref{fig:SAMcompare}).  
  The top panels show the log 
difference of measured-to-true SFRs.   The derived SFRs  are systematically higher
than the true values, with high scatter.   The middle panels show the ratio of the
measured-to-true stellar masses. The stellar masses are systematically lower.  
 The bottom panels show that the derived SFRs and stellar
masses do not well recover the SFR--mass relation in the models when using exponentially rising
star formation \refereetwo{histories.}   \vspace{3mm}}
\label{fig:AppendixLightcone}
\end{figure*}

The SFRs and stellar masses used in this work are derived using SPS
models with constant SFRs.   In contrast, one of the main results of
this work is that the high-redshift galaxies experience SFRs
 that increase monotonically with time, $\Psi \propto
t^{1.4\pm0.1}$.   Naively, it would seem these statements are in
conflict.  Here we discuss how the star formation history affects the
model interpretation, and we offer justification for the use of
models with constant SFRs.  

First, we tested SED fits with our Bayesian formalism that marginalize
over a range of star formation histories, including those that rise
with time, with $e$-folding times, $\tau = 50$, 100, 300 Myr, and  100
Gyr (the long $e$-folding time approximates a constant SFR and keeps all
the models handled identically in normalization by the BC03 software), 
where the star formation history is then defined as
$\Psi \sim \exp^{t/\tau}$.    We then use our Bayesian formalism to
derive model parameters using the synthetic photometry from the
galaxy mock catalog using the Somerville et al. SAM.

Figure \ref{fig:AppendixLightcone} shows the SFRs and stellar masses
measured from the synthetic photometry compared to the true model
values using this large range of model parameters.  This figure can be
compared directly to Figure~\ref{fig:SAMcompare} above, which used
models assuming only a constant star formation history.    In comparison,
the model parameters derived using the suite of histories that include
rising SFRs produce stellar masses that are skewed low and SFRs that are skewed high
compare to the true values.   This appears to be unphysical. 

\begin{figure}
\epsscale{1.1}
\centerline{\includegraphics[scale=0.45]{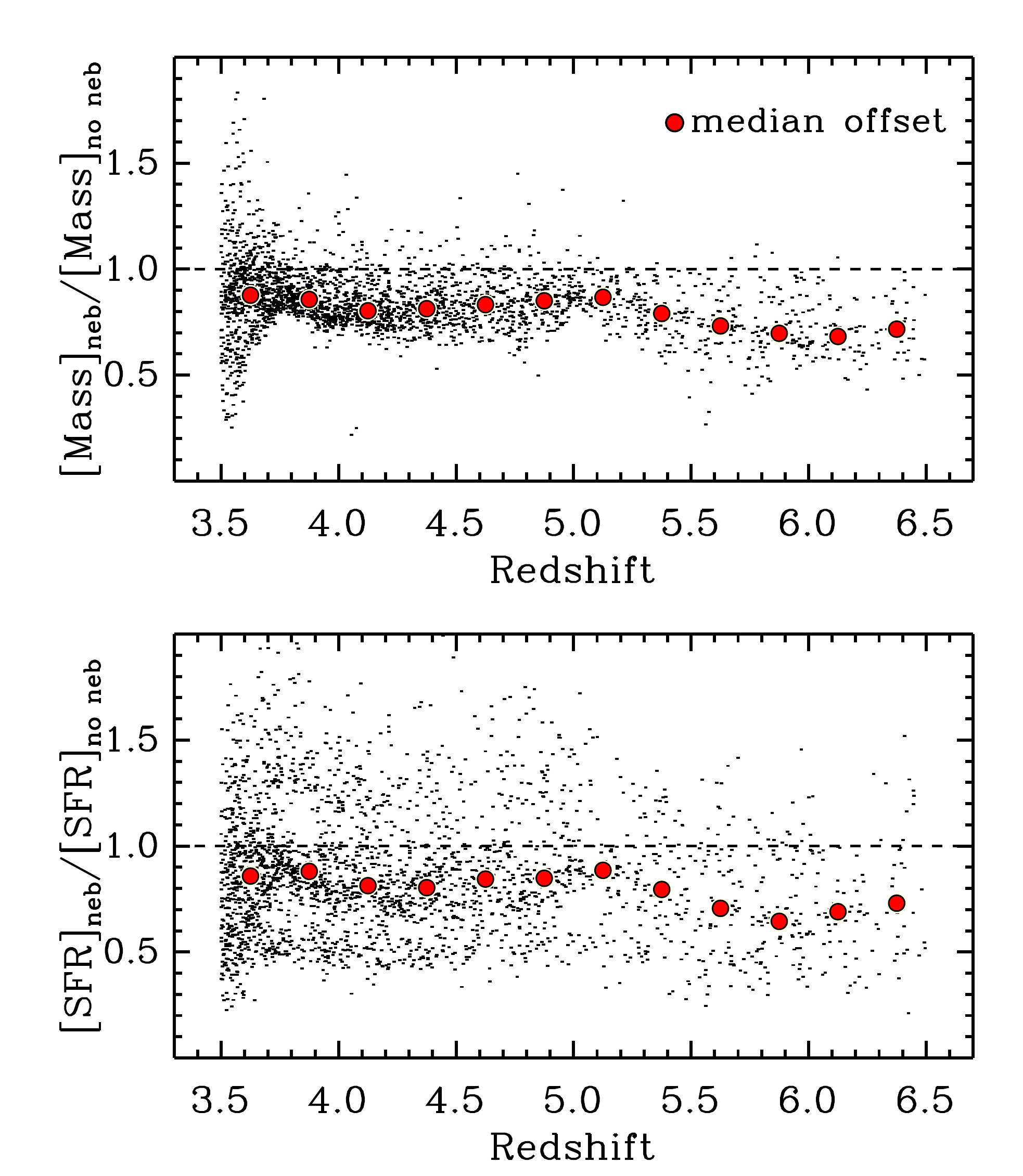}}
\caption{\emph{Top:} The ratio of the best-fit stellar
masses from model fits that include the effects of nebular emission
lines (e.g., $f_\mathrm{esc}=0$) and exclude emission lines (e.g.,
$f_\mathrm{esc}=1$) for the objects in our CANDELS sample.  Black
points show the ratio of the best-fit values as a function of
redshift, and the large red points show the medians in bins of
redshift.  \emph{Bottom:} Ratio of best-fit SFRs from model fits with
and without nebular emission lines.  For galaxies with  redshifts that
place strong emission lines in one or more of the bandpasses (see
Fig.~\ref{fig:photbyredshift}), the best-fit SFRs and stellar masses can be reduced
by up to 0.5 dex (factor of 3), with a median of 0.3 dex (factor of
2).  In contrast, our Bayesian formalism finds that including nebular
emission has only a small effect on the derived stellar masses and
SFRs (Fig.~\ref{fig:CompareBestFitMarg}).   \vspace{3mm}}
\label{fig:massbyredshift}
\end{figure}

In hindsight, the reason for this is the following.  In the BC03
models, the star formation history parameterization uses $e$-folding
timescales (motivated by the exponentially decreasing models
pioneered by \cite{Tinsley80}) and therefore the
SFRs rise \textit{exponentially} in time.  Therefore, older stellar
populations have unphysically increasing SFRs. At these late times,
the models increase much faster than supported by observations 
\citep[e.g.,][]{Papovich01} or seen in simulations \citep[e.g.,][]{Finlator06}.
As a result, models with exponentially increasing SFRs must
underproduce the stellar mass and overproduce the SFR to match the
observed SED of galaxies.     In contrast, the star formation
histories of galaxies in our redshift range and in simulations can be
approximately accurately as constant when averaged over the past
 $\sim$ 100 Myr, which includes the recently formed stars that dominate the
luminosity-weighted age \citep[see, e.g.,][]{Finlator06}. As a result, we do not consider the
assumption of  a constant star formation history to be a significant factor on the
conclusions of this work. We note parenthetically that
this is only true for the galaxies observed in this work because they are all
heavily star-forming with high sSFRs;  quiescent galaxies
would require declining star formation histories.   

Furthermore, our tests (discussed above in \S~\ref{sec:SAMcompareSEC} and
Figure~\ref{fig:SAMcompare}) show that SED fits using models with constant
star formation reproduce accurately the SFRs and stellar masses from
the models.   Therefore, this work fixes
the star formation history in fitting templates to be constant during the analysis of the 
CANDELS data.  As the conclusions show, the galaxy populations require a
star formation history with a rising SFR, but this evolution is slow
and monotonic, $\Psi \propto t^\gamma$ (where $\gamma= 1.4  \pm   0.1$ 
from \S~\ref{sec:cnd} and Figure~\ref{fig:CNDresult}), 
which is not currently supported in a simple
parameterization in the BC03 models.  Nevertheless, in a future work, we will
explore possible improvements in parameters using models with
star formation histories that increase as a power law in time, 
as well as other more complex star formation histories. 

\section{B. Stellar population synthesis fitting: Comparing Best-Fit Results}\label{sec:Appendix_bestfits}
We compare between using the best-fit results from SED fitting procedures 
to using our Bayesian formalism to derive parameters from their posterior probability
densities.  In  each subsection, we observe the offsets under different fitting template 
assumptions, including the effect of
nebular emission lines and dust-attenuation prescription.

\subsection{B.1 Effect of including nebular emission}\label{sec:nebeffect}

The effects of including effects of nebular emission to the stellar templates are
largely mitigated when using our Bayesian formalism (see
Figure~\ref{fig:CompareBestFitMarg}).   In contrast, the nebular emission lines can
strongly affect stellar masses and SFRs derived from best-fit model
parameters.  Figure~\ref{fig:massbyredshift} shows that in the
redshift range of our galaxy sample, the presence (or absence) of an
emission line in the \ks, [3.6], and [4.5] bandpasses  results in
best-fit models that typically have lower SFRs and stellar masses by
up to 0.5 dex  (factor of 3).   This is because a galaxy with a redder
rest-frame UV-optical color requires either an older stellar
population or heavier dust attenuation, with higher $M_\star/L$ ratios.
Models where emission lines are present in the redder passbands
reproduce the redder rest-frame UV-optical colors with lower stellar
masses and SFRs.  The effects of nebular emission are strongest in the redshift
range that strong emission lines are present (see Fig.~\ref{fig:photbyredshift}).
The median decrease is up 0.3 dex (factor of 2) in both SFR and stellar mass for 
$5.5 < z < 6.5$.  As discussed in the literature, this can affect the
interpretation of the galaxies \citep{ Schaerer09,Schaerer10,Schaerer13,Ono10, 
Finkelstein11, Curtis-Lake13, deBarros14, Reddy12, Stark13, Tilvi13}.


It is worth noting that strength of the nebular emission lines is
highly uncertain for several reasons. One clear reason is that the
model must use some escape fraction of ionizing photons, which is
allowed to range between $0 \leq f_{\text{esc}} \leq 1$.  Another
uncertainty is that in our parameterization, the strength of line
emission is tied to the age of the model stellar population, and age is
less constrained in SED fitting (see, e.g., \cite{Papovich01} and
Fig.~\ref{fig:MargPDF} above).   Fits to galaxies using models with a starburst-like
attenuation prescription \citep{Calzetti00} produce age distributions skewed
heavily to the low ages (in some cases unphysically low as the
ages are much less than a dynamical time, \citep[e.g.,][]{Yan06, Eyles07,
Schaerer09, deBarros14,Reddy12}).   Both the escape fraction 
and inferred ages are poorly known quantities which cause the effect of nebular
emission lines to likewise be poorly constrained, \citep[see
also,][]{Verma07, Oesch13}.    For these reasons it is advantageous to
use SED fitting that is not strongly influenced by the presence or
absence of such lines.  As we show above (\S~\ref{sec:CompareMargDustNeb} 
and Fig.~\ref{fig:CompareBestFitMarg}), our Bayesian formalism is less affected 
by nebular emission lines in the models.  Therefore, for the analysis in this
paper we adopt models with $f_{\text{esc}} = 0$.  In a future work, we
will consider fully marginalizing over a range of escape fraction,
although it seems unlikely to change the conclusions here.

\begin{figure*}
\epsscale{1.1}
\centerline{\includegraphics[scale=0.44]{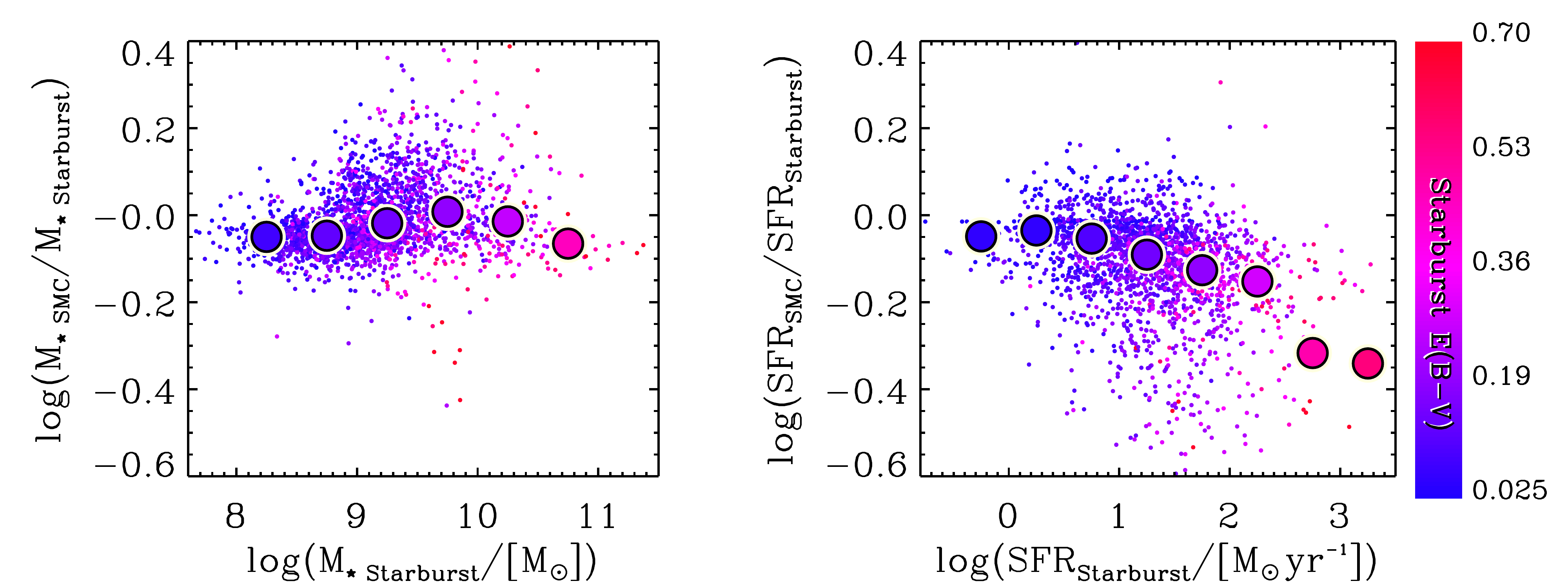}}
\caption{\emph{Left:}The ratio of stellar masses from best-fit models
that assume SMC-like attenuation to those that assume a starburst-like
attenuation as a function of best-fit mass.  Small points show
individual galaxies, where the color denotes the attenuation from the
best-fit model assuming starburst-like dust.  The large points show
medians in bins of $d \log  M_\star/\mathrm{M}_\odot$ = 0.5 dex.  (We have excluded
showing objects with best-fit models that have zero reddening,
$A_\mathrm{UV}$=0, as these lie on the unity line.)  At lower masses, the
extinction prescription has little effect on the best-fit masses.  At higher
masses, $\log  M_\star/\mathrm{M}_\odot >$ 10 dex, there is weak trend in that the best-fit
models with SMC-like dust have lower stellar masses ($\sim$0.1 dex at
$\log  M_\star/\mathrm{M}_\odot\ \gtrsim$ 10.8 dex).  This is likely related to the fact
that higher-mass galaxies appear to have higher reddening (so
presumably lower mass, dusty galaxies would also have the same trend
in mass).  \emph{Right:}  Ratio of the SFRs for the same best-fit
models.    Here there is a strong trend in SFRs from the best-fit
models:  as the SFR increases, there the models with starburst-like
dust have significantly higher SFRs, by up to a 0.5 dex, with a median
of $\sim$0.3 dex (factor of 2).  Clearly the choice of dust
attenuation prescription affects the interpretation of galaxy SFRs in the
best-fit models.  Nevertheless, as we show above in 
Figure~\ref{fig:CompareBestFitMarg},
in our Bayesian formalism these differences are mitigated, and the
dust attenuation \refereetwo{prescription} has negligible impact on the results here.    
\vspace{3mm} }
\label{fig:SMCMassSFR} 
\end{figure*}

%
%
%

\subsection{B.2 Dependence on attenuation prescription}\label{sec:SMCeffect}
Figure~\ref{fig:SMCMassSFR} shows the effects of the dust attenuation
prescription on the stellar masses and SFRs from the best-fit models.  The
choice of dust attenuation prescription has a weak effect on stellar mass,
where models using the SMC-like attenuation prescription have lower stellar
masses by $\sim$0.1 dex in the median compared to the starburst-like
dust attenuation (although the spread is larger, up to $\pm$0.2 dex).  

The choice of dust attenuation prescription has a much stronger  effect on the
SFRs derived from best-fit models. Figure~\ref{fig:SMCMassSFR} shows
that there is a strong trend in SFRs from the best-fit models:  as the
SFR increases, the models with starburst-like dust have
significantly higher SFRs, by up to a 0.5 dex, with a median of
$\sim$0.3 dex (factor of 2) compared to the best-fit solutions using an
SMC-like dust.  This is due to a combination  of effects. First, there
are high degeneracies between the  inferred attenuation and age that
arise  from broadband SED fitting (discussed in the previous Appendix
subsection).   The assumed SFR depends on the stellar population age,
especially at lower ages, where this leads to much higher ratios of
in the SFR/$L_{1500}$ ratio \citep[see Appendix of][]{Reddy12}
compared to the value typically assumed in the \cite{Kennicutt98}
relation which assumes ages greater than $10^8$ yr.   

Dust extinction and age are degenerate (negatively covariant)
in the SED modeling and models with starburst dust typically
have lower best-fit ages (Papovich et al.\ 2001). Therefore, the effects of
higher dust attenuation and higher SFR$/L_{1500}$ ratio both
contribute to a larger SFR for the case of starburst-like dust.   The
differences between starburst-like and SMC-like dust models are
highest for models with highest SFRs, as shown in
Figure~\ref{fig:SMCMassSFR}.  For objects with the  highest SFR, the
difference between the two  prescriptions can be as high as $\sim 0.7$
dex.  Therefore, for best-fit models \emph{an assumed SMC-like
attenuation causes starburst-attenuation-derived young, dusty, and
high-SFR objects to be older, less dusty, and with  lower SFR}.  This
also reinforces the result of nebular emission in the Appendix above
 that simple template assumptions  can significantly
impact the best-fit SFR.   Nevertheless, as we show above in
Figure~\ref{fig:CompareBestFitMarg}, in our Bayesian formalism these differences are
mitigated, and the dust attenuation \refereetwo{prescription} has negligible impact on the
results here. 

For these reasons, this work assumed an SMC-like attenuation
for all objects in our sample. If we otherwise had
very reliable age estimates we might assume, as
by \cite{Reddy12}, an ``age''-dependent attenuation prescription, 
such that younger galaxies have SMC and older have starburst
attenuation. However, the effects of nebular emission and
assumed attenuation are strongest when adopting best-fit values.  In
contrast, we mitigate these effects by using the results from our
Bayesian analysis, where the effects of changing dust attenuation 
prescription are minimized (see \S~\ref{sec:PDF}).

\subsection{B.3 Difference of Marginalized SFR compared to traditional methods}\label{sec:AppendixSFR}

\begin{figure}
\epsscale{1.1}
\centerline{\includegraphics[scale=0.44]{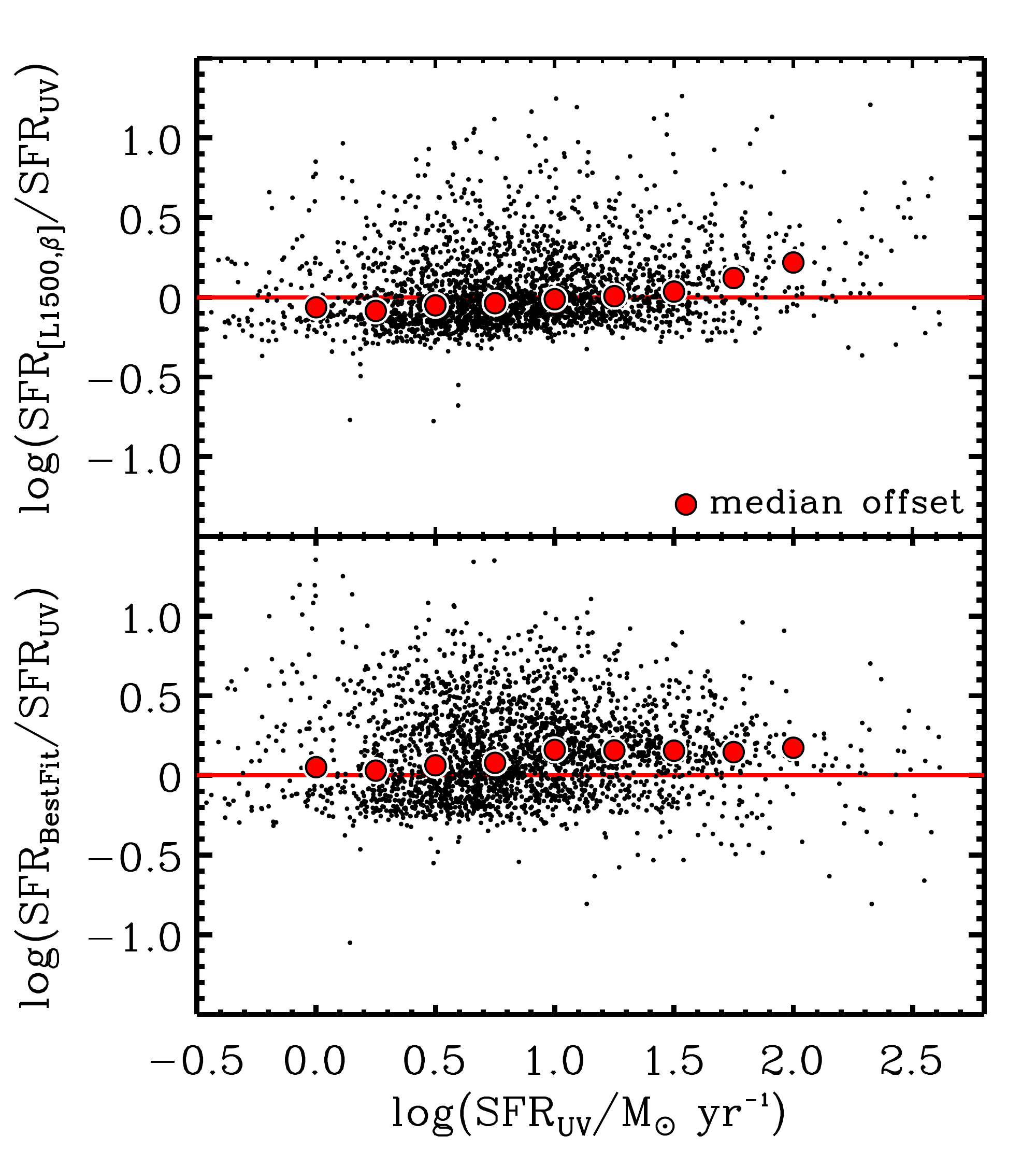}}
\caption{A comparison between different methods to compute the
SFR. The abscissa shows our adopted  method, which uses the SFR from
the attenuated luminosity of the photometric measurement closest to
rest-frame  1500~\AA, corrected for dust attenuation using the median
$A_\mathrm{UV}$ from the marginalized posterior PDF (see Equation
\ref{equ:sfr}).  The ordinate of the top panel compares our SFR to the
SFR derived from the UV luminosity at  L$_{1500}$ and the UV slope,
$\beta$, to correct for dust  attenuation.  The ordinate of the bottom
panel compares our SFR toe the SFR from the best-fit stellar
population model.  In both cases the alternative SFRs show a large
scatter, which can lead to significant biases.  We favor using our
method because our method
reproduces the SFRs from the SAMs (see \S~\ref{sec:SAMcompareSEC} and
 Fig.~\ref{fig:SAMcompare}), reducing some of this bias.  
\vspace{3mm}}
\label{fig:SFRcompare}
\end{figure}

The method used in the work to derive UV SFRs for high redshift
galaxies is different from the typical methods found in the literature.
The common methods include using a dust correction based on the UV
spectral slope, $f_\lambda \sim \lambda^\beta$, \citep[][]{Meurer99, 
Madau98, Kennicutt98} and using the instantaneous SFR from the
best-fit stellar population model.  As shown in Figure~\ref{fig:SFRcompare}, 
both of these alternative methods show high scatter when compared to 
the marginalized SFR from the method of this work.

The large scatter in Figure~\ref{fig:SFRcompare} is due to the fact
that the scale of SFR is predominantly dependent on the treatment of 
the dust correction to UV luminosity, which is an unconstrained
quantity in traditional SED fitting methods at high redshift. The
Bayesian approach has the advantage in producing realistic ages,
thereby reducing degeneracies with dust corrections. 
The median age for the SAM mock catalogs
is $\log (t_{\text{age}})$= 8.48 $\pm$ 0.22 dex which 
resembles the distribution of marginalized ages found in this work
for observed galaxies, $\log (t_{\text{age}})$= 8.73 $\pm$0.14 dex.
Conversely, the distribution of best-fit ages is typically very dissimilar 
from SAM and simulation  predictions. This is because best-fits typically 
find lowest $\chi^2$ at the extreme end of parameter space (youngest ages, 
highest  extinction). As a result, the median best-fit ages are lower with higher
scatter $\log (t_{\text{age}})$= 8.06 $\pm$ 0.94 dex. This
scatter results from the degeneracies between the young, dusty
and old, dust-free solutions of a given SED, and photometric
uncertainties (which affect the accuracy of measuring the UV spectral
slope, $\beta$) thus producing a wide range of acceptable SFRs.  

In summary, when marginalizing over other nuisance parameters, the posterior on age returns 
more physical ages on the order of  $\sim$350--750 Myr, reducing degeneracies in the
derived dust corrections and thereby reducing the uncertainty in SFR. 
In addition, the method used in this work reproduces SFRs from
semi-analytic models (see \S~\ref{sec:SAMcompareSEC}), and is
relatively unaffected by model variations such as extinction prescription
and/or nebular emission lines. This is additional 
evidence to favor the Bayesian approach to derive physical properties.

\begin{figure*}
\epsscale{1.1}
\centerline{\includegraphics[scale=0.65]{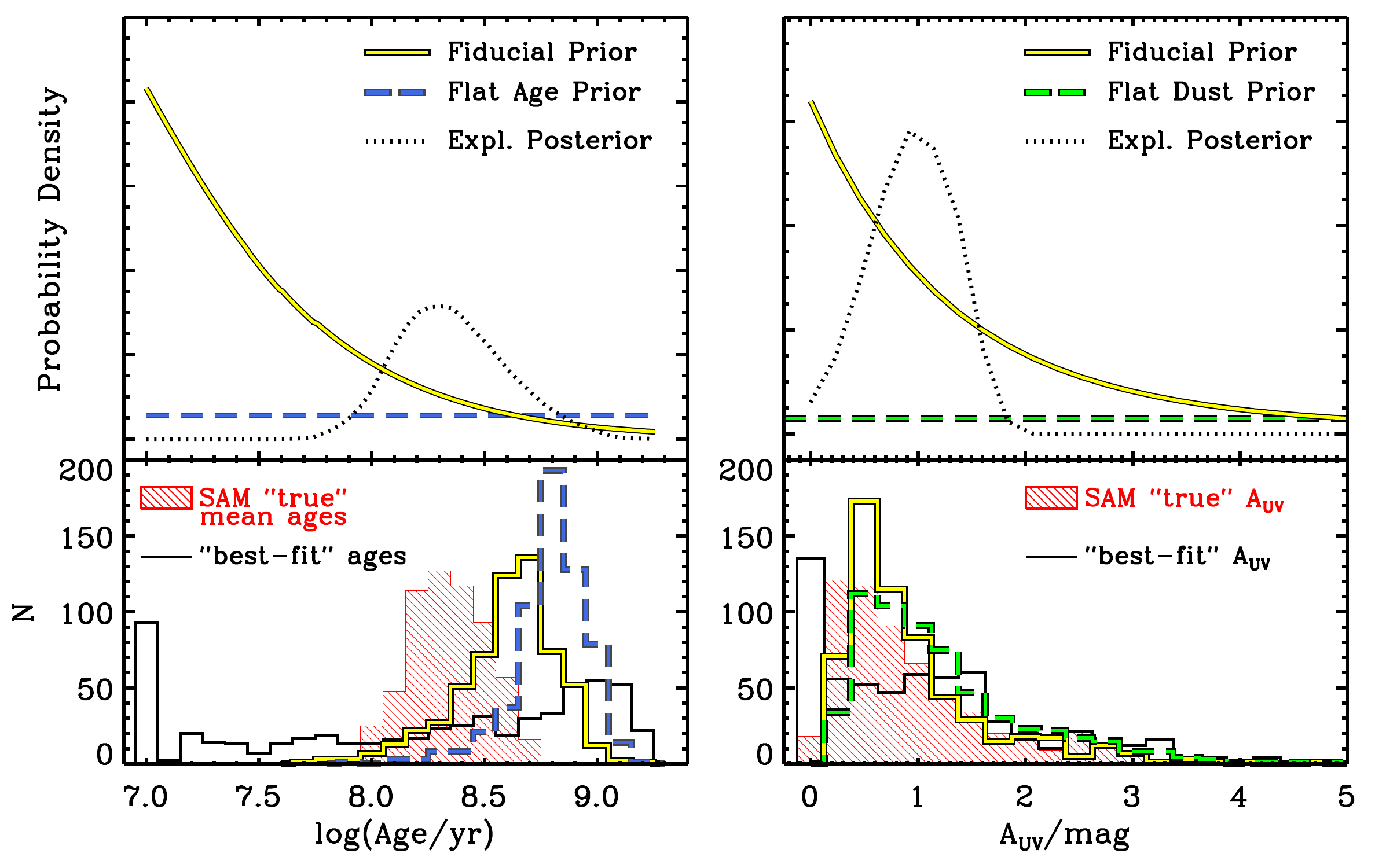}}
\caption{\refereetwo{ \emph{Top:} the shapes of priors as a function of $\log$(age) (left) and 
UV attenuation (right). Our ``fiducial'' prior is what we adopt for the results of this work, while
the flat priors are tests to study how the choice of prior affects the results. The dotted lines show 
the posterior of a given parameter for a single, example object assuming our fiducial prior. 
\emph{Bottom} histograms of the inferred stellar population ages for
a sample of 600 control objects from the Somerville et al. SAM. The red distribution is the ``true''
distribution from the SAM, while the yellow (solid, fiducial) and blue (dashed, flat) lines show 
the recovered distribution after fitting to the SAM fluxes assuming different priors. The thin black
line shows the recovered values assuming the maximum likelihood solution, or ``best-fit''. This figure
emphasizes that best-fit solutions do not well recover the distribution of input ages and attenuations, and
that our fiducial prior is preferred over a flat prior to recover stellar ages.}
\vspace{3mm}} 
\label{fig:Appendix_prior_hist}
\end{figure*} 

\begin{figure*}
\epsscale{1.1}
\centerline{\includegraphics[scale=0.65]{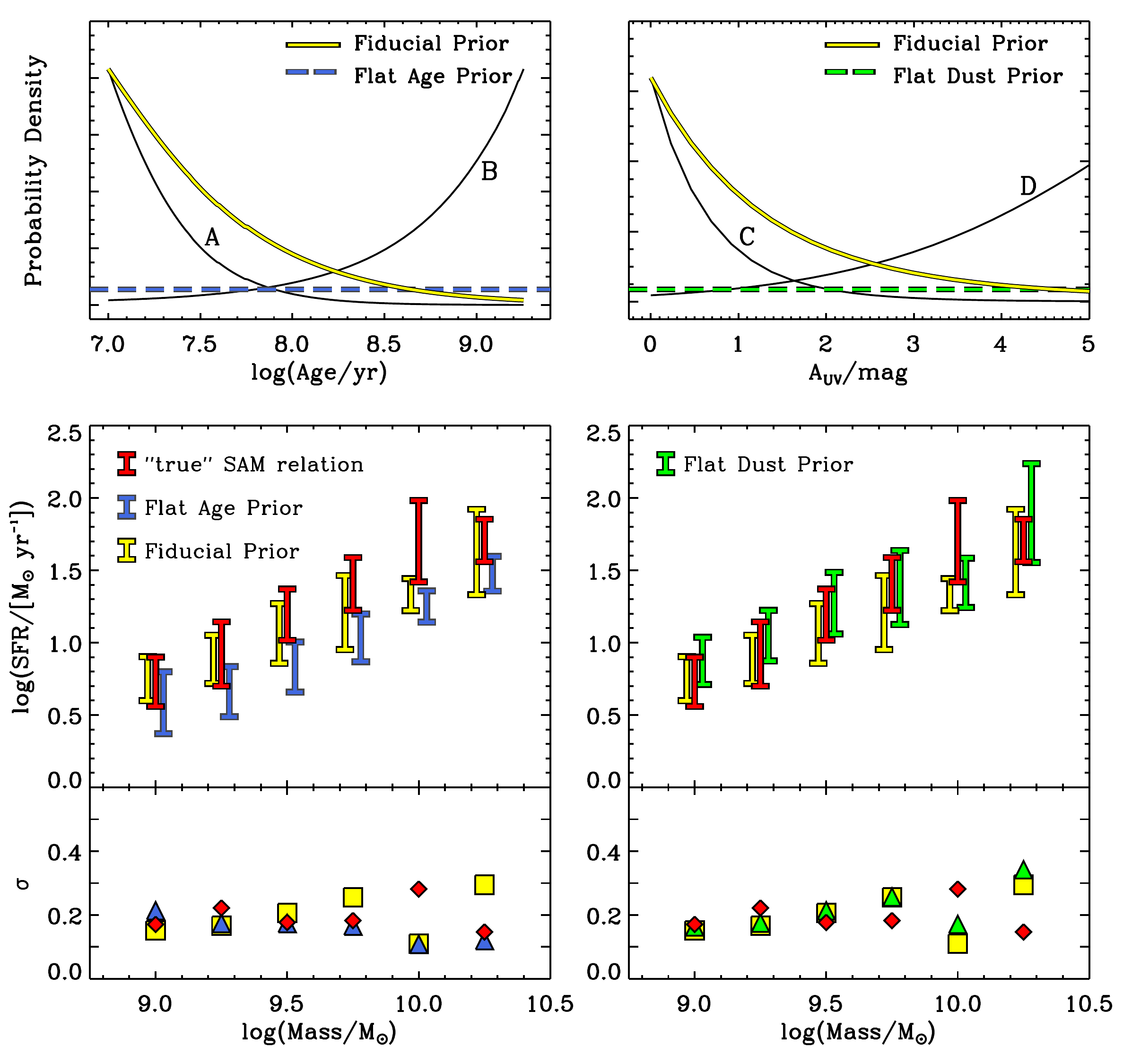}}
\caption{\refereetwo{\emph{Top:} the same as the top of Fig.~\ref{fig:Appendix_prior_hist}, shown for reference. 
Four additional test priors labeled A--D are shown and referred to in the text.
\emph{Middle} The slope and $\sigma$ scatter of the SFR--stellar mass relation for a sample of 600 control objects from the 
Somerville et al. SAM. The red is the ``true'' scatter from the SAM, while the yellow (solid, fiducial) 
and blue (dashed, flat) is the scatter after fitting to the SAM fluxes assuming different priors 
on age (left) and attenuation (right). 
\emph{Bottom:} the $\sigma$ scatter of SFR in each stellar mass bin. Squares
and triangles represent the scatter assuming our fiducial prior or a flat prior, respectively, while the diamonds 
represent the scatter of the input objects. This figure shows the recovery of the SFR--stellar mass relation 
under different priors, and that our fiducial prior is preferred over flat priors to recover the
input SFR--stellar mass relation.} 
\vspace{3mm}} 
\label{fig:Appendix_prior_sfrmass}
\end{figure*} 

\section{C. Derivation of Prior}\label{sec:AppendixPrior}

Here we describe the prior used in our SED fitting procedure and 
discuss tests to validate its use. The prior used in equation~\ref{equ:Bayes} 
was chosen  to allow for an analytic derivation to the posterior probability densities 
that is straight-forward to calculate for each of the stellar population 
parameters, i.e., $p(t_{\text{age}} | D)$.
Our prior-likelihood pair is therefore more easily computable (time efficient) 
and does not require more sophisticated methods such as a Markov Chain Monte Carlo.
\refereetwo{Our ``fiducial'' prior, which is used for the results of this work,
 is can be expressed} as a sum over $i$ bands as,

\begin{equation}
p(\Theta')\ =\ \left(\sum_{i=1}^nf_{\Theta}^2(\lambda_i)/\sigma_i^2\right)^{1/2}
\label{equ:prior}
\end{equation}

where $\Theta'$ represents the 
entire set of parameters, 
${ \Theta'\ = (\Theta\{t_{\text{age}}, \tau, A_{\text{UV}}\}, \Mstar)}$, and $f_{\Theta}$
is the model flux, unscaled by stellar mass.  We express $\Theta$ as a separate parameter set
from stellar mass because the prior in equation~\ref{equ:prior} is dependent on the fluxes 
of the gridded set of models, and independent of mass. 

This prior is ``flat'' with respect to stellar mass, but spans the range of masses 
that the stellar population models can produce for a given set of data.  Formally, the prior 
does depend on the other model parameters and the photometric uncertainties 
that are used to construct the stellar population synthesis models (age, dust, and 
star formation history).  This is because the models are constructed 
using a discretized grid of stellar population parameters and 
a normalization that gives the mass. \refereetwo{Since the prior is dependent on the 
fluxes of the gridded set of model parameters but not stellar mass, we distinguish 
$\Theta$ and \Mstar\ separately in our equations.
The shape of this fiducial prior is shown in 
the top panels of Figures~\ref{fig:Appendix_prior_hist} and \ref{fig:Appendix_prior_sfrmass}. }

\refereetwo{In order to confirm that our posterior is constrained by the data and not 
dominated by the prior, we conducted several tests. First, we steadily increased the photometric uncertainties
on the ``mock'' data from the SAM (lowered the S/N) and used that data as input to our procedure
(a similar process as in S~\ref{sec:SAMcompareSEC}). This allows us to search for a characteristic
S/N or stellar mass at which the SAM input values are poorly recovered, where here we adopt 
``poorly recovered'' to mean systematically discrepant by a factor of 3 (0.5 dex) as compared 
to the input SAMs. We find that at S/N = 1 the recovered stellar masses of galaxies with 
log \Mstar/\Msun = 9.75 are systematically higher by 0.5 dex.  Such low S/N represents 
the scenario where the data have no power and the posterior reverts to the prior. 
Since all detected bands are typically of S/N $\gg$ 1, this test confirms that it is the likelihood 
computed from the data that is driving the shape of the posterior, and provides evidence that the 
conclusions of this work are not dominated by the prior.}

\refereetwo{In another test, we explored the effects of changing the assumed prior. We conducted 
this test on a control sample of 600 SAM objects that span the stellar masses and SFRs in the 
SAM for $z >$3.5. We then observed how well we could recover the age and attenuation distributions 
of the SAM when we changed the assumed prior to be flat in age or flat in dust. 
Figure~\ref{fig:Appendix_prior_hist} shows that using alternative priors shift the distributions of each parameter,
but the prior does not overwhelm the likelihood from the data. For example, the flat age prior pushes the 
recovered age distribution away from the input age distribution because the flat age prior assigns more 
weight to older solutions than the fiducial prior. For both priors, the age distributions are old compared 
to the ``true'' values from the SAMs, but this is possibly a result of the differences between our assumed 
(slowly varying) star formation histories and the ``physical'' ones from the SAM.  We consider the 
agreement between the ``true'' ages and recovered ages to be good.}

\refereetwo{The attenuation distributions are much less sensitive to the choice of prior.  
Figure~\ref{fig:Appendix_prior_hist} shows that there are only subtle differences between 
the distribution in $A_{\text{UV}}$ using our fiducial prior compared to those when using a flat prior. 
The figure also shows an example (``Expl.'') posterior for one object in the sample. This object 
shows that the probability density does not follow the prior.  Moreover, this figure shows how 
either the fiducial or flat priors (for either parameter) do better at recovering the input distributions 
than the common method of taking the maximal likelihood, or ``best-fit'', model.}

\refereetwo{Finally, we test how the above priors impact the recovery of the slope and scatter of the 
SFR--stellar mass relation for the control sample of SAM objects. 
Figure~\ref{fig:Appendix_prior_sfrmass} shows the recovered $\sigma_{\text{MAD}}$ scatter about
the median SFR in bins of stellar mass, calculated in the same manner as in Figure~\ref{fig:massSFR},
but assuming different priors. 
The flat age prior shifts the distribution to older ages, and this 
effect propagates to SFR--stellar mass relation, shifting the SFR distribution to lower values. 
The flat age prior produces an SFR--stellar mass relation that is tightened and lower in normalization. 
One reason we disfavor the flat age prior is that the scatter in the SFR--stellar mass relation is even tighter 
than for our fiducial prior, and therefore the results from the fiducial prior are more conservative. 
Switching to a flat dust-prior has little effect on the slope and scatter in the SFR--stellar mass 
relation, given the scatter. We see similar effects on the results from the data when
switching to these priors. }

\refereetwo{The top panels of Figure~\ref{fig:Appendix_prior_sfrmass} also show a suite of alternative
priors (labeled A-D) that were applied to the SAM control sample. 
A prior younger than our fiducial, such as prior A, assigns more weight to 
the likelihood of high SFRs at a given mass. This creates a scenario 
where galaxies are preferred to be young, maximal starbursts, causing the 
SFR--stellar mass relation to be artificially higher in normalization by $\sim$0.25 dex than 
the input SAM relation with an inflated scatter ($\langle \sigma \rangle >$ 0.05 dex) 
due to a wider range in mass-to-light ratios. Conversely, prior C and a ``flat''
age prior assign more weight to old age solutions than our fiducial prior. 
This results in a more narrow range of mass-to-light ratios, and therefore 
these priors produce a tighter SFR--stellar mass relation (${\langle \sigma \rangle 
\sim  0.16  }$ dex) with 
lower normalization (lower SFR at a given mass than the input SAMs 
by $\sim$0.4 dex). On the other hand, our fiducial prior assigns more weight 
to younger age solutions, therefore avoiding an artificial decrease in 
SFR--stellar mass scatter, while not being too strong such that the recovered
SFRs from the SAM are overestimated. We also find our procedure to be 
robust against changes to the attenuation prior, and find little change to the normalization
slope and scatter when assuming the fiducial, flat, C, and D priors. }

\refereetwo{In summary, our tests indicate that the SFR-mass scatter is insensitive to 
the prior on dust and mildly sensitive to the prior on age. However, we conclude 
that our fiducial prior in age best recovers the age distribution and the SFR-stellar mass 
scatter of the SAM control sample and is straightforward to implement. Other priors that 
better reproduce the age distribution or the slope or scatter in SFR-stellar mass
have SFRs at a given mass that are significantly off in normalization 
from the input SAMs. }

\clearpage

\bibliographystyle{apj}
\bibliography{ms}
Gabor14
\end{document}